\documentclass[11pt,a4paper]{article}
\usepackage[T1]{fontenc}
\usepackage[utf8]{inputenc}
\usepackage{lmodern}
\usepackage[margin=25mm,headheight=14pt]{geometry}
\usepackage{amsmath,amssymb,amsthm,mathtools}
\usepackage{booktabs,array,tabularx,enumitem}
\usepackage{microtype,xcolor,fancyhdr,needspace}
\usepackage{soul}
\sethlcolor{yellow}
\newcommand{\mathrev}[1]{\begingroup\setlength{\fboxsep}{1pt}%
  \mathchoice{\colorbox{yellow}{$\displaystyle #1$}}%
             {\colorbox{yellow}{$\textstyle #1$}}%
             {\colorbox{yellow}{$\scriptstyle #1$}}%
             {\colorbox{yellow}{$\scriptscriptstyle #1$}}\endgroup}
\usepackage{tikz}
\usetikzlibrary{arrows.meta,positioning,fit,calc}
\usepackage[colorlinks=true,linkcolor=blue!45!black,citecolor=blue!45!black,urlcolor=blue!55!black]{hyperref}
\usepackage{bookmark}
\hypersetup{pdftitle={Precedence Constraint Scheduling with Unit-Time Jobs on a Fixed Number of Machines},pdfsubject={Klaus Jansen, Univ. Kiel}}
\definecolor{ink}{HTML}{17364D}
\definecolor{accent}{HTML}{087E8B}
\setlist{leftmargin=*,topsep=4pt,itemsep=3pt}
\newtheorem{theorem}{Theorem}[section]
\newtheorem{lemma}[theorem]{Lemma}
\newtheorem{proposition}[theorem]{Proposition}
\newtheorem{corollary}[theorem]{Corollary}
\theoremstyle{definition}
\newtheorem{definition}[theorem]{Definition}

\newtheorem{remark}[theorem]{Remark}
\newtheorem{algorithm}{Algorithm}

\let\originaltextsc\textsc
\renewcommand{\textsc}[1]{\textnormal{\originaltextsc{#1}}}

\usepackage[most]{tcolorbox}
\newif\ifshowrevisions
\showrevisionstrue
\definecolor{revisionyellow}{RGB}{255,245,153}
\definecolor{revisionred}{RGB}{190,0,0}
\let\revisionhl\hl
\newcommand{\revisionmath}[1]{\begingroup\setlength{\fboxsep}{1pt}%
 \mathchoice{\colorbox{revisionyellow}{$\displaystyle #1$}}%
             {\colorbox{revisionyellow}{$\textstyle #1$}}%
             {\colorbox{revisionyellow}{$\scriptstyle #1$}}%
             {\colorbox{revisionyellow}{$\scriptscriptstyle #1$}}\endgroup}
\newcommand{\revchange}[2]{%
 \ifshowrevisions
  \ifmmode {\color{revisionred}#1}\mkern4mu\revisionmath{#2}%
  \else {\color{revisionred}#1}\hspace{0.3em}%
   \begingroup\sethlcolor{revisionyellow}\revisionhl{#2}\endgroup\fi
 \else #2\fi}
\newcounter{revisionSavedEquation}
\newenvironment{revisionold}{%
 \setcounter{revisionSavedEquation}{\value{equation}}%
 \begin{tcolorbox}[enhanced,breakable,sharp corners,
  colback=white,colframe=revisionred,coltext=revisionred,
  colbacktitle=white,coltitle=revisionred,title={Text to remove},
  fonttitle=\bfseries,boxrule=0.4pt,left=3pt,right=3pt,top=3pt,bottom=3pt,
  before skip=6pt,after skip=6pt]
 }{\end{tcolorbox}\setcounter{equation}{\value{revisionSavedEquation}}}
\newenvironment{revisionnew}{%
 \begin{tcolorbox}[enhanced,breakable,sharp corners,
  colback=revisionyellow,colframe=revisionyellow!65!black,
  boxrule=0.4pt,left=3pt,right=3pt,top=3pt,bottom=3pt,
  before skip=6pt,after skip=6pt]
 }{\end{tcolorbox}}

\definecolor{repairgreen}{RGB}{220,245,220}

\newcommand{\repairtext}[1]{\begingroup\sethlcolor{repairgreen}%
  \repairhl{#1}\endgroup}
\newcommand{\repairmath}[1]{\begingroup\setlength{\fboxsep}{1pt}%
  \mathchoice{\colorbox{repairgreen}{$\displaystyle #1$}}%
             {\colorbox{repairgreen}{$\textstyle #1$}}%
             {\colorbox{repairgreen}{$\scriptstyle #1$}}%
             {\colorbox{repairgreen}{$\scriptscriptstyle #1$}}\endgroup}
\newenvironment{repairblock}{%
 \begin{tcolorbox}[enhanced,sharp corners,
  colback=repairgreen,colframe=repairgreen,boxrule=0pt,
  left=3pt,right=3pt,top=2pt,bottom=2pt,
  before skip=4pt,after skip=4pt]
 }{\end{tcolorbox}}

\renewcommand{\hl}[1]{#1}
\renewcommand{\mathrev}[1]{#1}

\ifdefined\showrevisionsfalse
  \showrevisionsfalse
\fi

\renewcommand{\hl}[1]{#1}
\renewcommand{\mathrev}[1]{#1}

\renewcommand{\repairtext}[1]{#1}
\renewcommand{\repairmath}[1]{#1}

\begin{document}
\thispagestyle{empty}
\begin{center}
{\LARGE\bfseries\color{ink} Precedence Constraint Scheduling on a Fixed Number of Machines}\\[8pt]
{\large Klaus Jansen, University of Kiel}
\end{center}

\ifshowrevisions
\begin{center}
\small Revision key: {\color{revisionred}red text is to be removed};
\colorbox{revisionyellow}{yellow highlighting marks new text}.\\[3pt]
\colorbox{repairgreen}{Light green marks the DeepReview repairs of 2 November 2026.}
\end{center}
\fi

\begin{abstract}
We give a deterministic exact algorithm for
$P_m\mid\mathrm{prec},p_j=1\mid C_{\max}$ for every fixed $m\ge3$.
The algorithm constructs an optimum schedule in
\[
 O(L)+m^{O(m^2)}\mathrev{n^{\revchange{6m^2+26m+1}{6m^2+13m+1}}}\mathrev{\log n}
\]
elementary RAM operations, where $n \ge 2$ is the number of jobs and $L$ is
the binary input length. Its word size is $O(m^2\mathrev{\log n})$ bits.
A direct implementation on a deterministic Turing machine with a
fixed number of tapes uses
\[
 O(L)+m^{O(m^2)}n^{6m^2+13m+1}\log^2 n
\]
steps. Thus both models have polynomial degree $94$ for three
machines, apart from their respective logarithmic factors.
\end{abstract}

\section{Problem and main theorem}
\label{af:introduction}

We are given a directed acyclic graph $G=(V,E)$ and $m$ identical
machines. Each vertex is a job of processing time one. A job can start
only after all its predecessors have finished. Jobs cannot be interrupted.
The objective is to minimize the time at which the last job finishes.
This is the problem
\[
 P_m\mid\mathrm{prec},p_j=1\mid C_{\max}.
\]
Write $n=|V|$, and let $L$ be the binary input length. Vertices
$1,\ldots,n$ are individually listed, including isolated vertices;
arcs are distinct and use standard binary vertex labels. A deadline,
when supplied, is a nonnegative binary integer.

\begin{theorem}[An exact algorithm for fixed machine capacity]
\label{af:main}
  For every integer $m\ge3$ and $n \ge 2$, there are algorithms to decide
deadline feasibility and construct an optimum schedule using
\begin{equation}
 O(L)+m^{O(m^2)}\mathrev{n^{\revchange{6m^2+26m+1}{6m^2+13m+1}}}\mathrev{\log n}
 \label{af:main-bound}
\end{equation}
elementary RAM operations in the model specified in
  Section~\ref{af:sec-runtime}. A single deadline test needs no
  logarithmic search factor. In particular, the simpler uniform bound
\[
 O(L)+m^{O(m^2)}\mathrev{n^{\revchange{15m^2}{11m^2}}}
\]
 holds for optimization as well.
On a deterministic Turing machine with a fixed number of tapes,
optimization and construction of an optimum schedule take
\[
 O(L)+m^{O(m^2)}n^{6m^2+13m+1}\log^2 n
\]
steps. A single deadline test, including construction of a feasible
schedule when one exists, uses $\log n$ instead of $\log^2 n$.
These multitape bounds are proved in
Proposition~\ref{af:multitape}.
Thus the running time is polynomial for each fixed $m$.
The algorithm is uniform: all constants, families, and loops
are specified as functions of its input.
\end{theorem}

The cases $m=1$ and $m=2$ have the familiar direct and Coffman--Graham
solutions, respectively~\cite{cg}. Our explicit procedures focus on
$m\geq3$, which is the range of the generalization.


\paragraph{The central algorithmic idea.}
A subset dynamic program can have $2^N$ job-set entries. The full-slot
algorithm retains only sets represented by prescribed short formulas.
For a candidate $W$, a state $(L,Z,R)$ separates an earlier prefix,
a full marked slot and the remaining suffix. Previously stored schedules
for the initial prefix, successive gaps and final suffix are concatenated
through the marks. Exact ideal and antichain tests make every accepted
certificate sound. The structural theorem supplies a complete certificate
for the root, while increasing prefix cardinality permits a single
acyclic state-path evaluation.

\paragraph{Previous results and status.}
The complexity of
$P_m\mid\mathrm{prec},p_j=1\mid C_{\max}$ has been investigated
for several decades; see the surveys by Lawler, Lenstra, Rinnooy Kan,
and Shmoys~\cite{llrs1993}, Chen, Potts, and
Woeginger~\cite{cpw1998}, and Mnich and van
Bevern~\cite{mnichvanbevern2018}. For $m=1$ the problem is immediate,
while for $m=2$ polynomial-time algorithms were given by Fujii,
Kasami, and Ninomiya~\cite{fkn1969} and Coffman and Graham~\cite{cg}.
Ullman~\cite{ullman1975} proved NP-hardness when the number $m$ of
machines is part of the input. This does not establish NP-hardness
for any fixed $m\ge 3$. Indeed, the exact complexity for three
machines remained a prominent open problem for many years;
Lenstra and Rinnooy Kan~\cite[pp.~518--519]{lrk1984} explicitly
raised the fixed-machine question, also discussed in the later
survey~\cite{mnichvanbevern2018}.

\paragraph{Special precedence orders.}
For unit-time jobs, polynomial-time exact algorithms were already known
for several restricted precedence classes. With an arbitrary number of
machines, these include in-forests and out-forests~\cite{hu1961},
interval orders~\cite{py1979}, quasi-interval
orders~\cite{moukrim1999}, and overinterval
orders~\cite{chardonmoukrim2005}. For opposing forests and level
orders, the cited algorithms require a fixed number of
machines~\cite{gjty1983,dw1985}; the bounded-height result requires
both the height and the number of machines to be
fixed~\cite{dw}. All these statements concern unit-processing-time
jobs.

\paragraph{Approximation algorithms.}
For arbitrary processing times and precedence constraints,
Graham's list scheduling achieves ratio $2-1/m$~\cite{graham1966}.
For background on list scheduling and approximation algorithms,
see Williamson and Shmoys~\cite[Section~2.3 and Exercise~2.3]{williamsonshmoys2011}.
In the unit-job case, Coffman--Graham scheduling and the analyses
of Lam and Sethi, and Braschi and Trystram, yield stronger bounds,
including the approximation ratio $4/3$ for three
machines~\cite{cg,lamsethi1977,braschitrystram1994}.
For $m\ge4$, Gangal and Ranade~\cite{gangalranade2008} obtained
ratio $2-7/(3m+1)$; see also the open questions collected by
Schuurman and Woeginger~\cite{schuurmanwoeginger1999}.
For unit-time jobs under arbitrary precedence constraints, Levey and
Rothvo{\ss}~\cite{LeveyRothvoss2021} obtained a
$(1+\varepsilon)$-approximation through an LP hierarchy, with a
super-quasipolynomial running-time bound. Garg~\cite{Garg2018}
gave a quasipolynomial-time approximation scheme, and
Li~\cite{Li2021} improved the running time to
$n^{O((m^4/\varepsilon^3)(\log\log n)^3)}$.
Das and Wiese~\cite{DasWiese2022} subsequently gave a simpler
combinatorial quasipolynomial-time approximation scheme.
These are approximation algorithms; their displayed running-time
bounds remain superpolynomial in $n$ when $m$ and $\varepsilon$
are fixed. In particular, they do not give polynomial-time exact
algorithms for general precedence orders.

\paragraph{Exact algorithms for general precedence orders.}
Nederlof, Swennenhuis, and W\k{e}grzycki~\cite{nsw}
obtained an exact algorithm with running time
$(1+n/m)^{O(\sqrt{nm})}$;
\repairtext{the result appeared at SODA 2025, and an accepted journal
version was published online in \emph{ACM Transactions on Algorithms}
in January 2026.}
For every fixed $m$ this is subexponential in $n$, but it is not a
polynomial bound. A September 2026 OpenAI
manuscript~\cite{openai} established a polynomial-time exact
algorithm for the three-machine problem
$P_3\mid\mathrm{prec},p_j=1\mid C_{\max}$.
The present paper gives a uniform algorithm for every fixed
$m\ge3$, with explicit description families, correctness proofs,
and bounds for both RAM and deterministic fixed-tape computation.

\paragraph{Relation to the OpenAI construction and the new ideas.}
The OpenAI construction~\cite{openai} for three machines uses
normalized intervals, global priority lists, and changes of frame;
its structural analysis involves lists of at most five entries,
of which at most three are injections. Our extension to arbitrary
fixed $m\ge3$ uses a boundary-exchange argument. At each strict
promotion, a spare job in the ending slot cannot be the successor
witness of a surviving high-priority birth group. Counting the
remaining witnesses yields at most $m-1$ surviving birth groups
and at most $m$ active entries at a node. When $m=3$, these bounds
are two groups and three entries. Historical extensions prevent
information from accumulating through successive frame reversals.
The proof distinguishes exact global descriptions of nodes from
carry identities valid only after restriction to a node.

For the algorithm, separate formula budgets for node sets and
prefix cuts, exact binary-trie lookup, and an acyclic evaluation
of prefix states yield the RAM bound
\[
 O(L)+m^{O(m^2)}\mathrev{n^{\revchange{6m^2+26m+1}{6m^2+13m+1}}}\mathrev{\log n}.
\]
The polynomial degrees in $n$ are $\revchange{133}{94}$, $\revchange{201}{149}$, and $\revchange{281}{216}$ for
$m=3,4,5$, respectively. Theorem~1.1 of the OpenAI
manuscript~\cite{openai} states a deterministic fixed-tape
bound $O((L+2)^{150020})$ for three machines. 
For the same input-length measure, our fixed-tape bound from
Proposition~\ref{af:multitape} implies $O((L+2)^{95})$ at $m=3$.
This compares upper bounds in the same formal computation model,
not implementation running times.
Unlike the cited three-machine theorem, our bound applies to every
fixed $m\ge3$.

\section{Schedules and a finite description language}
\label{af:representation}

\textbf{Road map: from explicit candidates to a structural certificate.}
We first specify the objects that the algorithm can compute, then prove
that they suffice. Section~\ref{af:representation} introduces integral
slots, padding by isolated jobs, and full schedules. It defines the
boundary sets and their predecessor and successor cones $P_A,D_A$,
the weak cones $\widehat P_A,\widehat D_A$, and a fixed topological
rank function with rank suffixes $R_h$. These globally defined sets
are the atoms of the Boolean language. The family $\mathcal F_K$
contains all distinct subsets of the padded universe $J$ represented
with at most $K$ atomic occurrences. Algorithm~\ref{af:alg-family},
\textsc{GenerateFamily}, constructs this family by set operations
and exact duplicate detection, without knowing a feasible schedule.

Section~\ref{af:sec-algorithm} gives the scheduling algorithm.
It processes candidate node sets in increasing cardinality and joins
previously stored schedules through selected full slots. A separate
cut family supplies candidate prefixes, and a directed state graph
records which pieces can be combined while respecting precedence.
Algorithm~\ref{af:alg-deadline} turns this evaluation into an exact
deadline test by padding and subsequently deleting the dummy jobs.
Algorithm~\ref{af:alg-optimum} uses binary search on the integer
deadline to obtain an optimum makespan and an explicit schedule.
Section~\ref{af:worked-example} illustrates the records, the stored
schedule, and the longer certificate used to justify acceptance.

Section~\ref{af:sec-correctness} proves correctness. After establishing
padding, exact family generation, and soundness of the schedule
combinations, it introduces a recursive hierarchy described by two
families $(\mathcal F,\mathcal F_{\mathrm{cut}})$. Marked full slots
split a node into smaller open gaps, which become its children.
Every node set must belong to $\mathcal F$, whereas a prefix before
a marked slot is required to have the form $W\cap C$ for a global
cut $C\in\mathcal F_{\mathrm{cut}}$. Induction on node size then
shows that such a hierarchy forces acceptance of the root. The
existence of the required hierarchy is stated here and proved in
Section~\ref{sec:str-structure}.
Section~\ref{af:sec-runtime} counts the candidates and the work needed
to generate, store, and evaluate them, including unsuccessful tests,
schedule reconstruction, and binary search. It gives the RAM bound
and a deterministic multitape implementation. This analysis uses
only the specified family budgets; its operation counts can therefore
be read before the structural proof.

Section~\ref{sec:str-structure} contains the structural heart of the
polynomial-time result. Starting from a feasible witness schedule,
it constructs the hierarchy using normalization, separators,
priority lists, and changes of frame. The central task is to prevent
inherited information from growing along the recursion and to turn
it into short global Boolean descriptions. The resulting linear
leaf budgets are $\revchange{4m+12}{4m+5}$ for nodes and $\revchange{m+6}{m+3}$ for cuts, giving
$\mathcal F=\mathcal F_{\revchange{4m+12}{4m+5}}$ and
$\mathcal F_{\mathrm{cut}}=\mathcal F_{\revchange{m+6}{m+3}}$.
These are bounds on formula size, not constant bounds on family
cardinality: for fixed $m$, they make the enumerated search space
polynomial in the number of jobs. The proof distinguishes exact
global node descriptions from cut identities valid after intersection
with the local node. It completes Theorem~\ref{af:structure}, thereby
discharging the structural premise of completeness and completing,
together with Section~\ref{af:sec-runtime}, the proof of the main
Theorem~\ref{af:main}.

\subsection{Slots, induced precedence, and full schedules}

The following representation is sufficient for the original scheduling
model: integral slots lose no optimality, and isolated dummy jobs can
fill unused positions of a deadline schedule. These statements are
proved in Lemmas~\ref{af:integral} and~\ref{af:padding}, after the
complete algorithm has been specified.

Compute the transitive closure of $G$. Write $x\prec y$ when there is
a directed path of positive length from $x$ to $y$. Every subset below
inherits this \emph{global} order: a precedence remains even if a path
establishing it uses jobs outside that subset.

A slot schedule of deadline $T$ assigns each job a slot
$\tau(x)\in\{1,\ldots,T\}$, with at most $m$ jobs per slot and
$\tau(x)<\tau(y)$ whenever $x\prec y$. Slot $t$ means execution during
$[t-1,t)$. A job's machine can be assigned arbitrarily within its slot,
using distinct machines for the jobs of that slot.

For a relevant deadline $\lceil n/m\rceil\leq T<n$, add $mT-n$
isolated jobs and call the resulting set $J$. Its size is $N=mT\leq mn$.
The algorithm tests whether these jobs have a full schedule of $T$ slots.

\begin{definition}[Full schedule, ideal, and antichain]
\label{af:basic-definitions}
A \emph{full schedule} of $W\subseteq J$ is an ordered partition of
$W$ into antichains of exactly $m$ jobs that respects the induced order
on $W$. It has $|W|/m$ slots. The empty set has the empty full schedule.
A set $U\subseteq W$ is an \emph{ideal of $W$} if $y\in U$, $x\in W$,
and $x\prec y$ imply $x\in U$. A set is an \emph{antichain} if no two
of its jobs are comparable.
\end{definition}

Both tests inspect entries of the precomputed transitive precedence
matrix. To test whether $U$ is an ideal of $W$, inspect every ordered
pair $(x,y)\in W\times U$ and reject if $x\prec y$ but $x\notin U$.
To test whether $Z$ is an antichain, inspect every ordered pair of
distinct jobs in $Z$ and reject if one precedes the other.
Each matrix entry and membership entry can be accessed in constant
time, so either test uses at most $O(N^2)$ RAM operations.

\paragraph{A small padding example.}
Let $m=3$ and consider five jobs with only $a_1\prec a_2$ and
$b_1\prec b_2$, together with an independent job $c$. For deadline
two, add one isolated job $d$. A full padded schedule is
\[
 \{a_1,b_1,c\},\qquad \{a_2,b_2,d\}.
\]
Removing $d$ gives the original five-job schedule. A table entry for
three jobs asks for one full slot; an entry for six jobs asks for two.
There is no separate time coordinate in a table entry.

\subsection{Atoms are sets computed from the input order}

All descriptions denote subsets of the entire padded set $J$. For an
$m$-element set $A\subseteq J$, define
\[
\begin{split}
 P_A&=\{x\in J:\exists a\in A,\ x\prec a\},\qquad
 D_A=\{x\in J:\exists a\in A,\ a\prec x\},\\
 \widehat P_A&=P_A\cup A,\qquad
 \widehat D_A=D_A\cup A.
\end{split}
\]
Choose one topological ordering by repeatedly removing the smallest
identifier among available minimal jobs. Its rank function
$\operatorname{rank}:J\to\{1,\ldots,N\}$ satisfies
$x\prec y\Rightarrow\operatorname{rank}(x)<\operatorname{rank}(y)$.
For $1\leq h\leq N+1$, put
\[
 R_h=\{x\in J:\operatorname{rank}(x)\geq h\}.
\]
In particular, $R_1=J$ and $R_{N+1}=\varnothing$.

\begin{definition}[Atoms and the family $\mathcal F_K$]
\label{af:language}
The atomic sets are
\[
 J,\quad\varnothing,\quad R_h,
 \quad A,\ P_A,\ D_A,\ \widehat P_A,\ \widehat D_A
 \quad (|A|=m).
\]
A Boolean formula uses these atoms, binary union and intersection,
and unary complement relative to $J$. Its leaf count is the number
of atomic occurrences, counted with multiplicity. Unary complement
does not add a leaf. The family $\mathcal F_K$ contains the distinct
sets represented by formulas with at most $K$ leaves.
\end{definition}

The enumeration uses every $m$-set $A$, even if it is not an antichain.
Only an actual full slot must pass the antichain test. Every atom is
computed directly from the order and rank table. No schedule is used
to generate this family.

Represent a set by its $N$-bit incidence vector. Equality, inclusion,
union, intersection, complement, and cardinality are then explicit
finite operations on vectors. When two formulas have the same vector,
the algorithm stores the vector once. It need not store the formulas.

\subsection{Generating the family directly by set operations}

The following procedure avoids enumerating tree shapes. It groups sets
by an upper bound on their leaf count. It is exact, not a heuristic
selection of promising subsets.

\paragraph{Exact storage by incidence vectors.}
  For duplicate detection, use a \emph{binary trie}. Its root represents
  the empty bit prefix. At depth $r=0,\ldots,N-1$, the next edge is labelled by the
membership bit of job $N-r$; a complete root-to-leaf path therefore
  reads the $N$ bits in decreasing job-label order. Missing edges are
  created during insertion. A terminal node stores a pointer to the
  record for that vector. An existing terminal identifies an exact
  duplicate. Insertion and lookup inspect at most $N$ edges and cost
$O(N)$ RAM operations; no hashing or probabilistic equality test is
  used. Each generated layer has its own trie and list of distinct
  vectors. A traversal taking edge $0$ before edge $1$ lists its vectors
 in lexicographic order.

\par\medskip\noindent\begin{minipage}{\textwidth}
\begin{algorithm}[\textsc{GenerateFamily}$(J,m,K)$]
\label{af:alg-family}
\textbf{Input:} the padded global order on $J$, its fixed rank order,
and integers $m,K\geq1$.\quad
  \textbf{Output:} all distinct members of $\mathcal F_K$ as bit vectors,
 together with the intermediate layers $F_1,\ldots,F_K$ and their tries.
\begin{enumerate}
\item Build the atomic list in the order $J,\varnothing,R_1,\ldots,R_{N+1}$,
followed by $A,P_A,D_A,\widehat P_A,\widehat D_A$ for each $m$-set $A$
in lexicographic identifier order.
  \item Create an empty list $F_1$ and its binary trie. For each atom $A$
  in the atomic list, insert $A$ and $J\setminus A$ into the trie. Append
  the vector to $F_1$ exactly when its terminal node is new. Each
 insertion, including exact duplicate detection, costs $O(N)$ operations.
\item For $k=2,3,\ldots,K$, do the following:
\begin{enumerate}[label=(\alph*)]
  \item Create an empty list $F_k$ and a new trie. Insert every vector of
 $F_{k-1}$ into this trie and append it to $F_k$.
  \item For $i=1,2,\ldots,k-1$, and then for each $X\in F_i$ and each
$Y\in F_{k-i}$ in their stored order, insert $X\cup Y$ and $X\cap Y$
 into $F_k$, using its trie for the same exact duplicate test.
\end{enumerate}
All source lists in step (b) have index smaller than $k$ and remain fixed
during the iteration.
  \item Return $F_K$, retaining the intermediate lists and tries for reuse.
\end{enumerate}
\end{algorithm}
\end{minipage}\par\medskip

Lemma~\ref{af:family-exact} proves that the returned family is
exactly $\mathcal F_K$. The size bound in Lemma~\ref{af:family-count}
and the operation count in Lemma~\ref{af:family-cost} explain its cost.

\paragraph{Why the loop has a specified final layer.}
  The procedure runs through $K=\revchange{4m+12}{4m+5}$ layers and also retains
 layer $\revchange{m+6}{m+3}$. A single unchanged layer
does not prove that all later layers will be unchanged: formulas with
  two more expensive subexpressions can first appear later. An independently
sufficient condition for early termination is that every one of the
 $2^N$ possible bit vectors has already been generated.
This optimization is unnecessary for the proof or its upper bound.

\paragraph{Two budgets for two purposes.}
 Use
 \[
\begin{aligned}
 K_{\mathrm{node}}&=\revchange{4m+12}{4m+5},&
 \mathcal F=\mathcal F_{\mathrm{node}}&=\mathcal F_{\revchange{4m+12}{4m+5}},\\
 K_{\mathrm{cut}}&=\revchange{m+6}{m+3},&
 \mathcal F_{\mathrm{cut}}&=\mathcal F_{\revchange{m+6}{m+3}}.
\end{aligned}
\]
 The first family supplies candidate job sets; the second supplies global
  cuts defining state prefixes. The structural theorem will prove these
 two description budgets separately.

\paragraph{Four different sets in one computation.}
The following distinctions are useful when reading the algorithm.
\begin{center}
\begin{tabularx}{\textwidth}{@{}p{23mm}>{\raggedright\arraybackslash}X>{\raggedright\arraybackslash}p{45mm}@{}}
\toprule
Notation & Meaning & Family requirement\\
\midrule
  $W$ & Current candidate whose acceptance is being computed & $W\in\mathcal F$.\\
  $C$ & Global cut used to describe a state & $C\in\mathcal F_{\mathrm{cut}}$.\\
  $L=W\cap C$ & Prefix represented by that state & Need not itself belong to either family.\\
$G$ & Gap queried between two successive marks & An absent nonempty set has table value false.\\
\bottomrule
\end{tabularx}
\end{center}
The structural theorem supplies family memberships for the gaps of
one successful decomposition. It does not assert that every intersection
or difference of candidate vectors is another candidate.
For an internal candidate, a table value of zero means that this
recurrence has not constructed a schedule; it need not certify that
an arbitrary candidate set is infeasible. Completeness is proved for
the full root instance by the structural hierarchy theorem.

\section{The exact algorithm}\label{af:sec-algorithm}

For a deadline $T$, padding gives $N=mT$ jobs. The algorithm generates
$\mathcal F=\mathcal F_{\revchange{4m+12}{4m+5}}$ and computes one Boolean value for each
$W\in\mathcal F$:
 \[
 a(W)=1\quad\Longleftrightarrow\quad
 \text{the algorithm has constructed a full schedule of }W.
\]
Whenever it sets a value to one, it also stores that schedule explicitly.
The empty set starts with value one and the empty schedule. Sets are
processed by increasing cardinality. Thus the basic operation is to join
\emph{already stored} schedules of smaller sets, with a full slot inserted
between successive pieces.

The smaller family $\mathcal F_{\mathrm{cut}}=\mathcal F_{\revchange{m+6}{m+3}}$ supplies candidate cuts.
Given a parent $W$ and a cut $C$, the set $L=W\cap C$ is a candidate prefix.
  The intersection is computed directly as a bit vector. It need not itself
 belong to either family.

\subsection{Records, states, and elementary predicates}

Fix job labels $1,\ldots,N$. If $b_j$ indicates membership of job $j$,
order incidence vectors lexicographically as $b_N\cdots b_1$, with
$0<1$; equivalently, order their integer values $\sum_j b_j2^{j-1}$.
  The cut family is traversed in this lexicographic vector order.
Order $m$-subsets by their increasing
lists of job labels. These orders fix every choice made below. A full
schedule is stored as a list of $m$-element slots, with the labels within
each slot increasing.

The family table has one row
\[
 (X,a(X),\sigma(X))\qquad(X\in\mathcal F).
\]
For each candidate set \(X\in\mathcal F\), the table stores a Boolean value \(a(X)\in \{0,1\}\), together with a schedule
field \(\sigma(X)\). The value \(a(X)=1\) means that the algorithm has constructed and stored a feasible full schedule of \(X\): every job of \(X\) occurs exactly once, each slot contains exactly \(m\) jobs, and all precedence relations induced on \(X\) are respected.
Note that $\sigma(X)$ is either an explicitly stored full schedule or the symbol
$\bot$. Rows are sorted by the pair $(|X|,b_N\cdots b_1)$ and retain
that order throughout evaluation. Initially all
values are zero and all schedules are $\bot$, except
$a(\varnothing)=1$ and $\sigma(\varnothing)=()$.

For $X\subseteq W$, define the two tests
\[
\begin{split}
 \operatorname{Ideal}(X,W)
 &\iff \forall x\in X\ \forall y\in W:
             (y\prec x\Rightarrow y\in X),\\
 \operatorname{Antichain}(X)
 &\iff \forall x,y\in X:\ \neg(x\prec y).
\end{split}
\]
These are finite tests against the precomputed global transitive order.
In particular, an induced precedence is retained even if its path in the
input DAG passes through jobs outside $W$.

\begin{definition}[State of a candidate set]\label{af:def-state}
  For a nonempty $W\in\mathcal F$ with $m\mid |W|$, a state is a partition
\[
 W=L\mathbin{\dot\cup}Z\mathbin{\dot\cup}R
\]
such that $L=W\cap C$ for some $C\in\mathcal F_{\mathrm{cut}}$, $|Z|=m$,
$m\mid |L|$, $m\mid |R|$, $Z$ is an antichain, and
 $L$ and $L\cup Z$ are ideals of $W$.
\end{definition}

The two ideal conditions state exactly that no job in a later part of
  $(L,Z,R)$ precedes a job in an earlier part. A state record stores
$(L,Z,R)$, the already computed prefix cardinality $|L|$, a reachability
bit, and either a predecessor-state index,
 $\mathsf{START}$, or $\mathsf{UNREACHED}$. Different cuts can produce the
same triple; retaining these duplicate records is harmless and simplifies
the specification.

The following algorithm uses two operations.
\textsc{Lookup}$(X)$ accepts any query set $X\subseteq J$, whether or not $X$ belongs to the generated family.
  It follows the incidence vector of $X$ through the family trie's
 $N$ levels to its table-row pointer, or detects a missing edge.
  Thus exact lookup costs $O(N)$ operations. It returns
\[
\operatorname{Lookup}(X)=
\begin{cases}
a(X), & X\in\mathcal F,\\
0, & X\notin\mathcal F.
\end{cases}
\]
\textsc{Fetch}$(X)$ retrieves the schedule whose existence is certified by a successful lookup. It is defined only when \textsc{Lookup}$(X)=1$, and it returns a copy of the explicitly stored slot list $\sigma(X)$.

\par\medskip\noindent\begin{minipage}{\textwidth}
\begin{algorithm}[Exact lookup and schedule retrieval]
\label{af:alg-lookup}
\textsc{Lookup}$(X)$ and \textsc{Fetch}$(X)$ operate on the current family table.
\begin{enumerate}
  \item Read the membership bits of $X$ in decreasing job-label order,
 following the correspondingly labelled edges of the family trie.
 A missing edge means there is no matching row; a complete path gives
 the terminal node's pointer to that row.
\item If the matching row is found, \textsc{Lookup} returns its stored bit
$a(X)$. If that bit is one, \textsc{Fetch} returns a copy of its stored
schedule $\sigma(X)$.
\item If there is no matching row, \textsc{Lookup} returns zero.
Calling \textsc{Fetch} on a missing or zero-valued row is forbidden; the
subsequent procedures only fetch sets whose lookup has returned one.
\end{enumerate}
\end{algorithm}
\end{minipage}\par\medskip

Thus a lookup never solves a new scheduling problem. It only reads
 the value stored in the matching row.

\par\medskip\noindent\begin{minipage}{\textwidth}
\begin{algorithm}[Enumerate every state of one parent]
\label{af:alg-states}
  \textsc{BuildStates}$(W,\mathcal F_{\mathrm{cut}},m)$ returns an ordered list $\mathcal S_W$.
\begin{enumerate}
\item Start with the empty list.
  \item For each $C\in\mathcal F_{\mathrm{cut}}$ in lexicographic vector order, set $L=W\cap C$.
If $m\nmid |L|$, proceed to the next cut.
\item For each $m$-subset $Z$ of $W\setminus L$ in the fixed subset order,
set $R=W\setminus(L\cup Z)$. Append $(L,Z,R)$ to the list exactly when
\[
 m\mid |R|,\qquad \operatorname{Antichain}(Z),\qquad
 \operatorname{Ideal}(L,W),\qquad
 \operatorname{Ideal}(L\cup Z,W).
\]
The loops in this step are executed separately for each cut in Step 2.
\item Number the appended records $1,\ldots,s$ in their append order,
initialize all reachability bits to zero and predecessor fields to
$\mathsf{UNREACHED}$, and return the list.
\end{enumerate}
\end{algorithm}
\end{minipage}\par\medskip

Write $(L_i,Z_i,R_i)$ for record $i$. Conceptually, there is an edge
$i\longrightarrow j$ when
\begin{equation}
 L_i\cup Z_i\subseteq L_j
 \quad\text{and}\quad
 \textsc{Lookup}\bigl(L_j\setminus(L_i\cup Z_i)\bigr)=1.
 \label{af:eq-edge}
\end{equation}
An initial record satisfies $\textsc{Lookup}(L_i)=1$, and a final record
satisfies $\textsc{Lookup}(R_i)=1$. The next algorithm finds a path from an
initial record to a final record. It computes the edge test when needed;
 storing a separate edge table is unnecessary. Every edge satisfies
  $|L_j|\ge |L_i|+m$. Thus increasing prefix cardinality gives an acyclic
 evaluation order.

\par\medskip\noindent\begin{minipage}{\textwidth}
\begin{algorithm}[Evaluate the finite family]
\label{af:alg-evaluate}
  \textsc{EvaluateFamily}$(J,\mathcal F,\mathcal F_{\mathrm{cut}},m)$ returns the family table.
\begin{enumerate}
  \item Sort the distinct members of $\mathcal F$ by
$(|X|,b_N\cdots b_1)$, using a lexicographic trie traversal followed
  by stable cardinality buckets. Bind each trie terminal to its table row.
Initialize the table rows in this order as
specified above, including the empty schedule in the row for
$\varnothing$.
\item Process the nonempty rows in that order. Skip a row if its
cardinality is not a multiple of $m$; its value remains zero. For each
remaining parent $W$, perform Steps 3--6.
  \item Set $\mathcal S_W=\textsc{BuildStates}(W,\mathcal F_{\mathrm{cut}},m)$ and let
 $s=|\mathcal S_W|$. For $i=1,\ldots,s$, if
$\textsc{Lookup}(L_i)=1$, set $\mathrm{reached}[i]=1$ and
$\mathrm{pred}[i]=\mathsf{START}$.
  \item Stably bucket the state indices by $|L_i|$, preserving their index
  order within a bucket. Process these buckets in increasing order.
  For each reached source $i$, inspect the target indices $j=1,\ldots,s$
  once. If $\mathrm{reached}[j]=0$ and \eqref{af:eq-edge} holds, set
 \[
 \mathrm{reached}[j]=1,\qquad
 \mathrm{pred}[j]=i.
\]
  Updates take effect immediately. Each possible edge goes to a later
 prefix-cardinality bucket, so no bucket needs to be processed again.
\item Inspect records $i=1,\ldots,s$ in order. If none satisfies
$\mathrm{reached}[i]=1$ and $\textsc{Lookup}(R_i)=1$, leave the row of $W$
at $(W,0,\bot)$.
Otherwise choose the first such record $t$, set $a(W)=1$, and store
\[
 \sigma(W)=\textsc{Reconstruct}(t,\mathcal S_W).
\]
\item Discard the temporary state list for this $W$ and continue with
the next parent in Step 2. After all parents have been processed, return
the table.
\end{enumerate}
\end{algorithm}
\end{minipage}\par\medskip

The state list can be empty; in that case there are no sources to process and the
 parent is rejected. A path consisting of one initial and final state is
allowed. Once a record is reached, its predecessor field is never changed.

\paragraph{Meaning of a zero entry.}
  After a row has been processed, zero means that no schedule certificate
  was found using these two families. Some feasible candidate sets can have
 value zero when a family omits needed descriptions. Soundness will establish
that every value one has a valid stored schedule. The structural theorem
will establish completeness for the final root decision. Unprocessed
rows also start at zero; the proper-query lemma below ensures that they
are never consulted during an earlier parent's evaluation.

\par\medskip\noindent\begin{minipage}{\textwidth}
\begin{algorithm}[Recover and store one full schedule]
\label{af:alg-reconstruct}
\textsc{Reconstruct}$(t,\mathcal S_W)$ is called only for a reached final state.
\begin{enumerate}
\item Follow predecessor fields backward from $t$ until a record with
predecessor $\mathsf{START}$ is reached. Reverse the recorded index list
to obtain a path $i_1,\ldots,i_k$ in forward order.
\item Start the output list with a copy of
$\textsc{Fetch}(L_{i_1})$.
\item For $r=1,\ldots,k$, append the full slot $Z_{i_r}$.
If $r<k$, append a copy of
\[
 \textsc{Fetch}\bigl(L_{i_{r+1}}\setminus(L_{i_r}\cup Z_{i_r})\bigr).
\]
\item Append a copy of $\textsc{Fetch}(R_{i_k})$ and return the resulting
list of full slots.
\end{enumerate}
\end{algorithm}
\end{minipage}\par\medskip

\subsection{Deadline feasibility and optimum makespan}

The following two wrappers complete the algorithm for the original jobs.
The graph is read and its global transitive order computed once. A
deterministic topological order is obtained by repeatedly choosing the
smallest-labelled available job. The padded order extends this order by
isolated dummy jobs; choose the smallest-labelled available job again
when computing the rank used by \textsc{GenerateFamily}.

In the next algorithm, \textsc{Deadline}$(G,m,T)$, $G=(V,E)$ denotes the original directed acyclic precedence graph, $m$ is the number of machines, and $T$ is the proposed nonnegative integer deadline. Thus, the algorithm asks whether the jobs represented by $G$ admit a feasible schedule with makespan at most $T$.

\par\medskip\noindent\begin{minipage}{\textwidth}
\begin{algorithm}[Exact deadline test]
\label{af:alg-deadline}
\textsc{Deadline}$(G,m,T)$ returns $\mathsf{NO}$ or a schedule of all
original jobs with makespan at most the nonnegative integer $T$.
\begin{enumerate}
\item If $n=0$, return the empty schedule. If $T<\lceil n/m\rceil$,
return $\mathsf{NO}$. If $T\geq n$, return the serial topological schedule.
\item Add $mT-n$ isolated dummy jobs with fresh consecutive labels.
  Let $J$ be the padded set, $N=mT$, and $K=\revchange{4m+12}{4m+5}$.
  \item Generate $\mathcal F=\textsc{GenerateFamily}(J,m,K)$ by the
specified Boolean-family procedure, retain
$\mathcal F_{\mathrm{cut}}=F_{\revchange{m+6}{m+3}}$, and run
 $\textsc{EvaluateFamily}(J,\mathcal F,\mathcal F_{\mathrm{cut}},m)$.
\item If $a(J)=0$, return $\mathsf{NO}$. Otherwise delete dummy jobs from
the stored full schedule $\sigma(J)$, discard trailing empty slots,
and return the remaining slot schedule.
Within each slot, assign its original jobs in increasing label order to
machines $1,2,\ldots$.
\end{enumerate}
\end{algorithm}
\end{minipage}\par\medskip

\par\medskip\noindent\begin{minipage}{\textwidth}
\begin{algorithm}[Exact optimization]
\label{af:alg-optimum}
\textsc{Optimum}$(G,m)$ returns an optimum schedule of the original jobs.
\begin{enumerate}
\item If $n=0$, return the empty schedule of makespan zero. Otherwise set
\[
 \ell=\lceil n/m\rceil-1,\qquad u=n,
\]
and let $\sigma_{\mathrm{best}}$ be the serial topological schedule.
\item While $u-\ell>1$, set $T=\lfloor(\ell+u)/2\rfloor$ and call
$\textsc{Deadline}(G,m,T)$.
If it returns $\mathsf{NO}$, set $\ell=T$.
If it returns a schedule $\sigma$, set $u=T$ and
$\sigma_{\mathrm{best}}=\sigma$.
\item Return $u$ and $\sigma_{\mathrm{best}}$.
\end{enumerate}
\end{algorithm}
\end{minipage}\par\medskip

All inputs to the scheduling calls are obtained from explicit job sets,
bit-vector operations, and previously computed table rows. In particular,
the algorithm never calls a procedure to find a normalized schedule or to
recognize a genuine structural history. Those objects will be used only
in the completeness proof.

\section{A small example: the stored schedule and a longer certificate}
\label{af:worked-example}

Consider $m$ disjoint chains of length three,
\[
 j_{i,1}\prec j_{i,2}\prec j_{i,3}\qquad(1\leq i\leq m),
\]
with no relations between different chains. Put
$Z_t=\{j_{i,t}:1\leq i\leq m\}$ for $t=1,2,3$ and
$W=Z_1\mathbin{\dot\cup}Z_2\mathbin{\dot\cup}Z_3$.
For deadline three no dummy jobs are needed. A feasible schedule is
$(Z_1,Z_2,Z_3)$, and a chain of length three excludes every smaller
deadline. We now trace how the algorithm obtains a schedule.

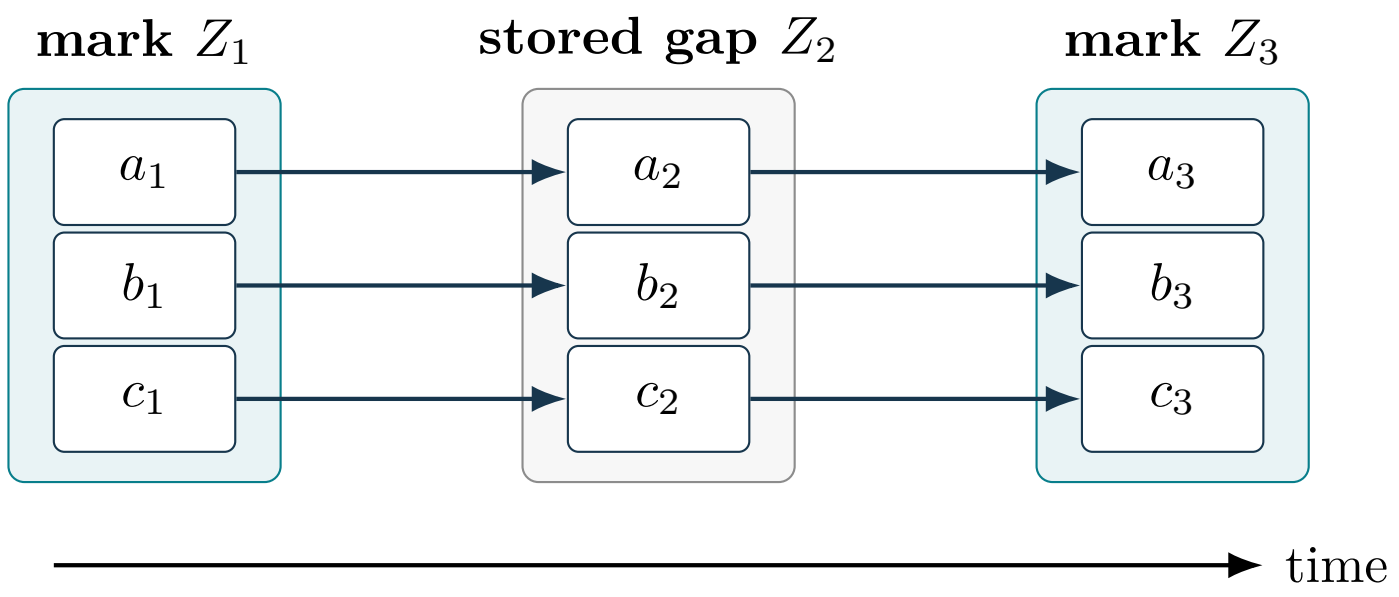
\begin{figure}[htbp]
\centering
\begin{tikzpicture}[>=Latex,x=1cm,y=1cm,
 job/.style={draw=ink,rounded corners=2pt,fill=white,minimum width=12mm,minimum height=7mm,font=\small}]
\filldraw[fill=accent!9,draw=accent,rounded corners=3pt] (-0.9,0.55) rectangle (0.9,-2.05);
\filldraw[fill=black!3,draw=black!45,rounded corners=3pt] (2.5,0.55) rectangle (4.3,-2.05);
\filldraw[fill=accent!9,draw=accent,rounded corners=3pt] (5.9,0.55) rectangle (7.7,-2.05);
\foreach \row/\letter in {0/a,1/b,2/c}{
 \node[job] (a\row) at (0,-0.75*\row) {$\letter_1$};
 \node[job] (b\row) at (3.4,-0.75*\row) {$\letter_2$};
 \node[job] (c\row) at (6.8,-0.75*\row) {$\letter_3$};
 \draw[->,thick,ink] (a\row) -- (b\row);
 \draw[->,thick,ink] (b\row) -- (c\row);
}
\node[above,font=\small\bfseries] at (0,0.58) {mark $Z_1$};
\node[above,font=\small\bfseries] at (3.4,0.58) {stored gap $Z_2$};
\node[above,font=\small\bfseries] at (6.8,0.58) {mark $Z_3$};
\draw[->,thick] (-0.6,-2.6) -- (7.4,-2.6) node[right,font=\small] {time};
\end{tikzpicture}
\caption{Three disjoint chains illustrate $m=3$; use $m$ chains for
general capacity. The longer certificate uses the middle full slot
as an already accepted gap between two marked slots.}
\label{af:figure-chains}
\end{figure}

\paragraph{Step 1: the required vectors are generated.}
Each $Z_t$ is an $m$-set and hence an atom. The global cut
$P_{Z_3}=Z_1\cup Z_2$ is another atom, as is
  $D_{Z_1}=Z_2\cup Z_3$. Both families also contain
  $\varnothing$ and $W=J$. Thus all the vectors below occur in both
 generated families without knowing the displayed schedule.

\paragraph{Step 2: single-slot sets are accepted.}
When the algorithm processes $Z_2$, it generates
$(\varnothing,Z_2,\varnothing)$. Its initial and final queries both
ask for the empty set, whose stored schedule is empty. It therefore
stores $\sigma(Z_2)=(Z_2)$. The same holds for $Z_1$ and $Z_3$.
The two-slot set $Z_2\cup Z_3$ is then accepted with the state
$(\varnothing,Z_2,Z_3)$, and stores the schedule $(Z_2,Z_3)$.

\paragraph{Step 3: the parent has two useful states.}
Using the cuts $\varnothing$ and $P_{Z_3}$ gives
\[
 s=(\varnothing,Z_1,Z_2\cup Z_3),\qquad
 t=(Z_1\cup Z_2,Z_3,\varnothing).
\]
Both are compatible. State $s$ is initial and $t$ is final. The
transition test calculates the actual vector
\[
 G=L_t\setminus(L_s\cup Z_s)
  =(Z_1\cup Z_2)\setminus Z_1=Z_2.
\]
Its lookup succeeds because step 2 stored a schedule of this exact set.
Thus $s\to t$ is a valid edge of the conceptual state graph. In this
small example, $t$ is also initial: its prefix $Z_1\cup Z_2$ is itself
an accepted candidate. An already initial state retains its
$\mathsf{START}$ pointer.

\paragraph{Step 4: the deterministic execution uses its first final state.}
With the fixed orders, $s$ is the first state: its cut is empty and
$Z_1$ is the only full slot consisting of minimal jobs. Its suffix
$Z_2\cup Z_3$ was accepted in step 2. Therefore $s$ is already both
initial and final. The procedure reconstructs the one-state path,
outputting the empty initial schedule, $Z_1$, and the stored schedule
$(Z_2,Z_3)$. Its result is $(Z_1,Z_2,Z_3)$.

\paragraph{How to read the nonempty-gap edge.}
The longer path $s\to t$ is also a valid accepting certificate. Along
that path, the pieces would be the empty initial schedule, $Z_1$,
the stored schedule $\sigma(Z_2)$, $Z_3$, and the empty final schedule.
This explains the nonempty gap in Figure~\ref{af:figure-chains}; it
does not assert that $t$ receives predecessor $s$ in the deterministic
execution. The soundness theorem applies to both certificates. The
structural theorem guarantees a suitable certificate for every
feasible instance, extending the example beyond disjoint chains.

\begin{figure}[htbp]
\centering
\begin{tikzpicture}[x=1cm,y=1cm,>=Latex,
 block/.style={draw=ink,minimum height=10mm,font=\small,align=center}]
\node[block,fill=black!4,minimum width=19mm] (g0) at (0,0) {$A_0$};
\node[block,fill=accent!12,minimum width=14mm] (z1) at (2.0,0) {$Z_1$};
\node[block,fill=black!4,minimum width=23mm] (g1) at (4.15,0) {$A_1$};
\node[block,fill=accent!12,minimum width=14mm] (z2) at (6.3,0) {$Z_2$};
\node[block,fill=black!4,minimum width=19mm] (g2) at (8.3,0) {$A_2$};
\draw[->] (g0) -- (z1); \draw[->] (z1) -- (g1);
\draw[->] (g1) -- (z2); \draw[->] (z2) -- (g2);
\draw[thick,accent] (-0.9,-0.8) -- (-0.9,-1.0) -- (0.9,-1.0) -- (0.9,-0.8);
\node[below,font=\small] at (0,-1.03) {$L_1=A_0$};
\draw[thick,accent] (-0.9,0.8) -- (-0.9,1.0) -- (5.28,1.0) -- (5.28,0.8);
\node[above,font=\small] at (2.2,1.03) {$L_2=A_0\cup Z_1\cup A_1$};
\end{tikzpicture}
\caption{In general, the queried edge gap is
$A_1=L_2\setminus(L_1\cup Z_1)$. The current prefix can grow through
several marks. A table entry is required for each gap, not for every
intermediate prefix.}
\label{af:figure-transition}
\end{figure}
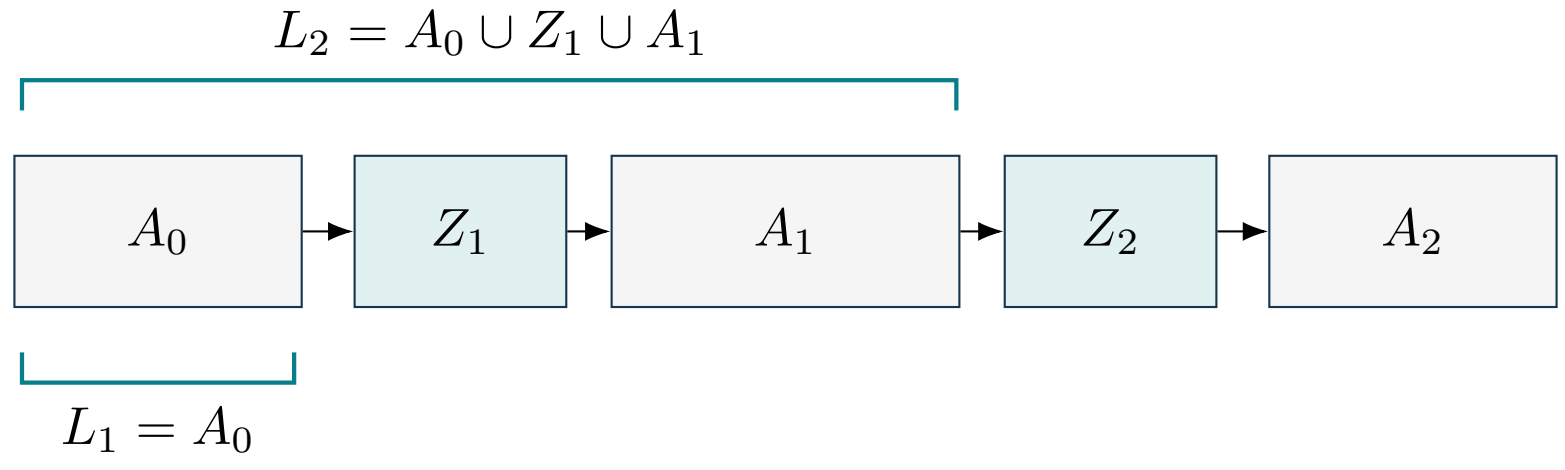

Figure \ref{af:figure-transition} illustrates a general decomposition; its marked-slot labels are independent of the labels in the preceding example. To clarify the construction, consider a feasible full schedule \((S_1,S_2,S_3,S_4,S_5,S_6)\), where each \(S_i\) is a slot containing exactly \(m\) jobs, and mark slots \(S_2\) and \(S_5\). In Figure \ref{af:figure-transition}, these marks are denoted by \(Z_1=S_2\) and \(Z_2=S_5\). The three gaps are then \(A_0=S_1\), \(A_1=S_3\cup S_4\), and \(A_2=S_6\). Thus, a gap may contain several consecutive slots. The prefixes strictly before the marks are \(L_1=S_1\) and \(L_2=S_1\cup S_2\cup S_3\cup S_4\), so the transition between the corresponding states queries precisely \(L_2\setminus(L_1\cup Z_1)=S_3\cup S_4=A_1\). The complete schedule is reconstructed by concatenating the stored schedule of \(A_0\), the marked slot \(S_2\), the stored schedule of \(A_1\), the marked slot \(S_5\), and the stored schedule of \(A_2\). The preceding three-slot example is the special case with empty initial and final gaps: its first and third slots are marked, and its middle slot is the intervening gap. Consequently, the set called \(Z_2\) in that example plays the role of \(A_1\), while its \(Z_3\) plays the role of the second mark labeled \(Z_2\) in Figure \ref{af:figure-transition}.

\section{Correctness}\label{af:sec-correctness}

There are two reasons the algorithm is correct. First, every accepted
table row stores a valid schedule; this property holds for every finite
  choice of node and cut families. Second, the bounded-description theorem
ensures that the generated families contain enough sets and cuts to accept
 a feasible instance.

\subsection{Discrete schedules, padding, and exact family generation}

\begin{lemma}[Why integral slots suffice]\label{af:integral}
Every feasible nonpreemptive unit-job schedule with nonnegative real
start times can be replaced by a slot schedule of no larger makespan.
\end{lemma}
\begin{proof}
Add an arc from every job to the next job on its assigned machine.
Along every original or added arc, the old start time increases by at
least one, so the augmented graph is acyclic. Process it topologically,
starting each job at the maximum of zero and the completion times of
its augmented predecessors. Induction gives integer start times no
later than their old values. The added arcs prevent overlap on a
machine, and the original arcs preserve precedence. Number the slots
by start time plus one.
\end{proof}

\begin{lemma}[Padding a deadline]\label{af:padding}
For $n>0$, a deadline below $\lceil n/m\rceil$ is infeasible, and a
deadline at least $n$ is feasible. For a remaining deadline $T$, add
$mT-n$ isolated jobs. If $J$ is the resulting job set and $N=|J|=mT$,
the original instance is feasible by $T$ if and only if $J$ can be
scheduled in $T$ slots containing exactly $m$ jobs each. Moreover,
$N\leq mn$.
\end{lemma}
\begin{proof}
There are $mT$ positions in $T$ slots. Fill the $mT-n$ unused positions
of any original schedule with the isolated jobs. Conversely, deleting
those jobs from a padded schedule preserves capacity and precedence.
Capacity proves the lower bound on $T$, and a serial topological
schedule proves the upper bound. The empty original instance has
optimum zero and is handled directly.
\end{proof}

\begin{lemma}[Exactness of the generation procedure]
\label{af:family-exact}
After layer $k$, the stored list $F_k$ contains exactly
$\mathcal F_k$. In particular, each $F_k$ is closed under complement
relative to $J$, although it need not be closed under arbitrary union
or intersection within the same layer.
\end{lemma}
\begin{proof}
De Morgan's laws and cancellation of double complements turn every
formula into one with complements only on atoms, without increasing
its leaf count. Hence a one-leaf formula is a signed atom, giving the
claim for $k=1$.

Assume the claim for smaller layers. Every copied set uses at most
$k-1$ leaves. Every newly inserted set combines a formula with at most
$i$ leaves and one with at most $k-i$ leaves, so it uses at most $k$.
This proves that no extra set is inserted.

Conversely, consider a normalized formula with at most $k$ leaves.
A signed atom is already in $F_1$. Otherwise its root is union or
intersection. If its left subtree has $i$ leaves, then $1\leq i<k$
and its right subtree has at most $k-i$ leaves. By induction their
sets occur in $F_i$ and $F_{k-i}$; the appropriate operation inserts
the resulting set. Every formula is therefore covered. Complementing
a formula preserves its leaf count, so exactness also proves complement
closure of $F_k$.
 The trie suppresses only repeated copies of exactly the same incidence
 vector, so duplicate detection changes none of these set memberships.
\end{proof}

\subsection{Progress, reachability, and soundness}

\paragraph{Purpose of the progress bound.}

The evaluator processes candidate sets in increasing order of their
cardinality. To justify this order, we must show that every call to
\(\operatorname{Lookup}\) or \(\operatorname{Fetch}\) made while
processing a nonempty set \(W\) concerns a strictly smaller set.
We must also show that state transitions make strict progress, so that
a path cannot return to a state with the same or a smaller prefix.
Both facts follow from the presence of a marked slot containing
exactly \(m\) jobs.

\begin{lemma}[All queries are proper and all state edges progress]
\label{lem:sec5-expanded-proper}
\label{af:proper-queries}
 While processing a nonempty candidate \(W\), every set passed to
\(\operatorname{Lookup}\) or \(\operatorname{Fetch}\) has cardinality
at most \(|W|-m\). Moreover, every state edge \(i\longrightarrow j\)
satisfies
\[
|L_j|\geq |L_i|+m.
\]
\end{lemma}

\begin{proof}
Rows whose cardinalities are not divisible by \(m\) are skipped and
make no such calls. Consider a row \(W\) for which the evaluator
constructs the state list. Every state record \(i\) is a disjoint
partition
\[
W=L_i\mathbin{\dot\cup} Z_i\mathbin{\dot\cup}R_i,
\qquad |Z_i|=m.
\]
We examine all types of queries.

The initial-state test queries \(L_i\), and the final-state test
queries \(R_i\). Both sets are contained in \(W\setminus Z_i\).
Consequently,
\[
|L_i|\leq |W|-m,
\qquad
|R_i|\leq |W|-m.
\]

For a proposed transition from \(i\) to \(j\), the queried gap is
\(
G_{ij}:=L_j\setminus(L_i\cup Z_i).
\)
Since \(L_j\subseteq W\) and this difference excludes every job of
\(Z_i\), we have
\(
G_{ij}\subseteq W\setminus Z_i.
\)
Thus \(|G_{ij}|\leq |W|-m\). This bound holds whenever the gap is
queried, irrespective of the answer returned by
\(\operatorname{Lookup}\).

Reconstruction fetches only an initial prefix, the gaps on the
selected path, and a final suffix. These are precisely the types of
sets just considered, so the same bound applies to every call to
\(\operatorname{Fetch}\).

Finally, an actual state edge \(i\longrightarrow j\) satisfies
\(L_i\cup Z_i\subseteq L_j\). As \(L_i\) and \(Z_i\) are disjoint,
\[
|L_j|
\geq |L_i\cup Z_i|
=|L_i|+|Z_i|
=|L_i|+m.
\]
This proves both assertions.
\end{proof}

\paragraph{Consequences for evaluation and reconstruction.}
Every queried set that belongs to the family has an earlier table
row, since its cardinality is strictly smaller than \(|W|\); the
empty row is already accepted at initialization. A lookup for a missing
row returns zero. Hence all lookup answers used to define
the state graph remain fixed while \(W\) is processed.

The prefix-size inequality also makes the state graph acyclic.
Following a predecessor pointer backward decreases prefix size by
at least \(m\). More generally, a path with \(r\) state records
satisfies
\[
(r-1)m
\leq |L_{\mathrm{last}}|-|L_{\mathrm{first}}|
\leq |W|-m,
\]
and therefore \(r\leq |W|/m\).

The cuts enumerated by Algorithm~3 are not the queried subproblems
  in this lemma. A cut \(C\in\mathcal F_{\mathrm{cut}}\) is used through its incidence
  vector to form \(L=W\cap C\). Its own cardinality may be as large as
  \(|J|\). No acceptance value for the cut is consulted.
The proper-query bound applies to the arguments of
\(\operatorname{Lookup}\) and \(\operatorname{Fetch}\).

\paragraph{Why one pass is sufficient.}
 The family trie identifies an exact candidate row, and
\textsc{Lookup} returns its stored bit $a(X)$, which may be zero.
 Step~4 processes sources in increasing order of $|L_i|$ and reads
a source's current reachability bit when its bucket is visited.
 Every edge satisfies $|L_j|\ge |L_i|+m$, so every newly reached
target belongs to a later bucket.
 Consequently, a source cannot become reachable after its bucket
has been passed, and one pass suffices.

\begin{lemma}[The pass computes exactly the reachable states]
\label{af:reachability}
At the end of Step 4 of Algorithm~\ref{af:alg-evaluate}, a record is
marked exactly when it is reachable from an initial record by edges
\eqref{af:eq-edge}. Its predecessor fields describe such a path.
\end{lemma}
\begin{proof}
  All queried table rows are smaller than $W$ and stay fixed while $W$
 is processed, so the edge relation is fixed during Step~4.
Initially only initial records are marked. Every later marking uses a
marked source and a valid edge, so induction on marking events proves
the forward implication and validates the stored predecessor.

For the converse, induct on the prefix-cardinality buckets processed
  in Step~4. Every initial record is marked before any bucket is processed.
 If a reachable record $j$ is not initial, choose a path to it and let
  $i$ be its predecessor on that path. By
Lemma~\ref{af:proper-queries}, $|L_i|<|L_j|$, so the bucket of $i$
  was processed earlier. The induction hypothesis says that $i$ was
  already marked when it was processed. Its scan of all target indices
  therefore marked $j$, unless $j$ had already been marked. Thus every
 reachable record is marked before its own bucket is processed.

There are no edges within one bucket. Equal prefix sizes and duplicate
 state triples therefore cause no difficulty. Predecessor pointers cannot cycle, since their
edges strictly increase prefix size by Lemma~\ref{af:proper-queries}.
\end{proof}

\begin{theorem}[Soundness for every explicit family]
\label{af:soundness}
Every row assigned value one by Algorithm~\ref{af:alg-evaluate} stores
  a feasible full schedule of its set. This holds for every family
$\mathcal F$ containing $\varnothing$ and $J$, and every finite cut family
 $\mathcal F_{\mathrm{cut}}\subseteq2^J$.
\end{theorem}
\begin{proof}
Induct on the cardinality of processed sets. The empty row is correct.
By Lemma~\ref{af:proper-queries}, every successful nonempty query while
processing $W$ is to an earlier row, whose stored schedule is feasible
by the induction hypothesis.

An initial record has a stored schedule of $L_i$. Suppose a reached
record $i$ certifies a full schedule of $L_i$ and leads to record $j$.
Let $G=L_j\setminus(L_i\cup Z_i)$. The edge test supplies a full
schedule of $G$. Since $L_i$ is an ideal of $W$, no job of $Z_i$
precedes a job of $L_i$. Since $L_i\cup Z_i$ is an ideal, no job of
$G$ precedes a job of $L_i\cup Z_i$. Thus the prefix schedule, the
single antichain slot $Z_i$, and the schedule of $G$ concatenate to a
feasible full schedule of $L_j$.

At a reached final record $t$, the same ideal conditions permit
appending $Z_t$ and the accepted schedule of $R_t$. The result is a full
schedule of $W$. Lemma~\ref{af:reachability} supplies precisely the
path followed by Algorithm~\ref{af:alg-reconstruct}, so the schedule
actually stored by the algorithm is the one justified by this argument.
\end{proof}

\subsection{Why every feasible root is accepted}

\paragraph{The global job set and a local hierarchy node.}
Throughout this subsection, \(J\) denotes the fixed padded job set,
and \(\mathcal F\subseteq 2^J\) is a finite family containing
 \(\varnothing\) and \(J\). The cut family \(\mathcal F_{\mathrm{cut}}\subseteq2^J\)
 may be chosen independently. Every subset inherits the restriction of
the global transitive precedence relation. The symbol \(W\) denotes
the job set of the node currently under consideration. At the root,
\(W=J\); at a proper descendant, \(W\) is a nonempty subset of its
parent's job set. The global universe remains \(J\) throughout.

A hierarchy describes how to assemble schedules of larger sets from
schedules of smaller sets and selected full slots. The algorithm
processes every candidate in \(\mathcal F\). The completeness proof
singles out the candidate sets that occur in one suitable hierarchy.

\begin{definition}[A hierarchy described by two families]
\label{def:sec5-expanded-hierarchy}
\label{af:def-hierarchy}
 A \emph{hierarchy} for a feasible full instance on \(J\) is a finite
rooted tree whose nodes are labelled by job sets, with root \(J\).
At each nonempty node \(W\), choose a feasible full schedule
\(
\mathcal S_W=(S_1,\ldots,S_q),
\quad q=\frac{|W|}{m},
\)
and choose marked positions
\(
1\leq t_1<\cdots<t_k\leq q,
\quad 1\leq k\leq q.
\)
The corresponding marked slots are \(Z_i:=S_{t_i}\).

Set \(t_0:=0\) and \(t_{k+1}:=q+1\). For \(i=0,\ldots,k\), define
the \emph{open gap}
\[
A_i:=\bigcup_{t_i<t<t_{i+1}} S_t,
\]
where an empty union is \(\varnothing\). The schedule is thus divided
into the ordered pieces
\[
A_0,\ Z_1,\ A_1,\ Z_2,\ \ldots,\ Z_k,\ A_k.
\]
Here a gap denotes the set of jobs in the indicated block of
consecutive slots. The children of \(W\) are exactly its nonempty
gap sets \(A_i\), one child for each such gap. The same construction
is made at every nonempty child. An empty root has no children.

The hierarchy is \emph{described by \((\mathcal F,\mathcal F_{\mathrm{cut}})\)} if every node
set belongs to \(\mathcal F\) and, at every nonempty node \(W\),
the prefix
\(
L_i:=\bigcup_{t<t_i}S_t
\)
before each marked slot \(Z_i\) satisfies
\(
L_i=W\cap C_i
\)
 for some global cut \(C_i\in\mathcal F_{\mathrm{cut}}\).
\end{definition}

\paragraph{Open gaps, marked slots, and the number of children.}
For \(1\leq i<k\), the gap \(A_i\) contains exactly the jobs scheduled
strictly between \(Z_i\) and \(Z_{i+1}\). The boundary marked slots
are excluded. The set \(A_0\) contains the jobs before the first
mark, and \(A_k\) contains those after the last mark. These gaps are
blocks of occupied slots; the word \emph{gap} refers to their position
between marks. A gap can contain several full slots, or can be empty.

All pieces are disjoint, every \(|A_i|\) is divisible by \(m\), and
\(
|W|=mk+\sum_{i=0}^{k}|A_i|.
\)
Consequently, every child satisfies
\[
|A_i|\leq |W|-mk\leq |W|-m<|W|.
\]
This strict decrease permits induction on node size.

The integer \(k=k(W)\) counts marked slots at this particular node.
It may vary between nodes and need not be bounded by \(m\).
There are \(k+1\) gaps and at most \(k+1\) children. A node with all
its slots marked has only empty gaps and is a leaf, even if it
contains more than one slot.

Each node has its own witness schedule and its own local slot
indices. A child may choose any feasible full schedule of its job
set; it need not use the order induced by its parent's witness.

\paragraph{The states associated with the marked slots.}
At a fixed node \(W\), let
\(
R_i:=W\setminus(L_i\cup Z_i).
\)
The triple \((L_i,Z_i,R_i)\) consists of all jobs before the mark,
the marked slot itself, and all jobs after the mark. In terms of
the gap decomposition,
\[
\begin{aligned}
L_i&=
\left(\bigcup_{j=0}^{i-1}A_j\right)
\cup
\left(\bigcup_{j=1}^{i-1}Z_j\right),\\
R_i&=
\left(\bigcup_{j=i}^{k}A_j\right)
\cup
\left(\bigcup_{j=i+1}^{k}Z_j\right).
\end{aligned}
\]
All unions in these formulas are disjoint. In particular,
\(
L_1=A_0,\quad R_k=A_k,
\)
and, for \(1\leq i<k\),
\(
L_{i+1}=L_i\mathbin{\dot\cup}Z_i\mathbin{\dot\cup}A_i,
\quad
A_i=L_{i+1}\setminus(L_i\cup Z_i).
\)
Thus the \(k\) selected marked slots will give \(k\) state records
and \(k-1\) connecting edges in the completeness proof. Their
record indices in the algorithm need not be \(1,\ldots,k\).

\paragraph{Global cuts and local prefixes.}
Every \(C_i\) is a subset of the fixed universe \(J\).
At the root, \(W=J\), so \(L_i=J\cap C_i=C_i\).
At a smaller node, the cut may contain jobs outside \(W\); only
its intersection with \(W\) determines the local prefix.
For a fixed \(L\subseteq W\), the sets satisfying \(W\cap C=L\)
are precisely
\(
C=L\cup U,\qquad U\subseteq J\setminus W.
\)
  The description requirement asks that at least one such set belong
 to \(\mathcal F_{\mathrm{cut}}\). Cuts need not be unique or be ideals of \(J\).
The local prefix is an ideal of \(W\), because it is a temporal
prefix of a feasible schedule.

\paragraph{Example: three chains and seven full slots.}
Let \(m=3\), and consider three independent precedence chains
\[
\begin{aligned}
a_1&\prec a_2\prec\cdots\prec a_7,\\
b_1&\prec b_2\prec\cdots\prec b_7,\\
c_1&\prec c_2\prec\cdots\prec c_7.
\end{aligned}
\]
There are no precedence relations between distinct chains. Put
\[
B_t:=\{a_t,b_t,c_t\}\quad (1\leq t\leq 7),
\qquad
J=\bigcup_{t=1}^{7}B_t.
\]
The blocks \(B_t\) are pairwise disjoint antichains of size three, and
\(
(B_1,B_2,B_3,B_4,B_5,B_6,B_7)
\)
is a feasible full schedule of the 21 jobs. No dummy jobs are needed.

At the root \(W=J\), choose the second and sixth slots as marks:
\(
k=2,\quad Z_1=B_2,\quad Z_2=B_6.
\)
The resulting decomposition is
\[
\underbrace{B_1}_{A_0},\
\underbrace{B_2}_{Z_1},\
\underbrace{B_3,B_4,B_5}_{A_1},\
\underbrace{B_6}_{Z_2},\
\underbrace{B_7}_{A_2}.
\]
Thus
\(
A_0=B_1,\quad
A_1=B_3\cup B_4\cup B_5,\quad
A_2=B_7.
\)

The middle gap contains nine jobs in three full slots. The two
associated states are
\[
\begin{aligned}
(L_1,Z_1,R_1)
&=\left(B_1,\ B_2,\ \bigcup_{t=3}^{7}B_t\right),\\
(L_2,Z_2,R_2)
&=\left(\bigcup_{t=1}^{5}B_t,\ B_6,\ B_7\right).
\end{aligned}
\]
Their connecting gap is exactly
\[
L_2\setminus(L_1\cup Z_1)
=B_3\cup B_4\cup B_5=A_1.
\]

\paragraph{Example continued: a gap becomes a child.}
The root has the three children \(B_1\), \(A_1\), and \(B_7\).
Write
\(
W_\ast:=A_1=B_3\cup B_4\cup B_5
\)
for its middle child. The global set remains \(J\), while the
current node now contains only nine jobs. Give this child the
local schedule
\(
(S'_1,S'_2,S'_3)=(B_3,B_4,B_5).
\)
Mark its second local slot, so that
\(
k'=1,\quad t'_1=2,\quad Z'_1=B_4.
\)
Its decomposition and associated state are
\(
A'_0=B_3,\quad Z'_1=B_4,\quad A'_1=B_5,
\quad
(L'_1,Z'_1,R'_1)=(B_3,B_4,B_5).
\)
The children of \(W_\ast\) are therefore \(B_3\) and \(B_5\).

At each of the four leaves \(B_t\), \(t\in\{1,3,5,7\}\), mark its
only slot. This gives the state
\(
(\varnothing,B_t,\varnothing)
\)
and no nonempty child. The hierarchy can be recorded as follows.
\begin{center}
\begin{tabular}{lll}
\hline
Node set & Marked slots & Nonempty gap children\\
\hline
\(J\) & \(B_2,B_6\) & \(B_1,W_\ast,B_7\)\\
\(W_\ast\) & \(B_4\) & \(B_3,B_5\)\\
\(B_t,\ t\in\{1,3,5,7\}\) & \(B_t\) & none\\
\hline
\end{tabular}
\end{center}
The marked slots \(B_2,B_6\) are handled at the root, and \(B_4\)
is handled at \(W_\ast\). Only the nonempty gaps become children.

\paragraph{Example continued: global cuts for these local states.}
Recall that
\[
P_B=\{x\in J:\ x\prec y\text{ for some }y\in B\},
\qquad
D_B=\{x\in J:\ y\prec x\text{ for some }y\in B\}.
\]
For the root states, take
\(
C_1=P_{B_2}=B_1,
\quad
C_2=P_{B_6}=\bigcup_{t=1}^{5}B_t.
\)
Since \(W=J\) at the root, these cuts are exactly \(L_1\) and \(L_2\).

For the child \(W_\ast\), choose
\(
C'_1=P_{B_4}=B_1\cup B_2\cup B_3.
\)
This is a global set of nine jobs, whereas the desired local prefix
has only three jobs:
\[
W_\ast\cap C'_1
=(B_3\cup B_4\cup B_5)\cap(B_1\cup B_2\cup B_3)
=B_3=L'_1.
\]
The jobs in \(B_1\cup B_2\) lie outside the current node and disappear
under intersection. The set \(B_3\) is an ideal of \(W_\ast\), but
not of \(J\): for example, \(a_2\prec a_3\), while \(a_2\notin B_3\).

The alternative cut \(B_3\cup B_7\) has the same intersection with
\(W_\ast\). This also illustrates that a global cut need not be a
temporal prefix of the displayed global schedule.
At a leaf, the cut \(\varnothing\) gives the empty local prefix.

All node sets belong to \(\mathcal F=\revchange{\mathcal F_{24}}{\mathcal F_{17}}\),
and all displayed cuts belong to
\(\mathcal F_{\mathrm{cut}}=\revchange{\mathcal F_9}{\mathcal F_6}\), since \(\revchange{4m+12}{4m+5}\revchange{=24}{=17}\)
 and \(\revchange{m+6}{m+3}\revchange{=9}{=6}\) for \(m=3\).
The sets \(J,\varnothing,B_t,P_{B_t},D_{B_t}\) are atoms, and
\(
W_\ast=D_{B_2}\cap P_{B_6}
\)
is an exact global description using two atomic occurrences.
Indeed, \(D_{B_2}=\bigcup_{t=3}^{7}B_t\) and
\(P_{B_6}=\bigcup_{t=1}^{5}B_t\).
The alternative cut \(B_3\cup B_7\) also uses two atomic occurrences.
  Hence this is a hierarchy described by
 \((\revchange{\mathcal F_{24}}{\mathcal F_{17}},\revchange{\mathcal F_9}{\mathcal F_6})\).

Notice that \(|C'_1|=|W_\ast|=9\). This creates no dependency on a
same-size table value: Algorithm~3 uses the incidence vector of
\(C'_1\) to construct a state, without querying whether \(C'_1\)
has been accepted.

\begin{theorem}[A hierarchy with bounded global descriptions]
\label{af:structure}
  Let $m\geq3$. Every feasible full instance has a hierarchy
in the sense of Definition~\ref{af:def-hierarchy} such that every node
set is represented exactly, on the entire ground set $J$, by a Boolean
  formula with at most $\revchange{4m+12}{4m+5}$ atomic occurrences. For each marked slot, its
prefix in its node $W$ is $W\cap C$, where $C$ has a global Boolean
 description with at most $\revchange{m+6}{m+3}$ atomic occurrences. The atoms are the
boundary sets, global predecessor and successor cones, weak cones,
rank suffixes, and constants already used by \textsc{GenerateFamily}.
Every nonempty node has at least one marked full slot, and every child
has at most $|W|-m$ jobs.
\end{theorem}

\paragraph{Proof idea and purpose.}
A described hierarchy supplies a suitable sequence of marked slots
at every node. We show that this sequence gives an accepting path
in the algorithm's state graph. The proof proceeds by induction on
node size. Each gap is empty or is a smaller hierarchy node, so its
table entry has already been accepted. These entries certify the
initial prefix, the connecting gaps, and the final suffix.
  Theorem~5.8 establishes the existence of the required hierarchy
for the two families used by the algorithm; the following theorem explains
 how that hierarchy implies acceptance.

\begin{theorem}[Completeness from a described hierarchy]
\label{af:completeness}
\label{thm:sec5-expanded-completeness}
  Let \(\mathcal F\subseteq 2^J\) be a finite family containing
\(\varnothing\) and \(J\), and let
 \(\mathcal F_{\mathrm{cut}}\subseteq2^J\) be a finite cut family.
  If the full instance on \(J\) has a hierarchy described by
\((\mathcal F,\mathcal F_{\mathrm{cut}})\) in the sense of
Definition~\ref{def:sec5-expanded-hierarchy}, then Algorithm~4
 accepts every node set of the hierarchy. In particular, it accepts
the root \(J\).
\end{theorem}

\begin{proof}
We prove by strong induction on \(|W|\) that the row of every hierarchy
node \(W\) has value \(a(W)=1\) after it is processed. The empty row
is accepted at initialization, with its empty schedule; this also
settles an empty root.

Let \(W\) be a nonempty hierarchy node, and assume the claim for
  all smaller hierarchy nodes. Since the hierarchy is described by
 \((\mathcal F,\mathcal F_{\mathrm{cut}})\), its node set \(W\) belongs to \(\mathcal F\).
Its witness schedule is full, so \(m\) divides \(|W|\), and
Algorithm~4 does not skip this row.

Use the witness schedule at this node to obtain the ordered
decomposition
\[
A_0,Z_1,A_1,Z_2,\ldots,Z_k,A_k,
\qquad k\geq 1.
\]
Every nonempty \(A_i\) is a child of \(W\), belongs to \(\mathcal F\),
and satisfies
\[
|A_i|\leq |W|-mk\leq |W|-m<|W|.
\]
Its row has therefore already been processed and accepted by the
induction hypothesis. Empty gaps have the initialized empty row.
It follows that
\(
\operatorname{Lookup}(A_i)=1
\quad (i=0,\ldots,k).
\)
These values stay fixed during the processing of \(W\).

For each \(i=1,\ldots,k\), let \(L_i\) be the prefix before \(Z_i\)
and put \(R_i=W\setminus(L_i\cup Z_i)\).
  The description hypothesis provides a cut \(C_i\in\mathcal F_{\mathrm{cut}}\)
 such that \(L_i=W\cap C_i\).
We verify that Algorithm~3 enumerates the triple
\((L_i,Z_i,R_i)\).

First, the three sets partition \(W\), the marked slot has size
\(|Z_i|=m\), and both \(|L_i|\) and \(|R_i|\) are divisible by \(m\).
These assertions follow because the partition boundaries occur
between complete slots of the witness schedule. The set \(Z_i\)
is an antichain, since it is one feasible slot.
Second, \(L_i\) and \(L_i\cup Z_i\) are ideals of \(W\).
To see this, let \(U\) be either of these temporal prefixes.
If \(y\in U\), \(x\in W\), and \(x\prec y\), feasibility places
\(x\) in a strictly earlier slot than \(y\). Hence \(x\in U\).
Thus both ideal tests succeed.

Algorithm~3 considers the cut \(C_i\), obtains \(L_i\), and
enumerates the \(m\)-set \(Z_i\). All its tests succeed, so a record
with the required triple occurs. Choose one such record and denote
its actual index by \(r_i\). Different cuts may produce duplicate
records; any one record with the required triple is sufficient.

The first selected record is initial, because
\(
L_1=A_0
\quad\text{and}\quad
\operatorname{Lookup}(A_0)=1.
\)
For every \(i=1,\ldots,k-1\), the ordered decomposition gives
\(
L_{i+1}=L_i\mathbin{\dot\cup} Z_i\mathbin{\dot\cup} A_i.
\)
Consequently,
\(
L_i\cup Z_i\subseteq L_{i+1},
\quad
L_{i+1}\setminus(L_i\cup Z_i)=A_i.
\)
Since \(\operatorname{Lookup}(A_i)=1\), the edge test succeeds and
the state graph contains
\(
r_i\longrightarrow r_{i+1}.
\)
Finally,
\(
R_k=A_k
\quad\text{and}\quad
\operatorname{Lookup}(A_k)=1,
\)
so \(r_k\) is final.

Thus \(r_1,\ldots,r_k\) is a path from an initial record to a final
record. If \(k=1\), the single record is both initial and final,
and no edge is needed. Lemma~5.5 guarantees that the reachability
procedure marks \(r_k\). The final inspection in Algorithm~4
therefore finds a reached final record and sets \(a(W)=1\).

This completes the induction. Applying it to the root \(W=J\)
proves the theorem.
\end{proof}

\paragraph{Witness schedules and stored schedules.}
The witness schedule at \(W\) identifies the sets in the
decomposition and proves the required state conditions.
The schedule stored for a gap \(A_i\) may be a different feasible
full schedule of exactly that job set. It has the same length
\(|A_i|/m\). Substituting it within that gap preserves precedence
inside the gap by its own feasibility, and preserves precedence
between different pieces because their order agrees with the
feasible witness schedule. Hence
\(
\sigma(A_0),Z_1,\sigma(A_1),\ldots,Z_k,\sigma(A_k)
\)
is a feasible full-schedule certificate.

The proof exhibits an accepting path. The deterministic evaluator
may select another reached final record or other predecessor
pointers; Theorem~5.6 guarantees the validity of the schedule
reconstructed from its actual choices.

\paragraph{Acceptance in the seven-slot example.}
We now follow the induction for the hierarchy constructed after
Definition~\ref{def:sec5-expanded-hierarchy}.
The relevant node sizes are three at the leaves, nine at the
middle child \(W_\ast\), and 21 at the root.

For each leaf \(B_t\), \(t\in\{1,3,5,7\}\), the cut
\(\varnothing\) and marked slot \(B_t\) produce the state
\(
(\varnothing,B_t,\varnothing).
\)
Both its prefix and suffix lookups return one, since
\(\operatorname{Lookup}(\varnothing)=1\).
The state is simultaneously initial and final, so the leaf is
accepted with its one-slot schedule \((B_t)\).

At \(W_\ast=B_3\cup B_4\cup B_5\), the cut \(P_{B_4}\)
produces the state
\(
(B_3,B_4,B_5).
\)
The leaf rows \(B_3\) and \(B_5\) have already been accepted.
This state is therefore both initial and final, proving that
\(W_\ast\) is accepted. Its displayed certificate concatenates
\(
\sigma(B_3),\ B_4,\ \sigma(B_5)
=(B_3,B_4,B_5).
\)

At the root, let \(r_1,r_2\) be records with the two triples
\[
\left(B_1,B_2,\bigcup_{t=3}^{7}B_t\right),
\qquad
\left(\bigcup_{t=1}^{5}B_t,B_6,B_7\right).
\]
The first record is initial because
\(\operatorname{Lookup}(B_1)=1\). The edge from the first record
to the second is present because
\(
L_1\cup Z_1=B_1\cup B_2\subseteq L_2
\)
and
\(
\operatorname{Lookup}
\bigl(L_2\setminus(L_1\cup Z_1)\bigr)
=\operatorname{Lookup}(W_\ast)=1.
\)
The second record is final because
\(\operatorname{Lookup}(B_7)=1\).
Hence \(r_1\longrightarrow r_2\) is an accepting path.
The exhibited certificate concatenates
\(
\sigma(B_1),\ B_2,\ \sigma(W_\ast),\ B_6,\ \sigma(B_7),
\)
which expands to the full schedule
\(
(B_1,B_2,B_3,B_4,B_5,B_6,B_7).
\)

The required lookup sets are the three root gaps
\(B_1,W_\ast,B_7\), the two child gaps \(B_3,B_5\), and the empty
gaps at the leaves. Each is strictly smaller than the parent at
which it is queried, as guaranteed by
Lemma~\ref{lem:sec5-expanded-proper}.

\begin{remark}[Why no closure assumption is needed]
Although $L=W\cap C$ defines a state, the algorithm does not assume that
$L$ belongs to $\mathcal F$. Similarly, an edge difference need not
belong to $\mathcal F$. An absent nonempty query returns zero. In the
successful hierarchy, each nonempty queried gap has its \emph{own} node
description and therefore belongs to the family. The proof requires
those specific memberships, not closure under arbitrary intersections
or differences.
\end{remark}

\begin{lemma}[Reconstruction has no repeated recursive expansion]
\label{af:reconstruction}
An accepting path for $W$ contains at most $|W|/m$ marks. Its marked
slots and fetched gaps form an ordered disjoint partition of $W$.
Algorithm~\ref{af:alg-reconstruct} copies exactly $|W|$ job occurrences
into the stored schedule of $W$.
\end{lemma}
\begin{proof}
For a path $i_1,\ldots,i_k$, the edge inclusions yield
\[
\begin{split}
W={}&L_{i_1}\mathbin{\dot\cup} Z_{i_1}
\mathbin{\dot\cup}
\bigl(L_{i_2}\setminus(L_{i_1}\cup Z_{i_1})\bigr)
\mathbin{\dot\cup}\cdots\\
&\mathbin{\dot\cup} Z_{i_k}\mathbin{\dot\cup}R_{i_k}.
\end{split}
\]
All displayed gaps have successful lookups, and the soundness proof
permits their schedules to be concatenated in this order. There are
$k$ disjoint sets of $m$ marked jobs, so $k\leq |W|/m$. Every job
appears exactly once in the partition and therefore exactly once in
the copied output. State records are accessed directly by their indices.
  Locating a stored gap schedule by Algorithm~\ref{af:alg-lookup} follows
 its incidence vector through the node trie, costing $O(N)$ RAM operations.
This search cost and the cost of copying the schedule are included in Section~\ref{af:sec-runtime}.
\end{proof}

\begin{theorem}[Correctness of the deadline and optimization algorithms]
\label{af:correct}
For $m\geq3$, Algorithms~\ref{af:alg-deadline} and~\ref{af:alg-optimum} decide deadline
feasibility exactly and return an optimum schedule, respectively.
\end{theorem}
\begin{proof}
For $T<\lceil n/m\rceil$, capacity excludes feasibility; for $T\geq n$,
the serial topological schedule is feasible. In the remaining range,
adding $mT-n$ isolated jobs preserves feasibility in both directions:
fill all unused positions with dummy jobs, or delete them from a full
padded schedule, as proved in Lemma~\ref{af:padding}.

The generated families are exactly $\mathcal F_{\revchange{4m+12}{4m+5}}$ and
 $\mathcal F_{\revchange{m+6}{m+3}}$. If the evaluator
accepts the padded root, Theorem~\ref{af:soundness} gives a full schedule,
  and removing dummies gives a feasible original schedule. If the padded
root is feasible, Theorem~\ref{af:structure} supplies a hierarchy
 described by these two families. Theorem~\ref{af:completeness} then makes the
evaluator accept it. Thus every deadline answer is correct.

For optimization, initially $\ell=\lceil n/m\rceil-1$ is infeasible,
$u=n$ is feasible, and $\sigma_{\mathrm{best}}$ witnesses the upper
bound. Feasibility is monotone in the deadline. Each update therefore
preserves these facts and decreases $u-\ell$. At termination $u=\ell+1$,
so $u$ is the smallest feasible integer deadline. The stored schedule
has makespan at most $u$ and must be optimum. Lemma~\ref{af:integral}
shows that this integer
optimum also equals the continuous-time optimum.
\end{proof}

The proof separates the recurrence from its structural completeness
argument: Section~\ref{af:structural-appendix} proves the hypothesis
used in Theorem~\ref{af:completeness}. The
algorithm itself neither receives a hierarchy as input nor computes the
normalizations used to prove that a hierarchy exists.

\section{Running-time analysis}\label{af:sec-runtime}

This section analyzes Algorithms~\ref{af:alg-family}--\ref{af:alg-optimum}
and proves Theorem~\ref{af:main}.
\hl{For }$\mathrev{n\ge2}$\hl{, the simplifications used in the final bounds follow from}
\(
 \mathrev{n+2\le2n},\quad
 \mathrev{\log(n+2)\le\log(2n)=1+\log n\le2\log n}.
\)
\hl{These are upper bounds, not identities. Thus, for any }$\mathrev{a,b\ge0}$\hl{,}
\(
 \mathrev{(n+2)^a\log^b(n+2)\le2^{a+b}n^a\log^b n}.
\)
For $a=E(m)=6m^2+13m+1$ and $b\in\{0,1,2\}$, the additional
parameter-dependent factors are absorbed into $m^{O(m^2)}$.
The same argument applies to the simpler exponent $11m^2$. This leaves all displayed polynomial degrees unchanged. 
We retain the intermediate $N+2$ estimates below and simplify
them when passing from the padded size $N$ to $n$.

We count elementary operations on a random access machine (RAM).
Sets are stored as Boolean arrays of length $N$, and the transitive
precedence relation as an $N\times N$ Boolean matrix. Access to an
indexed array entry, an assignment, and arithmetic or comparison of
  indices and counters cost one operation. Thus a set operation or
comparison costs $O(N)$ operations, an ideal or antichain test costs
$O(N^2)$, and exact trie lookup in Algorithm~\ref{af:alg-lookup}
  costs $O(N)$, including an unsuccessful query. Exact insertion has
 the same cost. State records are accessed
  by their indices. Words are large enough to hold the working
addresses and counters; for the generated families,
 $O(m^2\log(N+2))$ bits suffice, which is $O(\log(N+2))$ for fixed $m$.
  No operation treats an entire $N$-bit set as one word.
  We discuss bit costs at the end.

We use
 \[
 K_{\mathrm{node}}=\revchange{4m+12}{4m+5},\qquad K_{\mathrm{cut}}=\revchange{m+6}{m+3},
 \qquad
 B_{\mathrm{node}}=|\mathcal F|,
 \quad B_{\mathrm{cut}}=|\mathcal F_{\mathrm{cut}}|.
\]

\subsection{Size and generation cost of the candidate families}

\begin{lemma}[Number of candidates]\label{af:family-count}
For $m\geq3$, $N\geq m$, and $K\geq1$, the family size satisfies
\[
 B:=|\mathcal F_K|\leq B_0:=
 2\cdot160^K(N+2)^{mK}.
\]
  For the node and cut budgets, respectively, this gives
 \[
 \begin{split}
 B_{\mathrm{node}}&\le B_{\mathrm{node},0}
 :=2\cdot160^{\revchange{4m+12}{4m+5}}(N+2)^{\revchange{4m^2+12m}{4m^2+5m}},\\
 B_{\mathrm{cut}}&\le B_{\mathrm{cut},0}
 :=2\cdot160^{\revchange{m+6}{m+3}}(N+2)^{\revchange{m^2+6m}{m^2+3m}}.
 \end{split}
\]
\end{lemma}
\noindent\textit{Proof idea.}
We count normalized formulas by their tree shapes, operation labels,
and signed atomic labels, and then sum over the allowed leaf counts.

\begin{proof}
A \emph{full ordered rooted binary tree} has either zero or two
children at each vertex, with left and right children distinguished.
A formula is \emph{normalized} if its internal vertices are labelled
by $\cup$ or $\cap$ and its leaves by atoms or their complements.
By De Morgan's laws and double-complement cancellation, every formula
in Definition~\ref{af:language} has an equivalent normalized formula with the same
number of atomic occurrences.

A full binary tree with $k$ leaves has exactly $k-1$ internal
vertices~\cite[Example~1.3.22, p.~43]{West2021}. The number $t_k$ of
ordered tree shapes is
\(
 t_k=C_{k-1}
     =\frac1k\binom{2k-2}{k-1}
     \le4^{k-1},
\)
by the classical bijection with Dyck words of length $2(k-1)$;
see \cite[Theorem~1.3.23, p.~44]{West2021} and
\cite[Theorems~1.4.1 and~1.5.1]{Stanley2015}.

Let $M$ denote the length of the indexed atomic list, allowing
different names to represent the same set. For a fixed tree with
$k$ leaves, there are $2^{k-1}$ choices of binary-operation labels
and $(2M)^k$ choices of signed atomic labels. Hence the number $f_k$
of normalized formulas with exactly $k$ leaves satisfies
\(
 f_k
 \le4^{k-1}\,2^{k-1}(2M)^k
 =\frac{(16M)^k}{8}
 \le(16M)^k.
\)
Every set in $\mathcal F_K$ has such a description with at most
$K$ leaves. Therefore, putting $q=16M\ge16$ and using the finite
geometric sum,
\[
 \begin{aligned}
 B
 &\le\sum_{k=1}^K f_k
 \le\sum_{k=1}^K q^k\\
 &=\frac{q^{K+1}-q}{q-1}
 \le q^K\frac{q}{q-1}
 \le2q^K.
 \end{aligned}
\]
Finally, the atomic list contains five names for every $m$-set,
the $N+1$ rank suffixes, and the two constants $J,\varnothing$.
Thus
\( M=5\binom Nm+N+3\le10(N+2)^m,
\)
and consequently
\(
 B\le2(16M)^K
   \le2\bigl(160(N+2)^m\bigr)^K
   =B_0.
\)
  Substituting $K_{\mathrm{node}}=\revchange{4m+12}{4m+5}$ and
 $K_{\mathrm{cut}}=\revchange{m+6}{m+3}$ proves the two displayed specializations.
  In each case, the coefficient $2\cdot160^K$ is $2^{O(m)}$.
\end{proof}

\begin{lemma}[Cost of direct family generation]\label{af:family-cost}
  For $m\ge3$, $N\ge m$, and $K\ge1$,
Algorithm~\ref{af:alg-family}, using binary tries for exact duplicate
detection and including order and atom preprocessing, uses
\[
 O\!\left(
 N^3+mN\binom Nm+N^2+
 K^2\,160^K(N+2)^{mK+1}
 \right)
\]
  RAM operations. Generating all layers through $K_{\mathrm{node}}$
 also supplies the cut layer $K_{\mathrm{cut}}$.
\end{lemma}

\noindent\emph{Proof idea.}
  Each exact insertion costs $O(N)$. To count the attempted unions
and intersections, use the separate bounds for their two source
 layers; their leaf budgets add to the output budget.

\begin{proof}
Let $\mathcal F_i$ denote the list constructed at layer $i$ of
 Algorithm~\ref{af:alg-family}. For a fixed layer $k$ and split
$i+(k-i)=k$, the algorithm considers all pairs in
 $\mathcal F_i\times\mathcal F_{k-i}$. Lemma~\ref{af:family-count} gives
 \[
 \begin{split}
 |\mathcal F_i|\,|\mathcal F_{k-i}|
 &\le
 \bigl(2\cdot160^i(N+2)^{mi}\bigr)
 \bigl(2\cdot160^{k-i}(N+2)^{m(k-i)}\bigr)\\
 &=4\cdot160^k(N+2)^{mk}.
 \end{split}
\]
  Hence the total number of attempted pairs is at most
 \[
 \begin{aligned}
 \sum_{k=2}^K\sum_{i=1}^{k-1}
       |\mathcal F_i|\,|\mathcal F_{k-i}|
 &\le4\cdot160^K(N+2)^{mK}\sum_{k=2}^K(k-1)\\
 &\le2K^2\,160^K(N+2)^{mK}.
 \end{aligned}
\]
Each pair produces two candidate vectors, its union and its
intersection, so the number of attempted insertions is at most
twice this quantity.

The source vectors are addressed by their indices. Forming their
  union and intersection costs $O(N)$ operations. A binary trie stores
 one path for each incidence vector.
  Inspecting or creating the path and detecting whether its terminal
  node was already occupied takes $O(N)$ operations. Thus exact
duplicate detection adds no factor proportional to the number of
  stored sets. The pair loops cost
 $O(K^2\,160^K(N+2)^{mK+1})$.

Initialization makes at most $2M$ insertion attempts, where
  $M=5\binom Nm+N+3$, and costs $O(MN)$ operations. Copying the preceding layer into the next layer,
 with its trie, costs $O(N|\mathcal F_{k-1}|)$ per layer.
  Both contributions fit the preceding bound.

It remains to account for preprocessing. Transitive closure can be
computed in $O(N^3)$ operations, and a deterministic topological
order in $O(N^2)$ operations. For each of the $\binom Nm$
choices of an $m$-set $A$, constructing
\[
 A,\quad P_A,\quad D_A,\quad
 \widehat P_A,\quad\widehat D_A
\]
uses $O(mN)$ operations: for each job, inspect its precedence
relation with the $m$ jobs of $A$ in the precomputed matrix.
All boundary atoms therefore cost $O(mN\binom Nm)$ operations.
The constants and the $N+1$ rank-suffix vectors cost $O(N^2)$.
  Adding these costs to the pair, initialization, and copying costs proves
  the claimed bound. The cut layer is already one of the generated
 layers, so retaining it does not require a second enumeration.
\end{proof}

Algorithm~\ref{af:alg-deadline} calls
Algorithm~\ref{af:alg-family} and passes the node and cut layers to
Algorithm~\ref{af:alg-evaluate};
Algorithm~\ref{af:alg-optimum} makes these deadline calls during
  binary search. The node family supplies candidate sets, and the cut
 family supplies cuts.
The precedence matrix is used to construct the atoms and to perform
the ideal and antichain tests in Algorithm~\ref{af:alg-states}.
The topological ranks define the rank-suffix atoms $R_h$.

\subsection{Evaluation, reconstruction, and optimization}

\begin{theorem}[Cost of the explicit evaluator]
\label{af:evaluator-cost}
  Let $B_{\mathrm{node}}=|\mathcal F|\ge1$ and
$B_{\mathrm{cut}}=|\mathcal F_{\mathrm{cut}}|\ge1$, with the distinct family members supplied as
 $N$-bit incidence vectors and the global transitive order precomputed.
  For $m\ge3$ and $N\ge m$,
Algorithms~\ref{af:alg-lookup}--\ref{af:alg-reconstruct} use
\[
 O\!\left(
 \operatorname{poly}(m)\,
 B_{\mathrm{node}}B_{\mathrm{cut}}^2(N+2)^{2m+1}
 \right)
\]
RAM operations, including exact dictionary construction and the
 stored schedules of all accepted sets.
\end{theorem}

\noindent\emph{Proof idea.}
First bound the number of state records belonging to one parent.
Then count the inspections of ordered pairs and the operations in
each inspection. Finally, account for constructing the states and
copying the already stored schedules during reconstruction.

\begin{proof}
  Put $Q=(N+2)^m$ and $S=B_{\mathrm{cut}}Q$. Fix a parent $W\in\mathcal F$
and let $s_W$ be the number of state records appended for $W$.
  Algorithm~\ref{af:alg-states} considers at most $B_{\mathrm{cut}}$ cuts $C$.
Once $C$ is fixed, the prefix $L=W\cap C$ is determined.
There are at most $\binom Nm$ choices for the marked slot $Z$,
and then $R=W\setminus(L\cup Z)$ is determined as well.
  Each choice produces at most one record, so
\( s_W\le B_{\mathrm{cut}}\binom Nm\le B_{\mathrm{cut}}(N+2)^m=S.
 \)
This count includes duplicate records: different choices of $C$
may yield the same triple, and we count each appended record.

Consider an ordered pair of records $(i,j)$. Their addresses,
reachability flags, and predecessor indices are accessed in
constant time. Testing $L_i\cup Z_i\subseteq L_j$ and forming
the gap $L_j\setminus(L_i\cup Z_i)$ require scanning incidence
arrays of length $N$, which costs $O(N)$ operations.
  Algorithm~\ref{af:alg-lookup} follows this gap's incidence vector
through the exact trie, using $O(N)$ operations whether the gap is
 found or is absent.
Updating the target's flag and predecessor field costs $O(1)$.
  Consequently, one pair inspection costs
 \begin{equation}
 D=O(N+2).
 \label{af:eq-pair-cost}
\end{equation}
The ideal and antichain conditions have already been verified
during state construction; they need not be tested again for
each pair.

For the parent $W$, Algorithm~\ref{af:alg-evaluate} processes
sources in increasing order of $|L_i|$, inspecting at most $s_W^2$
  ordered pairs in total. Every edge satisfies $|L_j|\ge|L_i|+m$,
so all incoming edges are processed before a state is used as a
  source. Lemma~\ref{af:reachability} shows why this one pass suffices
  to find all reachable records. The cost is at most $O(s_W^2D)$;
  failed guards only reduce it. Summing over the at most
$B_{\mathrm{node}}$ parents and substituting the bounds on $S$ and $D$
gives
 \[
 \begin{aligned}
 \sum_{W\in\mathcal F}O(s_W^2D)
 &\le O(B_{\mathrm{node}}S^2D)\\
 &=O\!\left(B_{\mathrm{node}}\,[B_{\mathrm{cut}}(N+2)^m]^2(N+2)\right)\\
 &=O\!\left(B_{\mathrm{node}}B_{\mathrm{cut}}^2(N+2)^{2m+1}\right).
 \end{aligned}
\]

We also count the work outside these pair loops. Across all parents,
 there are at most $B_{\mathrm{node}}B_{\mathrm{cut}}Q$ candidate states.
Constructing their vectors and checking cardinalities, the antichain
condition, and the two ideal conditions takes
  $O(\operatorname{poly}(m)(N+2)^2)$ per candidate. State construction
thus costs
 $O(\operatorname{poly}(m)B_{\mathrm{node}}B_{\mathrm{cut}}(N+2)^{m+2})$.
  There are at most two initial or final lookups per state, giving
\(
 O(B_{\mathrm{node}}\cdot S\cdot N)
 \subseteq O(B_{\mathrm{node}}B_{\mathrm{cut}}(N+2)^{m+1}).
 \)

Constructing the acceptance trie and computing family
 cardinalities costs $O(B_{\mathrm{node}}N)$.
 A lexicographic traversal of the trie, followed by stable bucketing
by cardinality, orders the parents in the specified order within
  the same bound. For each parent, the states can be bucketed by
 $|L|$ in $O(s_W+N)$ operations; their stored indices need not change.
  Equal-cardinality buckets have no internal edges.
 All ordering and initialization costs fit the state-construction
 bound.

For reconstruction, Lemma~\ref{af:reconstruction} bounds an
accepting path by $N/m$ marked slots and its gaps by $N/m+1$.
  There are therefore $O(N)$ searches for stored gap schedules,
 each costing $O(N)$. Following predecessor indices and forming
the gap vectors costs at most $O(N^2)$. The gaps and marked slots
partition $W$, so concatenation copies exactly $|W|\le N$ job
occurrences. The gap schedules are already stored; their construction
  is not recursively repeated. Reconstruction consequently costs
 $O(N^2)$ per accepted parent and $O(B_{\mathrm{node}}N^2)$ over all parents.

Since $B_{\mathrm{node}},B_{\mathrm{cut}}\ge1$ and $m\ge3$,
all these additional terms are bounded by
\[
 O\!\left(\operatorname{poly}(m)B_{\mathrm{node}}B_{\mathrm{cut}}^2(N+2)^{2m+1}\right),
\]
 which proves the claim.
\end{proof}

\begin{theorem}[Overall operation count]\label{af:runtime}
  Put $E(m)=\revchange{6m^2+26m+1}{6m^2+13m+1}$. With
$K_{\mathrm{node}}=\revchange{4m+12}{4m+5}$ and $K_{\mathrm{cut}}=\revchange{m+6}{m+3}$,
deadline feasibility and reconstruction of a schedule use
\[
 O(L)+m^{O(m^2)}\mathrev{n^{E(m)}}
\]
  RAM operations. Optimization, including construction of an optimum
schedule, uses
 \[
 O(L)+m^{O(m^2)}\mathrev{n^{E(m)}}\mathrev{\log n}.
\]
  In particular, for every $m\ge3$ the latter is at most
 \[
 O(L)+m^{O(m^2)}\mathrev{n^{\revchange{15m^2}{11m^2}}}.
\]
\end{theorem}

\noindent\emph{Proof idea.}
First combine family generation and evaluation for one padded
  deadline instance. Substitute the two family bounds, express the bound in terms of
 the original number $n$ of jobs, and then account for binary search.

\begin{proof}
\hl{Instances with }$\mathrev{n\le1}$\hl{ take }$\mathrev{O(L+1)}$\hl{ time, so assume }$\mathrev{n\ge2}$\hl{.}
  For one nontrivial deadline test,
Lemma~\ref{af:family-count} and
Theorem~\ref{af:evaluator-cost} give the computational bound
 \[
 O\!\left(
 \operatorname{poly}(m)\,
 B_{\mathrm{node},0}B_{\mathrm{cut},0}^2(N+2)^{2m+1}
 \right).
\]
  Substituting the bounds for $B_{\mathrm{node},0}$ and $B_{\mathrm{cut},0}$ gives

\ifshowrevisions
\begin{revisionold}
\renewcommand{\theHequation}{deleted.0013.\arabic{equation}}
 \[
 \begin{split}
 B_{\mathrm{node},0}B_{\mathrm{cut},0}^2(N+2)^{2m+1}
 &=8\cdot160^{K_{\mathrm{node}}+2K_{\mathrm{cut}}}
       (N+2)^{mK_{\mathrm{node}}+2mK_{\mathrm{cut}}+2m+1}\\
 &=8\cdot160^{6m+24}(N+2)^{6m^2+26m+1}.
 \end{split}
\]
\end{revisionold}
\begin{revisionnew}
\[
 \begin{split}
 B_{\mathrm{node},0}B_{\mathrm{cut},0}^2(N+2)^{2m+1}
 &=8\cdot160^{K_{\mathrm{node}}+2K_{\mathrm{cut}}}
       (N+2)^{mK_{\mathrm{node}}+2mK_{\mathrm{cut}}+2m+1}\\
 &=8\cdot160^{6m+11}(N+2)^{6m^2+13m+1}.
 \end{split}
\]
\end{revisionnew}
\else
\[
 \begin{split}
 B_{\mathrm{node},0}B_{\mathrm{cut},0}^2(N+2)^{2m+1}
 &=8\cdot160^{K_{\mathrm{node}}+2K_{\mathrm{cut}}}
       (N+2)^{mK_{\mathrm{node}}+2mK_{\mathrm{cut}}+2m+1}\\
 &=8\cdot160^{6m+11}(N+2)^{6m^2+13m+1}.
 \end{split}
\]
\fi

  For completeness, the powers of $N+2$ in the main costs are

\ifshowrevisions
\begin{revisionold}
\renewcommand{\theHequation}{deleted.0014.\arabic{equation}}
 \[
 \begin{array}{ll}
 \text{family generation:}       &4m^2+12m+1,\\
 \text{state construction:}      &5m^2+19m+2,\\
 \text{transition evaluation:}   &6m^2+26m+1,\\
 \text{schedule reconstruction:} &4m^2+12m+2.
 \end{array}
\]
\end{revisionold}
\begin{revisionnew}
\[
 \begin{array}{ll}
 \text{family generation:}       &4m^2+5m+1,\\
 \text{state construction:}      &5m^2+9m+2,\\
 \text{transition evaluation:}   &6m^2+13m+1,\\
 \text{schedule reconstruction:} &4m^2+5m+2.
 \end{array}
\]
\end{revisionnew}
\else
\[
 \begin{array}{ll}
 \text{family generation:}       &4m^2+5m+1,\\
 \text{state construction:}      &5m^2+9m+2,\\
 \text{transition evaluation:}   &6m^2+13m+1,\\
 \text{schedule reconstruction:} &4m^2+5m+2.
 \end{array}
\]
\fi

  The transition exponent is largest for $m\ge3$. The transitive
closure and atom preprocessing terms of
 Lemma~\ref{af:family-cost} are smaller as well.
  Thus generation, evaluation, and reconstruction together fit the
 displayed deadline bound in $N$.

\hl{Padding is used only for }$\mathrev{\lceil n/m\rceil\le T<n}$\hl{, and hence }$\mathrev{N=mT\le mn}$\hl{. For }$\mathrev{n\ge2}$\hl{, we have }$\mathrev{N+2\le m(n+2)\le2mn}$\hl{. Replacing }$\mathrev{N+2}$\hl{ by }$\mathrev{2mn}$\hl{ contributes the factor }$\mathrev{(2m)^{E(m)}}$\hl{. The full coefficient depending on }$\mathrev{m}$\hl{ is therefore bounded by}
\[
 \mathrev{\operatorname{poly}(m)\,8\,160^{\revchange{6m+24}{6m+11}}\,(2m)^{E(m)}
 =m^{O(m^2)}}.
\]
  Here $E(m)=O(m^2)$ and $m\ge3$.
  One nontrivial deadline call thus costs
 $m^{O(m^2)}\mathrev{n^{E(m)}}$ operations.

In Algorithm~\ref{af:alg-optimum}, the interval between the
infeasible lower bound and feasible upper bound has initial length
at most $n+1$ and is halved, up to rounding, after each call.
There are therefore $O(\mathrev{\log n})$ deadline calls. Every successful
call already returns a schedule, and the best one is retained,
so there is no additional search for an optimum witness.
  The total computational term is at most
\(
 m^{O(m^2)}\mathrev{n^{E(m)}}\mathrev{\log n}.
 \)

The original input is read once, contributing $O(L)$.
A supplied binary deadline can be compared with $n$ during this
reading while retaining only $O(\mathrev{\log n})$ bits and a flag for
$T\ge n$. In that case a serial topological schedule suffices;
otherwise the retained deadline fits the working word size.
The graph data are reused between optimization calls, so the
$O(L)$ input term is not multiplied by the number of calls.
Immediate rejection by the capacity bound, serial scheduling,
and deleting dummy jobs all fit the same overall bound.
  Consequently, the total number of RAM operations is
\[
 O(L)+m^{O(m^2)}\mathrev{n^{E(m)}}\mathrev{\log n},
\]
 including construction of an optimum schedule.

Finally, $\mathrev{\log n=O(n)}$ and

\ifshowrevisions
\begin{revisionold}
\renewcommand{\theHequation}{deleted.0015.\arabic{equation}}
 \[
 15m^2-\bigl(E(m)+1\bigr)
   =9m^2-26m-2
   =(m-3)(9m+1)+1\ge1
 \qquad(m\ge3).
\]
\end{revisionold}
\begin{revisionnew}
\[
 11m^2-\bigl(E(m)+1\bigr)
   =5m^2-13m-2
   =(m-3)(5m+2)+4\ge4
 \qquad(m\ge3).
\]
\end{revisionnew}
\else
\[
 11m^2-\bigl(E(m)+1\bigr)
   =5m^2-13m-2
   =(m-3)(5m+2)+4\ge4
 \qquad(m\ge3).
\]
\fi

 Absorbing the logarithm into one power of $\mathrev{n}$ therefore gives
 the simpler exponent $\revchange{15m^2}{11m^2}$.
\end{proof}

\ifshowrevisions
\begin{revisionold}
\renewcommand{\theHequation}{deleted.0016.\arabic{equation}}
\begin{center}
\begin{tabular}{@{}rrrrrr@{}}
\toprule
  $m$ & $K_{\mathrm{node}}$ & $K_{\mathrm{cut}}$
   & $mK_{\mathrm{node}}$ & $mK_{\mathrm{cut}}$ & $E(m)$\\
\midrule
  3&24&9&72&27&133\\
  4&28&10&112&40&201\\
  5&32&11&160&55&281\\
\bottomrule
\end{tabular}
\end{center}
\end{revisionold}
\begin{revisionnew}
\begin{center}
\begin{tabular}{@{}rrrrrr@{}}
\toprule
  $m$ & $K_{\mathrm{node}}$ & $K_{\mathrm{cut}}$
   & $mK_{\mathrm{node}}$ & $mK_{\mathrm{cut}}$ & $E(m)$\\
\midrule
  3&17&6&51&18&94\\
  4&21&7&84&28&149\\
  5&25&8&125&40&216\\
\bottomrule
\end{tabular}
\end{center}
\end{revisionnew}
\else
\begin{center}
\begin{tabular}{@{}rrrrrr@{}}
\toprule
  $m$ & $K_{\mathrm{node}}$ & $K_{\mathrm{cut}}$
   & $mK_{\mathrm{node}}$ & $mK_{\mathrm{cut}}$ & $E(m)$\\
\midrule
  3&17&6&51&18&94\\
  4&21&7&84&28&149\\
  5&25&8&125&40&216\\
\bottomrule
\end{tabular}
\end{center}
\fi

The last column is the degree before the logarithmic factor for
 optimization. These are conservative accounting bounds for the specified
implementation. The qualitative conclusion is the uniform bound
$m^{O(m^2)}\mathrev{n^{O(m^2)}}$, hence polynomial time for every fixed
number of machines. The smaller degree uses separate node and cut
budgets, exact trie lookup, and acyclic state evaluation, with the structural construction of Section~\ref{af:structural-appendix} unchanged.


\paragraph{Bit complexity.}
Let $w=O(m^2\log(N+2))$ be the working word length.
If word operations are charged by their operand bit lengths while
retaining random access, schoolbook arithmetic and comparisons give
a conservative overhead $O(w^2)$ per RAM operation.
Thus the preceding RAM bounds incur only an additional factor
polynomial in $m$ and $\log n$ in this bit-cost random-access model.
Set operations still scan their $N$ entries; an incidence vector is
not treated as a constant-size integer.
A multitape Turing machine must additionally account for sequential
memory access. The following implementation handles this explicitly.

\begin{proposition}[A multitape implementation retaining the RAM degree]
\label{af:multitape}
The algorithms of this paper admit an implementation on a deterministic
Turing machine with a fixed number of tapes.
Deadline feasibility, including construction of a feasible schedule
when one exists, takes
\[
 O(L)+m^{O(m^2)}n^{6m^2+13m+1}\log n
\]
steps. Optimization and construction of an optimum schedule take
\[
 O(L)+m^{O(m^2)}n^{6m^2+13m+1}\log^2 n
\]
steps.
\end{proposition}

\paragraph{Proof idea.}
A general simulation of a logarithmic-cost RAM by a deterministic
multitape Turing machine has quadratic overhead
\cite[Theorem~2(a)]{CookReckhow1973}.
For our word-RAM model, a direct simulation by sequentially stored
address--value records gives the conservative bound $O(R^2w^2)$
for $R$ RAM operations on $w$-bit words.
Applied to Theorem~\ref{af:runtime}, this gives the previous degree
$2E(m)$, where $E(m)=6m^2+13m+1$.
We avoid this doubling by implementing the evaluator directly:
batch exact set queries by parent cardinality, and batch
reachability updates by source-prefix cardinality.
Both operations use sorting and merging instead of individual
random-access lookups.

\begin{proof}
Consider a nontrivial padded deadline instance with $N$ jobs, and put
\[
 P=B_{\mathrm{node}},
 \qquad
 S=B_{\mathrm{cut}}(N+2)^m.
\]
There are at most $P$ parents, at most $S$ state records for each
parent, and at most $PS^2$ ordered candidate transitions in total.
We describe an implementation whose evaluation time is
\begin{equation}
 O\!\left(
   \operatorname{poly}(m)\,
   PS^2(N+2)\log(N+2)
 \right).
 \label{af:multitape-evaluation}
\end{equation}

\smallskip
\noindent
\emph{Family generation and state construction.}
Enumerate the normalized Boolean formulas counted in the proof of
Lemma~\ref{af:family-count}, evaluate their incidence vectors, and
sort and remove duplicate vectors.
This replaces online trie insertion by a sequential procedure and
produces exactly the same node and cut families.
In particular, the formula count, rather than the number of distinct
sets alone, bounds the number of generated records.
Atoms can be obtained by scans of the precomputed precedence matrix.
Even allowing $O(\operatorname{poly}(m)(N+2)^2)$ bit operations per
formula, the resulting polynomial degree is at most
\[
 mK_{\mathrm{node}}+2=4m^2+5m+2<E(m).
\]
Sorting these records also fits the final deadline bound.

Construct the states by sequential enumeration of parents, cuts,
and marked slots. Incidence vectors are copied and compared by scans.
For each candidate state, the ideal and antichain conditions can
be checked by a scan of the precedence matrix, using
$O(\operatorname{poly}(m)(N+2)^2)$ bit operations.
These costs remain below the final bound.
All lists and temporary records are stored on a fixed number of
tapes; additional lists do not require additional tape heads.

\paragraph{Physical representation of records.}
Represent each set by its $N$-bit incidence vector.
Assign each state an immutable identifier when it is
created, and retain that identifier throughout all
subsequent reorderings of tape files. A state record
stores its parent identifier, the vectors $L,Z,R$,
the prefix cardinality $|L|$, a reachability bit,
and a predecessor identifier. Query and transition
records carry the identifiers needed to return their
answers to the originating states. Each record has
$O(N+m^2\log(N+2))
 =O(\operatorname{poly}(m)(N+2))$ bits.
Sequential merge sorting can be implemented by
alternating runs on a fixed number of work tapes;
logical files and buckets are represented by
delimited tape regions, not by separate tapes.

\smallskip
\noindent
\emph{Batched exact set queries.}
Process all parents $W$ of the same cardinality $h$ together.
Store the state records in lexicographic order of
(parent identifier, original local state index).
For each parent, generate the initial-prefix queries $L_i$,
the final-suffix queries $R_i$, and the gap queries
\(
 G_{ij}=L_j\setminus(L_i\cup Z_i)
\)
for the ordered pairs satisfying
$L_i\cup Z_i\subseteq L_j$.

The ordered pairs are generated using two sequential copies
of the state file. For a fixed source record of a parent $W$,
scan only the contiguous target block belonging to $W$,
rewinding that block before processing the next source.
The source record is retained on a work tape during this scan.
If $s_W$ is the number of states of $W$, this costs
$O((s_W+s_W^2)b)$ tape steps for that parent, where
$b=O(\operatorname{poly}(m)(N+2))$ bounds the record length.
In particular,
\(
 \sum_W s_W\le PS,
 \,
 \sum_W s_W^2\le PS^2.
\)
No scan of the state blocks of other parents is performed
for an individual source.
Attach the parent identifier, original state indices,
and query type to each query. The total number of query
and transition records over all parent cardinalities is
$O(PS^2)$.

Sort each batch by its query incidence vector and merge it with
the family table of incidence vectors and acceptance bits.
Repeated queries receive the same answer; a vector absent from the
family receives answer zero.
The proper-query lemma proves that every queried set has cardinality
at most $h-m$.
Its acceptance value has therefore already been finalized.
The induction invariant is stronger than finality of
acceptance bits: before processing cardinality $h$,
every accepted family member $X$ with $|X|<h$
has an explicitly stored feasible full schedule
$\sigma(X)$, and these smaller rows remain unchanged.

After evaluating all parents of cardinality $h$,
complete their reconstruction and append their schedules
to the schedule table before starting the next
parent-cardinality batch.
This is possible because every reconstruction request
also concerns a set of cardinality at most $h-m$.
Consequently, all query answers for parents of cardinality $h$
can be determined before evaluating their reachability states.

\paragraph{Returning batched query answers.}
A query record contains the requested incidence
vector and the original parent, state or transition
identifier together with its query type.
After sorting by the requested vector and merging
with the finalized acceptance table, we sort the
answered records by their original identifiers.
The resulting sequential merge attaches the
answers to the appropriate state or transition
records. A query vector absent from the candidate
family is assigned answer zero. This return sort
is charged to the same total $O(PS^2)$
query-record volume and does not require random
access.

\paragraph{Persistent file order and bucket operations.}
Keep the state table ordered by
(parent identifier, original local state index).
This order is retained throughout the sweep.
Every state record also stores its parent cardinality
and prefix cardinality.
For a source-prefix bucket, sort its transition records
by their source-state identifiers and merge them with
one sequential scan of the state table. For every transition
whose source is reached and whose gap answer is one,
emit a proposal containing its target identifier and
original source index.

Sort the proposals by target identifier and then by
original source index. Merge them with another sequential
scan of the state table, writing the updated state table
to a second file in the same persistent order.
An already reached target retains its predecessor.
For an unreached target, retain the smallest proposing
source index.
Only the current bucket and its proposals are sorted;
the complete state table is scanned, not sorted,
for each bucket. A constant number of copies and
a constant number of work tapes suffice. The unread
transition stream remains on a separate tape with
its head at the beginning of the next bucket;
sorting the current bucket does not require copying
or rescanning that unread suffix.

\smallskip
\noindent
\emph{Batched reachability updates.}
Initialize each state's reachability bit from its resolved
initial-prefix query.
Sort the transition records by parent cardinality and then by
source-prefix cardinality, retaining the source and target
identifiers and the resolved gap-query answer.
For each source-prefix bucket, sort and merge the source identifiers
with the reachability table.
A transition produces a target proposal precisely when its source
is reachable and its gap query has answer one.
Sort these proposals by target identifier and merge them into the
reachability table, retaining one predecessor for each newly
reached target.

Every transition satisfies
\[
 |L_j|\ge |L_i|+m.
\]
Thus all incoming proposals have been processed before a state is
used as a source. One traversal of the source-prefix buckets
therefore computes exactly the required reachability relation.
The transition stream is consumed bucket by bucket: transitions
outside the current bucket are not repeatedly scanned.
Each transition participates in only a constant number of sorting
and merging operations.

\paragraph{Ordering invariant for the sweep.}
Before processing the source-prefix bucket of
cardinality $q$, the reachability bits of all
states with prefix cardinality at most $q$
are final. Indeed, every transition $i\to j$
satisfies
\(
|L_j|\ge |L_i|+m,
\)
so all predecessors of a state in the current
bucket belong to earlier buckets. In particular,
processing a bucket cannot create a new reachable
source in that same bucket.

We keep the transition stream ordered by
source-prefix cardinality and consume its
buckets sequentially, without rescanning the
full transition stream.

When several proposals first reach the same
target, retain the proposal with the smallest
original source-state identifier; a previously
stored predecessor, including $\mathsf{START}$,
is never replaced. This gives the same
predecessor choices as the original ordered
evaluator.

\smallskip
\noindent
\emph{Scan and sorting costs.}
There are $O(N^2)$ pairs consisting of a parent cardinality and a
source-prefix cardinality.
Even if the entire state table is scanned for each such pair,
charging $O(\operatorname{poly}(m)(N+2))$ bits per state record,
the total scan cost is at most
\[
 O\!\left(\operatorname{poly}(m)\,PS(N+2)^3\right).
\]
Since
\[
 S\ge (N+2)^m\ge (N+2)^2
 \qquad(m\ge3),
\]
this is bounded by
\[
 O\!\left(\operatorname{poly}(m)\,PS^2(N+2)\right).
\]
Scanning the family table once for each parent cardinality costs
$O(\operatorname{poly}(m)P(N+2)^2)$ bits and is smaller still.
These estimates also cover rewinding the scanned tables.

\paragraph{Why repeated state scans are affordable.}
The state table has at most $PS$ records,
whereas the transition stream has at most
$PS^2$ records.

With $O(N^2)$ pairs of parent cardinality
and source-prefix cardinality, scanning and
rewinding the entire state table for every
pair costs
\(
O\bigl(\operatorname{poly}(m)
PS(N+2)^3\bigr)
\)
tape steps.
Since
\(
S\ge(N+2)^m\ge(N+2)^2,
\)
this is at most
\(
O\bigl(\operatorname{poly}(m)
PS^2(N+2)\bigr).
\)

This argument does not license repeated scans
of the larger transition stream; its records
are sorted into buckets and processed in
sequential order.

A query or transition record contains a constant number of
$N$-bit incidence vectors and identifiers of
$O(m^2\log(N+2))$ bits.
Its length is therefore
\[
 O(N+m^2\log(N+2))
 =O(\operatorname{poly}(m)(N+2)).
\]
Merge sorting on a fixed number of tapes uses a logarithmic number
of sequential passes through the records.
The family-size bounds give
\[
 \log(PS^2+2)=O(m^2\log(N+2)).
\]
The total volume of transition and query records is
$O(\operatorname{poly}(m)PS^2(N+2))$ bits.
Summing the sorting costs over the batches therefore introduces
only one factor $\log(N+2)$.

\paragraph{Summing sorting costs over disjoint batches.}
If batch $t$ contains $M_t$ records, and every
record enters only a constant number of sorting
procedures, then
\(
\sum_t M_t=O(PS^2).
\)
Fixed-tape merge sorting a batch of $M_t$
records of length at most
\(
b=O(\operatorname{poly}(m)(N+2))
\)
takes
\(
O(M_t b\log(M_t+2))
\)
steps. Therefore
\(
\sum_t O(M_t b\log(M_t+2))
\le
O\bigl(PS^2 b\log(PS^2+2)\bigr).
\)

The number of separate batches does not
create another logarithmic factor.
This accounting includes the return sort
for query answers and the sorting of
target proposals.

Elementary operations on the identifiers fit the same budget;
for example, schoolbook arithmetic costs
$O(m^4\log^2(N+2))=O(\operatorname{poly}(m)(N+2))$
per record.
This proves~\eqref{af:multitape-evaluation}.

\smallskip
\noindent
\emph{Reconstruction.}
For each accepted parent, choose the smallest original
local state index among the reached states whose
resolved suffix query has answer one.
This selection is obtained by a sequential scan of
the state table in its persistent order and agrees
with Step~5 of Algorithm~\ref{af:alg-evaluate}.

Follow the selected predecessor paths in parallel, retrieving
predecessor records by sorting and merging identifiers.
Each path has at most $N/m$ states.
Thus at most $O(N)$ reconstruction rounds are needed for each
parent-cardinality batch, and their repeated state-table scans
fit the scan bound above.

\paragraph{Predecessor retrieval and its sorting charge.}
For each parent-cardinality batch, start with
one request containing the chosen final-state
identifier for each accepted parent.

Sort the requests by immutable state identifier,
merge them with the state table, and emit the
corresponding predecessor identifiers as the
next-round requests, dropping those equal to
$\mathsf{START}$.

Store the recovered states tagged by parent
and path position. Since every transition
increases prefix size by at least $m$, each
chain contains at most $N/m$ marked states.

Over all parent cardinalities there are at
most $O(N^2)$ rounds. Charging a complete
scan of the $PS$-record state table and a
sort of at most $PS$ requests per round gives
the conservative bound
\( O\bigl(\operatorname{poly}(m)
PS(N+2)^3\log(PS+2)\bigr).
\)

Because $S\ge(N+2)^m$ and $m\ge3$, this is
absorbed by \eqref{af:multitape-evaluation}.

The recovered paths specify requests for an initial-prefix
schedule, the intervening gap schedules, and a final-suffix
schedule. Each request carries the requested incidence
vector, the destination parent identifier, and the
position of the requested piece in the output schedule.
Empty requests are answered directly by the empty schedule.

Maintain the completed schedule table in lexicographic
order of its incidence vectors. Sort the nonempty requests
by their requested vectors and merge them with this table.
For a group of requests for the same set $X$, copy
$\sigma(X)$ to a scratch tape once and scan and rewind
that scratch copy for each destination request.
Thus repeated requests require explicit copying,
but no random access and no repeated dynamic-programming
computation.

A stored schedule has at most $N$ job labels and therefore
has $O(N\log(N+2))$ bits. There are $O(PN)$ piece requests
over all parents. Allowing each returned piece to have
the maximum schedule length gives total returned volume
\(
 O(PN^2\log(N+2)).
\)
Encode the marked slots as additional pieces and sort
all returned pieces by destination parent and path
position before concatenating them.

For completeness, schedule-table maintenance can be
implemented by merging the schedules completed at each
parent cardinality into the existing ordered table.
Using padded records of length
$O((N+2)\log(N+2))$, the request sorts, returned-piece
sorts, and schedule-table maintenance together cost at most
\(
 O\!\left(
 \operatorname{poly}(m)P(N+2)^2\log^2(N+2)
 \right)
\)
tape steps. Here
$\log(P(N+2)+2)=O(m^2\log(N+2))$.
Since $S\ge(N+2)^m$ and $m\ge3$, this bound is absorbed by
\eqref{af:multitape-evaluation}.

\paragraph{Why the concatenation is feasible.}
For a recovered path
$s_1\to\cdots\to s_r$, the returned pieces
are $\sigma(L_1)$, the accepted gap schedules
\(
\sigma\bigl(
L_{i+1}\setminus(L_i\cup Z_i)
\bigr),
\)
and $\sigma(R_r)$, interleaved with the full
marked slots $Z_1,\ldots,Z_r$.

The ideal conditions on each $L_i$ and
$L_i\cup Z_i$ prevent precedence arcs from
pointing backward across the boundaries
of these pieces.
Each gap schedule is feasible by its stored
acceptance certificate, and each marked slot
is an antichain of exactly $m$ jobs.
Thus their concatenation is a feasible full
schedule of the parent.
Each stored smaller schedule is retrieved by
a sorted request and copied, never recomputed
by a new recursive dynamic-programming call.
No smaller schedule is constructed recursively a second time.

\smallskip
\noindent
\emph{Substitution and optimization.}
The family-size estimates imply
\[
 \begin{aligned}
 PS^2(N+2)
 &=B_{\mathrm{node}}B_{\mathrm{cut}}^2(N+2)^{2m+1}\\
 &\le
 8\cdot160^{6m+11}(N+2)^{6m^2+13m+1}.
 \end{aligned}
\]
Preprocessing, family generation, and state construction have
smaller polynomial degrees and fit the resulting bound.
Using $N\le mn$ and absorbing parameter-only factors into
$m^{O(m^2)}$ gives
\[
 m^{O(m^2)}n^{E(m)}\log n
\]
steps for one nontrivial deadline test, including reconstruction.

Read the original input once, contributing $O(L)$.
A supplied deadline can be capped at $n$ while reading it, and
subsequent graph processing is included in the computational term.
Finally, the optimization algorithm makes $O(\log n)$ deadline
tests. Reusing the tapes between tests and retaining the best
returned schedule gives
\[
 O(L)+m^{O(m^2)}n^{E(m)}\log^2 n
\]
steps, as claimed.
\end{proof}

For $m=3$, the multitape optimization bound is
\[
 O(L)+O(n^{94}\log^2 n).
\]
Since jobs are individually listed, $n=O(L)$, and hence
$O((L+2)^{95})$ is also a valid simpler bound.
More generally, the multitape polynomial degrees for
$m=3,4,5$ are $94,149,216$, respectively, before the displayed
logarithmic factors.
The direct implementation therefore retains the RAM polynomial
degree while accounting explicitly for sequential memory access.


\begin{proof}[Proof of Theorem~\ref{af:main}]
Theorem~\ref{af:correct} proves feasibility testing and optimum
schedule construction.
   Its completeness input,
Theorem~\ref{af:structure}, is proved in
Section~\ref{af:structural-appendix}.
  Theorem~\ref{af:runtime} gives the RAM operation bound, including
 its simpler $\revchange{15m^2}{11m^2}$ consequence. All families and loops are determined by
the input, so the algorithm is uniform.
 Proposition~\ref{af:multitape} proves the additional multitape bounds
without changing the scheduling algorithm.
\end{proof}

\section{Structural induction and global descriptions}
\label{sec:str-structure}
\label{af:structural-appendix}

This section proves Theorem~\ref{af:structure}. We begin with a feasible
full schedule of the padded instance and use it as an existence witness.
A \emph{mark} is a selected full slot of this schedule. The jobs strictly
between consecutive marks, or between an exterior boundary and the
nearest mark, form the open temporal gaps. These gaps become the children
of the current node. Every nonempty node receives at least one marked
slot, so every child contains at least $m$ fewer jobs than its parent.
This makes recursion on the number of jobs possible.

The main difficulty is not merely to find smaller feasible gaps.
We must describe their job sets, and the prefixes before their marked
slots, by short formulas using the \emph{global} atoms available to the
algorithm. Here global means that a formula is evaluated on the entire
ground set $J$, with the original transitive precedence relation
interpreted in the current frame.

\paragraph{How the proof is organized.}
The construction has three connected tasks.
\begin{enumerate}
\item \emph{Choose useful marks.}
We specify a priority order, choose a normalized witness schedule,
and find separator slots. A separator relates a cutoff in that priority
order to the jobs occurring before a particular time slot.
Lemma~\ref{str:prefix} gives the resulting prefix formula, and
Lemma~\ref{str:walk} explains how to move a separator when the cutoff
changes.
\item \emph{Keep the stored information bounded.}
At each interval we maintain a list of global successor-closed sets.
The invariants in the next subsection allow this information to pass
to children. Later arguments bound the number of surviving entries
by $m$ and simplify their formulas so that one entry cannot accumulate
an unbounded description of its ancestors.
\item \emph{Obtain formulas that the algorithm can enumerate.}
Each child set must have an exact description on all of $J$.
For a marked slot in a node $W$, its prefix must have the form
$W\cap C$, where $C$ itself has a bounded global description.
The final counting argument bounds the number of atomic occurrences
and completes Theorem~\ref{af:structure}.
\end{enumerate}

\paragraph{The role of the witness schedule.}
A priority order is specified on the domain being scheduled.
Normalization means choosing the lexicographically smallest feasible
full schedule for that priority, after sorting the jobs within each
slot by priority; the next subsection makes this definition precise.
The feasible schedules form a finite nonempty collection, so such a
choice exists. This is an existence argument: we do not claim that
normalization can be performed efficiently. During the proof, a feasible
reordering of an interval may be substituted into its witness while
keeping exterior jobs fixed, as justified by Lemma~\ref{str:replacement}.
The algorithm already enumerates the truth sets of bounded formulas
and tests compatible states. Its completeness proof requires the
existence of a suitable hierarchy; it does not require the algorithm
to find the witness schedule first.

\paragraph{Temporal domains, priority cutoffs, and global sets.}
Three distinct objects occur throughout the proof.
A temporal domain $W=(a,b)$ consists of the jobs scheduled strictly
between boundary positions $a$ and $b$. A local cutoff $S\subseteq W$
is an initial segment of a specified priority order on $W$; it need not
be an initial segment in time. A global set $H\subseteq J$ represents
that local cutoff when
\[
                         H\cap W=S.
\]
This equality determines membership in $H$ only for jobs in $W$.
It neither asserts $H=S$ nor determines what $H$ contains outside $W$.
Whenever the proof uses this equality for a particular job, that job
must first be known to belong to $W$. The next subsection defines
frames, cones, closure properties, priorities, and the stored
information in prose; subsequent subsections introduce the additional
expressions needed by the construction.

\subsection{The information maintained at an interval}

\paragraph{Global precedence and full witness schedules.}
Fix an integer $m\ge3$ and a finite padded job set $J$ with $N=|J|$. Every job has processing time one. Write $x\prec_+ y$ if there is a directed path of positive length from job $x$ to job $y$ in the input precedence graph. This is a strict partial order: it is irreflexive and transitive. Thus a predecessor may be direct or indirect. The subscript $+$ identifies the original direction of precedence.

For the structural construction, assume that $J$ has a feasible full schedule. Such a schedule is an ordered partition
\(
\sigma=(S_1,\ldots,S_T),\quad T=N/m,
\)
in which every $S_t\subseteq J$ has exactly $m$ jobs and $x\prec_+ y$ implies $\tau_+(x)<\tau_+(y)$, where $\tau_+(x)=t$ means $x\in S_t$. In particular, every slot is an antichain: its jobs are pairwise incomparable. Slot $t$ executes during $[t-1,t)$, and its jobs execute simultaneously on distinct machines. A \emph{complete witness schedule} means such a schedule of all of $J$, including the jobs outside any current subproblem. It is used in the existence proof, not supplied to the scheduling algorithm.

\paragraph{Predecessor and successor cones.}
For any subset $A\subseteq J$, define its strict predecessor cone and strict successor cone by
\[
\begin{aligned}
P_A^+&=\{x\in J:\text{there exists }a\in A\text{ with }x\prec_+ a\},\\
D_A^+&=\{x\in J:\text{there exists }a\in A\text{ with }a\prec_+ x\}.
\end{aligned}
\]
The word \emph{cone} here simply names a set of jobs related by precedence; no geometric structure is intended. Membership requires a relation to \emph{at least one} member of $A$, not to every member. The weak cones also contain the boundary set itself:
\(
\widehat P_A^+=P_A^+\cup A,\quad
\widehat D_A^+=D_A^+\cup A.
\)
The adjective \emph{strict} refers to the strict relation $\prec_+$. If $A$ contains comparable jobs, a strict cone can contain members of $A$. When $A$ is a full antichain slot, both strict cones are disjoint from $A$.

\paragraph{Examples of cones and the meaning of a frame.}
For an example with $m=3$, let $J=\{a,b,c,d,e,f,g,h,i\}$, with the chain $a\prec_+b\prec_+c$ and six independent jobs $d,\ldots,i$. Take $A=\{b\}$. Then
\[
P_A^+=\{a\},\quad D_A^+=\{c\},\quad
\widehat P_A^+=\{a,b\},\quad\widehat D_A^+=\{b,c\}.
\]
The unrelated jobs $d,\ldots,i$ belong to none of these four sets. For another full-slot example with $m=3$, use the three chains
\[
a_1\prec_+ a_2\prec_+ a_3,\quad
b_1\prec_+ b_2\prec_+ b_3,\quad
c_1\prec_+ c_2\prec_+ c_3.
\]
Put $Z_i=\{a_i,b_i,c_i\}$. The schedule $(Z_1,Z_2,Z_3)$ is feasible and full. For the boundary slot $Z_2$, its predecessor cone is $Z_1$, its successor cone is $Z_3$, and its weak cones are $Z_1\cup Z_2$ and $Z_2\cup Z_3$.

A \emph{frame} specifies which of two consistent orientations is being used. In the \emph{forward frame}, denoted $f=+$, precedence is $\prec_+$ and time is read in the execution order $S_1,S_2,\ldots,S_T$. Thus \emph{forward time} means increasing slot indices: a job with a smaller slot index executes earlier.

In the \emph{reversed frame}, denoted $f=-$, reverse both precedence and the complete slot order. Formally,
\(
x\prec_- y\quad\Longleftrightarrow\quad y\prec_+ x,
\qquad
\tau_-(x)=T+1-\tau_+(x).
\)
The reversed witness schedule is $(S_T,S_{T-1},\ldots,S_1)$. It is feasible for $\prec_-$: from $x\prec_-y$ we obtain $\tau_+(y)<\tau_+(x)$ and hence $\tau_-(x)<\tau_-(y)$. This is a proof device; it does not instruct the original jobs to run backward in physical time. In the three-chain example, the reversed schedule is $(Z_3,Z_2,Z_1)$ and, for example, $a_3\prec_-a_2\prec_-a_1$.

\paragraph{Ranks and cone notation in a chosen frame.}
Fix once and for all a bijective topological rank function $\operatorname{rank}_+:J\to\{1,\ldots,N\}$ for the original order. It satisfies $x\prec_+y\Rightarrow\operatorname{rank}_+(x)<\operatorname{rank}_+(y)$. Define
\[
\operatorname{rank}_-(x)=N+1-\operatorname{rank}_+(x),\qquad
R_h^f=\{x\in J:\operatorname{rank}_f(x)\ge h\}\quad(1\le h\le N+1).
\]
In either frame $R_1^f=J$ and $R_{N+1}^f=\varnothing$. Rank is a position in a fixed topological list, not an execution time. Feasible rearrangements of the witness schedule do not change these ranks.

Define $P_A^f,D_A^f$ by the cone formulas above using $\prec_f$. Then
\(
P_A^-=D_A^+,\quad D_A^-=P_A^+,
\quad R_h^+=J\setminus R_{N+2-h}^-.
\)
The weak cones transform in the same way. A superscript $\mathrm{rev}$ on an expression means that the \emph{same subset of $J$} is rewritten using reversed-frame atoms. It does not mean complementation of the represented set. For example, the old set $P_{\{b\}}^+=\{a\}$ is written as $D_{\{b\}}^-=\{a\}$ after reversal. Once a frame is fixed, we omit its superscript and interpret every cone, rank, and time comparison in that frame.

\paragraph{Actual boundary slots, sentinels, and temporal intervals.}
Fix a complete witness schedule in the current frame. An \emph{actual boundary} is one of its full slots. If $a$ is actual, write $\operatorname{jobs}(a)$ for its set of $m$ jobs and $\operatorname{pos}(a)$ for its slot index. We identify $a$ with $\operatorname{jobs}(a)$ when writing cones or set expressions such as $P_a$ or $y\in a$.

We also allow two formal boundary symbols, called \emph{sentinels}. The left sentinel $\bot$ is placed before every job and the right sentinel $\top$ after every job:
\(
\operatorname{pos}(\bot)=0,\quad \operatorname{pos}(\top)=T+1,
\quad\operatorname{jobs}(\bot)=\operatorname{jobs}(\top)=\varnothing.
\)
Sentinels are not jobs, contain no jobs, consume no machine capacity, and are not extra schedule slots. Their cone values are assigned by convention:
\[
\begin{aligned}
D_{\bot}=\widehat D_{\bot}=J,&\qquad
P_{\bot}=\widehat P_{\bot}=\varnothing,\\
P_{\top}=\widehat P_{\top}=J,&\qquad
D_{\top}=\widehat D_{\top}=\varnothing.
\end{aligned}
\]
These are definitions for formal boundary symbols; they are \emph{not} obtained by applying the ordinary cone formula to the empty job set. The conventions make endpoint formulas work at the beginning and end without introducing artificial precedence constraints between real jobs. Reversing the frame exchanges the two sentinel roles.

Let $a,b$ be boundaries with $\operatorname{pos}(a)<\operatorname{pos}(b)$. Their \emph{temporal open interval} is the job set
\[
W=(a,b)=\{x\in J:\operatorname{pos}(a)<\tau_f(x)<\operatorname{pos}(b)\}.
\]
The word \emph{open} means that jobs in actual boundary slots are excluded. Thus $W$ is the union of the complete slots strictly between the boundaries; its size is a multiple of $m$, and it may be empty. An interval with consecutive boundaries is empty, and $(\bot,\top)=J$. The expression ``at or before $a$'' means $\tau_f(x)\le\operatorname{pos}(a)$.

For the three-chain schedule, $(Z_1,Z_3)=Z_2$, while $(\bot,Z_3)=Z_1\cup Z_2$. These identities concern positions in the chosen witness schedule. In general, $(a,b)$ is not the same set as $D_a\cap P_b$.
For example, schedule nine independent jobs in three full slots and choose the first and third as $a,b$. Then $(a,b)$ is the nonempty middle slot, but $D_a=P_b=\varnothing$. This example explains the temporal notation; an arbitrary interval need not satisfy the additional invariants imposed later in the construction.

\paragraph{The domain of an induced precedence relation.}
A \emph{domain} $U$ means a specified subset of $J$. Its induced strict order is the restriction of the global order:
\(
\prec_U\ =\ \prec_f\cap(U\times U).
\)
This restriction retains indirect precedence even when its original path uses jobs outside $U$. For instance, if $a\prec_f b\prec_f c$ and $U=\{a,c\}$, then $a\prec_U c$. Deleting $b$ from the domain does not remove this relation. The domains used for scheduling are unions of full slots; the definitions of closure and priority below also apply to arbitrary subsets $U\subseteq J$.

\paragraph{Predecessor closure, successor closure, ideals, and upsets.}
Let $U\subseteq J$ be a specified domain. A set $I\subseteq U$ is \emph{predecessor-closed in $U$}, or an \emph{ideal of $U$}, if
\(
\forall y\in I\ \forall x\in U:
\quad x\prec_f y\ \Longrightarrow\ x\in I.
\)
Thus including a job requires including all its direct and indirect predecessors that belong to the stated domain. Predecessors outside $U$ are not required to belong to $I$.

A set $K\subseteq U$ is \emph{successor-closed in $U$}, or an \emph{upset of $U$}, if
\(
\forall x\in K\ \forall y\in U:
\quad x\prec_f y\ \Longrightarrow\ y\in K.
\)
Thus including a job requires including all its successors in the stated domain. A \emph{global} ideal or upset means that the domain is $U=J$. Whenever the domain is omitted below, it is $J$. These definitions always use the current frame; reversal turns a global ideal into a global upset, and conversely.

For $J=\{a,b,c,d,e,f,g,h,i\}$ with $a\prec b\prec c$ and the remaining jobs independent, $\{a,b\}$ is a global ideal and $\{b,c\}$ is a global upset. The set $\{b\}$ is neither: it omits predecessor $a$ and successor $c$. Both $\varnothing$ and $J$ are ideals and upsets. The complement in $U$ of an ideal of $U$ is an upset of $U$, and conversely.
For a local example, take $U=\{b,c\}$. Then $\{b\}$ is an ideal of $U$, although it is not a global ideal. A global ideal extending it is $H=\{a,b\}$, because $H\cap U=\{b\}$. In general, an identity $H\cap U=S$ specifies agreement inside $U$, not equality of the global sets $H$ and $S$.

The strict and weak predecessor cones of an actual job set are global ideals; the corresponding successor cones are global upsets. For example, if $x\in D_A$ and $x\prec y$, then some $a\in A$ satisfies $a\prec x\prec y$, so $y\in D_A$. Rank suffixes $R_h$ are global upsets for the same reason: ranks increase along precedence.

\paragraph{A priority order has an explicitly specified domain.}
For a fixed domain $U\subseteq J$, a \emph{priority order on $U$} is a strict total order $\triangleleft_U$ on the jobs of $U$ that extends their induced precedence order. More explicitly, it is irreflexive and transitive; for distinct $x,y\in U$, exactly one of $x\triangleleft_U y$ and $y\triangleleft_U x$ holds; and
\[
x,y\in U,\quad x\prec_f y\quad\Longrightarrow\quad x\triangleleft_U y.
\]
The notation $x\triangleleft_U y$ means that $x$ has earlier or better priority. It says nothing about jobs outside $U$. Equivalently, list $U$ as $u_1\triangleleft_U\cdots\triangleleft_U u_s$ and define its priority position $\rho_U(u_i)=i$. This position can differ from both a job's numerical label and its fixed global topological rank. A priority order is an order for comparison, not an assignment of execution times.

\paragraph{Priority prefixes and their relation to time.}
For the priority list $u_1\triangleleft_U\cdots\triangleleft_U u_s$, its prefix of length $k$ is
\(
S_k=\{u_1,\ldots,u_k\},\qquad 0\le k\le s,
\)
with $S_0=\varnothing$. Every $S_k$ is an ideal of $U$: if $y\in S_k$ and $x\in U$ precedes $y$, then $x\triangleleft_U y$, so $x$ occurs among the first $k$ jobs as well. In the separator construction a chosen priority prefix is called the \emph{low set} or \emph{cutoff}; the remaining domain jobs are called high. Low therefore means earlier priority, not later or less important priority.

A priority prefix need not be a temporal prefix of the witness schedule. In the three-chain example, use the priority list
\(
a_1,a_2,a_3,b_1,b_2,b_3,c_1,c_2,c_3.
\)
It extends precedence. Its first three jobs form the ideal $\{a_1,a_2,a_3\}$, although these jobs occupy three different slots in $(Z_1,Z_2,Z_3)$. A temporal prefix, in contrast, consists of all jobs occurring strictly before a specified slot.

\paragraph{The list of global upsets and the meaning of ``qualifies.''}
A node carries an ordered finite list $\mathcal K=(K_1,\ldots,K_\ell)$, where every $K_i\subseteq J$ is a global upset in the current frame. Entries may overlap, may contain jobs outside the node, and are not required to partition $J$ or the node. The empty list is allowed. Define
\[
Q(\mathcal K)=\bigcup_{i=1}^{\ell}K_i,
\quad Q(())=\varnothing.
\]
A job $x\in J$ \emph{qualifies with respect to $\mathcal K$} if $x\in Q(\mathcal K)$. Such a job is called a \emph{qualifier}. These terms refer only to membership in the current list. They do not mean that the job is ready to run, has passed a scheduling test, or belongs to the current interval. For example, with $a\prec b\prec c$ and $\mathcal K=(\{b,c\})$, job $b$ qualifies even before its predecessor $a$ has completed. Job $a$ can be ready without qualifying.

For a qualifier $x$, define its largest membership index and its key by
\[
i_{\mathcal K}(x)=\max\{i:x\in K_i\},\qquad
\operatorname{key}_{\mathcal K}(x)=\bigl(i_{\mathcal K}(x),\operatorname{rank}(x)\bigr).
\]
These quantities are undefined for a nonqualifier. Compare keys lexicographically: the smaller index comes first, and equal indices are ordered by increasing rank. Thus a job belonging to several entries uses its \emph{last} entry for comparison. For independent jobs with ranks equal to labels and
\(
K_1=\{1,2,5\},\quad K_2=\{2,3,6\},\quad K_3=\{4,5,6\},
\)
we have $\operatorname{key}(2)=(2,2)$ and $\operatorname{key}(5)=(3,5)$, although both jobs also belong to $K_1$.

The key order is a strict total priority order on the domain $Q(\mathcal K)$. Indeed, if $x\prec y$ and $x$ qualifies, upward closure makes every entry containing $x$ also contain $y$. Hence $i_{\mathcal K}(x)\le i_{\mathcal K}(y)$; if the indices tie, $\operatorname{rank}(x)<\operatorname{rank}(y)$. Restricting the key order gives a priority order on $W\cap Q(\mathcal K)$. It does not, by itself, order the nonqualifying jobs of $W$.

\paragraph{Normalized schedules: the comparison domain and feasible choices.}
Fix a domain $U\subseteq J$ and a priority order $\triangleleft_U$ on it. When $m\mid|U|$, a feasible full schedule of $U$ is an ordered partition
\(
\pi=(A_1,\ldots,A_q),\quad q=|U|/m,
\)
into sets of size $m$, such that $x,y\in U$ and $x\prec_f y$ imply that the slot of $x$ is strictly earlier than the slot of $y$. This condition makes each slot an antichain. Denote the collection of these schedules by $\mathcal S_f(U)$. Set $\mathcal S_f(U)=\varnothing$ if $m\nmid|U|$ or if no such schedule exists. The collection $\mathcal S_f(\varnothing)$ consists of the empty schedule.

For $\pi\in\mathcal S_f(U)$, sort the jobs of each $A_t$ by increasing priority position $\rho_U$, and concatenate the resulting lists. Let $\operatorname{enc}_{\triangleleft_U}(\pi)$ be the resulting sequence of $|U|$ priority positions. Compare two such sequences lexicographically: at the first position where they differ, the sequence with the smaller number is earlier. A schedule $\pi^*\in\mathcal S_f(U)$ is \emph{normalized for $\triangleleft_U$} if
\[
\operatorname{enc}_{\triangleleft_U}(\pi^*)
\le_{\mathrm{lex}}
\operatorname{enc}_{\triangleleft_U}(\pi)
\qquad\text{for every }\pi\in\mathcal S_f(U).
\]
The feasible comparison class contains schedules of exactly $U$, on exactly $|U|/m$ full slots, using its induced global precedence order. No exterior jobs are included. If $U$ is a contiguous interval of a complete witness, these local slots are placed in its original temporal positions; Lemma~\ref{str:replacement} below shows that every feasible replacement preserves the exterior constraints.

For six independent jobs, $m=3$, and priority $1\triangleleft\cdots\triangleleft6$, compare
\(
\pi=(\{1,2,3\},\{4,5,6\}),\qquad
\pi'=(\{1,2,4\},\{3,5,6\}).
\)
Their encodings are $(1,2,3,4,5,6)$ and $(1,2,4,3,5,6)$. The first is smaller at the third position; in fact $\pi$ is normalized among all feasible full schedules in this example.

Normalization does not mean dividing a priority list into consecutive groups of $m$. For example, let the only precedence be $4\prec1$, let $m=3$, and use the priority list $(3,4,1,2,5,6)$. Its first three jobs cannot share a slot. The normalized full schedule is instead
\(
(\{3,4,2\},\{1,5,6\}),
\)
whose priority-position encoding is $(1,2,4,3,5,6)$. The first slot uses the earliest feasible three-job choice in this particular example.

Whenever $\mathcal S_f(U)$ is nonempty, a normalized schedule exists because it is a finite nonempty collection with a total lexicographic comparison. Sorting within a slot is only a comparison convention: its jobs still execute simultaneously. This convention is separate from storing job labels increasingly in the algorithm. Choosing a normalized witness is used only in the existence proof; no efficient procedure for finding it is assumed here.

\paragraph{The data stored at a node and the three invariants.}
A node consists of a current frame, a complete feasible witness schedule in that frame, boundaries $a,b$, their temporal interval $W=(a,b)$, and a list $\mathcal K$ of global upsets. The full schedule determines what ``before,'' ``after,'' and ``at a boundary'' mean even for jobs outside $W$. Distinct children may use distinct copies of that complete witness. The construction requires the following three invariants; they are additional properties of its annotated nodes, not consequences of merely being a temporal interval.

\textbf{(I1) Local cone cover.} Qualifiers inside $W$ precede the right boundary, and nonqualifiers inside $W$ descend from the left boundary:
\[
W\cap Q(\mathcal K)\subseteq P_b,
\qquad
W\setminus Q(\mathcal K)\subseteq D_a.
\]
In particular $W\subseteq P_b\cup D_a$. The two cones may overlap. This property does not say that every predecessor of $b$ qualifies or that every descendant of $a$ fails to qualify.

\textbf{(I2) Control of qualifying jobs in the past.} Every global qualifier occurring at or before the left boundary belongs to its weak predecessor cone:
\[
\{x\in Q(\mathcal K):\tau_f(x)\le\operatorname{pos}(a)\}\subseteq\widehat P_a.
\]
This quantifies over $J$, not just $W$. A qualifying job before $a$ must precede some job of $a$; a qualifying job in $a$ is included by the weak cone. The invariant will allow later global formulas to exclude all qualifying jobs in the temporal past.

\textbf{(I3) No improving exchange at the left boundary.} First choose \emph{any} feasible full reordering of the jobs of $W$ within exactly their original slots, keeping every job in $J\setminus W$ fixed. For every pair
\[
x\in W\setminus D_a,\qquad y\in\operatorname{jobs}(a)\cap Q(\mathcal K),
\qquad \operatorname{key}_{\mathcal K}(x)<_{\mathrm{lex}}\operatorname{key}_{\mathcal K}(y),
\]
exchanging the slots occupied by $x$ and $y$, while leaving all other jobs fixed, must make the complete schedule infeasible. Such an exchange moves the better-priority job $x$ to the earlier boundary slot and $y$ to $x$'s former slot. It preserves the number of jobs in each slot, so infeasibility here is a failure of a precedence constraint. The keys and the list remain fixed throughout this test. Property (I1) guarantees that the key of $x$ is defined: if $x\in W$ were a nonqualifier, it would belong to $D_a$.

The phrase \emph{any feasible reordering} is part of (I3). The condition must remain true after every permitted rearrangement of the interior, not only in the currently displayed witness. This strength is needed when descendants later choose different feasible schedules.

\paragraph{What ``vacuous'' means.}
A universally quantified condition is \emph{vacuously true} when there is no object satisfying the conditions to which it applies. At a left sentinel there is no job $y\in\operatorname{jobs}(a)$, so there is no pair to test in (I3); it is true for that reason. In addition, $D_a=J$ makes $W\setminus D_a=\varnothing$. The set quantified over in (I2) is also empty, since no real job occurs at or before slot position zero. Vacuous truth is a logical statement about an empty range of possibilities; it does not mean that the invariant is omitted or assumed without a condition.

\paragraph{An example satisfying the invariants.}
Use the three-chain witness $(Z_1,Z_2,Z_3)$, with actual boundaries $a=Z_1$, $b=Z_3$, interval $W=Z_2$, and list $\mathcal K=(J)$. Then every job qualifies, including jobs outside $W$. Since $P_b=Z_1\cup Z_2$, we have $W\cap Q(\mathcal K)=Z_2\subseteq P_b$. There are no nonqualifying jobs in $W$, so the other inclusion in (I1) holds. The jobs at or before $a$ are exactly $Z_1$, and $\widehat P_a=Z_1$, proving (I2). Finally $W\subseteq D_a$, so there is no $x\in W\setminus D_a$; (I3) is vacuously true.

At the root one uses sentinel boundaries and the list $(J)$. Then $W=J$, $P_b=J$, and (I1) holds; (I2) and (I3) are vacuous at the left sentinel. Later lemmas prove that the construction preserves all three properties at children.

\paragraph{How a priority order on the whole node is specified.}
The key order above applies only to qualifiers. When normalizing the entire node in the subsequent construction, let $O=W\cap Q(\mathcal K)$ and use the following three consecutive priority blocks:
\[
(W\setminus O)\cap P_b,\qquad O,\qquad W\setminus P_b.
\]
Order the first and third blocks by global rank and the second by qualifier key. By (I1), $O\subseteq P_b$, so these blocks partition $W$. The first block is an ideal of $W$, and the last is an upset of $W$; hence no precedence edge points from a later block to an earlier one. The internal orders also respect precedence. This specifies a strict total priority order on exactly $W$, including its nonqualifiers.

\begin{lemma}[Replacement and prefix minimality]\label{str:replacement}
Let $U$ be the jobs in a contiguous interval of a complete feasible full witness schedule in a fixed frame. Every feasible full schedule of $U$ can replace its current schedule within the same slots, with all jobs outside $U$ fixed.

Fix also a priority order on $U$, and suppose its current local schedule is normalized for that priority. Let $r$ be one of its slots. After any feasible reordering of the jobs in slots of $U$ strictly after $r$, keeping $r$ and all earlier slots unchanged, no feasible exchange can replace a job $y\in r$ by a later job $x\in U$ with $x\triangleleft_U y$.
\end{lemma}

\begin{proof}
For replacement, consider every precedence relation. Relations between exterior jobs are unchanged. Relations between interior jobs hold by feasibility of the replacement. If an exterior job precedes an interior job, it occurred before the whole interval: it cannot lie in an interior slot, and feasibility excludes its occurring after the interval. Similarly, an exterior successor of an interior job occurs after the whole interval. The replacement uses exactly the same slots, so both types of exterior relation remain satisfied. Each slot still has $m$ jobs.

For prefix minimality, suppose the asserted exchange were feasible after the allowed later reordering. Sort each slot by the fixed priority. Every slot before $r$ would agree with the original normalized schedule. At $r$, replacing $y$ by the better-priority job $x$ makes its sorted list lexicographically smaller: in the sorted union of the old and new slot sets, the first differing job is the newly included $x$, which precedes the removed $y$. Thus the resulting feasible schedule of $U$ has a smaller encoding than the normalized schedule, regardless of its later slots. This contradicts the definition of normalization.
\end{proof}

\subsection{A separator describes a temporal prefix}

We work in one fixed frame, so time, precedence, and all cones are
interpreted in that frame. Let $U\subseteq J$ be the jobs in a nonempty
contiguous interval of a feasible full witness schedule. Write its slots
in chronological order as
\[
 \sigma_U=(Z_1,\ldots,Z_q),\qquad
 U=Z_1\mathbin{\dot\cup}\cdots\mathbin{\dot\cup}Z_q,\qquad |Z_i|=m.
\]
Define the local time map $\tau_U:U\to\{1,\ldots,q\}$ by
$\tau_U(x)=i$ when $x\in Z_i$. For a slot $z=Z_r$, write
$\operatorname{pos}_U(z)=r$. Comparisons of slot positions, such as
$s<t$, refer to these chronological indices. Its strict temporal prefix is
\[
 \operatorname{Pref}_U(z)=\{x\in U:\tau_U(x)<r\}.
\]
Cones such as $P_z$ and $D_z$ are still computed on all of $J$.
They include indirect predecessors and successors as well as direct
ones, even when a witnessing precedence path leaves $U$.

\paragraph{What a cutoff means.}
Fix a strict total priority order
\(
 u_1\mathrel{\triangleleft_U}u_2\mathrel{\triangleleft_U}
 \cdots\mathrel{\triangleleft_U}u_n,\quad n=|U|,
\)
on precisely the domain $U$. It extends the induced precedence order:
if $x,y\in U$ and $x\prec y$, then $x\triangleleft_U y$.
For $0\leq k\leq n$, the \emph{priority cutoff at $k$} is the set
\(
                 S_k=\{u_1,\ldots,u_k\},\qquad S_0=\varnothing.
\)
Thus $k$ is a position in the priority list, whereas
$\tau_U(x)$ is a position in the schedule. The jobs of $S_k$ are
called \emph{low}, and the jobs of $U\setminus S_k$ are called
\emph{high}. Here low means a smaller position in the priority order:
low jobs have earlier, and hence better, priority.

Every cutoff $S_k$ is an ideal of the induced order on $U$.
Indeed, if $y\in S_k$ and $x\in U$ precedes $y$, then $x$ occurs
earlier in the priority list and also belongs to $S_k$.
Its complement $U\setminus S_k$ is an upset on $U$.
A cutoff need not be a temporal prefix: an early-priority job can
occur late in a feasible schedule. The definition of a separator below
also applies to an arbitrary ideal $S\subseteq U$, without first
specifying a priority order.

\begin{definition}[Separator]
Let $S$ be an ideal of the induced precedence order on $U$.
An actual slot $z=Z_r$ \emph{separates $S$ on $U$} if
\[
 \begin{aligned}
 &x\in U\setminus S,\ \tau_U(x)<r
       &&\Longrightarrow x\in P_z,\\
 &x\in S,\ \tau_U(x)>r
       &&\Longrightarrow x\in D_z.
 \end{aligned}
\]
We call these the \emph{earlier-high property} and the
\emph{later-low property}, respectively.
The exterior left boundary $a$ has local position $0$ and is a
separator when $S\subseteq D_a$. It may be an actual boundary slot
outside $U$ or the left sentinel, with the cone conventions of
Section~7.1.
\end{definition}

The earlier-high property says that every high job already scheduled
before $z$ precedes at least one job in $z$.
The later-low property says that every low job still scheduled after
$z$ has at least one predecessor in $z$.
Neither condition constrains jobs in $z$ itself. Nor does separation
require all low jobs to be before $z$, or all high jobs to be after it.
For the exterior left boundary there are no earlier domain jobs, so
only the condition $S\subseteq D_a$ remains.

\paragraph{Example: a cutoff crossing a separator.}
Let $m=3$, and let $J=U$ consist of three independent chains
\[
 a_1\prec a_2\prec a_3,\qquad
 b_1\prec b_2\prec b_3,\qquad
 c_1\prec c_2\prec c_3.
\]
Consider the feasible schedule
\[
 Z_1=\{a_1,b_1,c_1\},\quad
 Z_2=\{a_2,b_2,c_2\},\quad
 Z_3=\{a_3,b_3,c_3\}.
\]
Use the priority list
$(a_1,a_2,a_3,b_1,b_2,b_3,c_1,c_2,c_3)$ and the cutoff
$S=\{a_1,a_2,a_3\}$.
The slot $z=Z_2$ separates $S$: the earlier high jobs $b_1,c_1$
belong to $P_z$, and the later low job $a_3$ belongs to $D_z$.
The low job $a_2$ lies in the separator slot.
Thus the same cutoff has a low job before, at, and after the
separator. In particular,
$S\ne\operatorname{Pref}_U(z)=\{a_1,b_1,c_1\}$.

\paragraph{Turning separation into a global formula.}
Let $H\subseteq J$ be a global set that agrees with the low set on
the domain, meaning $H\cap U=S$. Define
\[
              B_z(H)=P_z\cup\bigl(H\setminus\widehat D_z\bigr),
              \qquad \widehat D_z=z\cup D_z.
\]
The term $P_z$ includes every earlier high job guaranteed by
separation. The second term includes low jobs, but removes the
separator slot itself and all its descendants.
Using the \emph{weak} cone is essential: subtracting only $D_z$
would leave low jobs in $z$, although the prefix is strictly before
$z$. This expression is global; the intersection with $U$ below
restricts its answer to the domain under discussion.

\begin{lemma}[The prefix selected by a separator]\label{str:prefix}
Suppose that $z$ is an actual separator of $S$ on $U$ and
$H\subseteq J$ satisfies $H\cap U=S$. Then
\[
                  \operatorname{Pref}_U(z)=U\cap B_z(H).
\]
Reversing both time and precedence and replacing $S$ by
$U\setminus S$ preserves separation.
If, in addition, $H$ is a global ideal, then $B_z(H)$ is a
global ideal.
\end{lemma}

\begin{proof}
\textbf{1. Every job before $z$ belongs to the formula.}
Take $x\in U$ with $\tau_U(x)<\operatorname{pos}_U(z)$.
If $x$ is high, separation gives $x\in P_z$.
If $x$ is low, then $x\in S=H\cap U$, so $x\in H$.
Moreover, $x\notin z$, and $x\notin D_z$: a successor of a job
in $z$ must be scheduled after $z$ in every feasible schedule.
Hence $x\in H\setminus\widehat D_z$.
Both cases give $x\in U\cap B_z(H)$.

\textbf{2. Every domain job selected by the formula is before $z$.}
Take $x\in U\cap B_z(H)$.
If $x\in P_z$, feasibility puts $x$ strictly before $z$.
Otherwise, $x\in H\setminus\widehat D_z$.
Because $x\in U$, the equality $H\cap U=S$ makes $x$ low.
It cannot be in $z$, since $z\subseteq\widehat D_z$.
It cannot be after $z$, since the later-low property would put
it in $D_z\subseteq\widehat D_z$.
Therefore it is before $z$.
Together, these two steps prove the stated equality.

\textbf{3. Simultaneous reversal preserves separation.}
In the reversed frame, the old high set $U\setminus S$ becomes
the new low set. It is an ideal in the reversed induced order,
because it was an upset in the original induced order.
An old low job after $z$ becomes a new high job before $z$;
its membership in the old $D_z$ becomes membership in the new
$P_z$. An old high job before $z$ becomes a new low job after
$z$; its membership in the old $P_z$ becomes membership in the
new $D_z$. These are exactly the two required conditions.

\textbf{4. The global formula is an ideal when $H$ is one.}
The set $P_z$ is a global ideal by transitivity.
The weak successor cone $\widehat D_z$ is a global upset, so its
complement $J\setminus\widehat D_z$ is a global ideal.
Under the additional hypothesis on $H$,
\[
 H\setminus\widehat D_z
       =H\cap(J\setminus\widehat D_z)
\]
is an intersection of ideals, and $B_z(H)$ is a union of ideals.
It is therefore a global ideal.
This additional hypothesis is not needed for the prefix equality.
\end{proof}

For an exterior left separator $a$, the same formula selects the
empty domain prefix. Indeed, feasibility gives $U\cap P_a=\varnothing$,
and $H\cap U=S\subseteq D_a$ gives
$U\cap(H\setminus\widehat D_a)=\varnothing$.
These statements also hold with the left sentinel conventions.

\paragraph{Checking the formula in examples.}
In the three-chain example, take $H=S$ and $z=Z_2$. Then
\(
 P_z=Z_1,\quad \widehat D_z=Z_2\cup Z_3,\quad
 H\setminus\widehat D_z=\{a_1\},
\)
so $U\cap B_z(H)=Z_1$, exactly as required.
To see why the $H$ term is needed in general, instead take six
independent jobs with schedule
$(\{1,2,3\},\{4,5,6\})$. The second slot $z=\{4,5,6\}$
separates $S=H=\{1,2,3\}$. Here $P_z=D_z=\varnothing$,
yet the temporal prefix is nonempty:
$B_z(H)=H\setminus z=\{1,2,3\}$.

The restriction to $U$ is equally necessary. Return to the three
chains, but now use the smaller domain $U'=Z_2\cup Z_3$,
the local ideal $S'=Z_2$, the global ideal $H'=Z_1\cup Z_2$,
and separator $z'=Z_3$.
The formula gives $B_{z'}(H')=Z_1\cup Z_2$ on the entire ground
set. Its restriction to $U'$ is $Z_2$, the desired local prefix.
The jobs of $Z_1$ are outside $U'$ and must not be counted.

\paragraph{Moving between consecutive priority cutoffs.}
Now fix a priority order on $U$ and normalize its full schedule.
For consecutive cutoffs
$S=S_k$ and $S^+=S_{k+1}=S\cup\{j\}$, the job $j=u_{k+1}$
changes from high to low. We call this change a \emph{promotion}.
It changes the classification of a job, not its time slot.
Likewise, the walk below moves a separator position through a
\emph{fixed} normalized schedule; it does not reschedule jobs.

For a current position $r$, call $x$ a \emph{low exception} for a
low set $L$ if
\(
 x\in L,\,
 \tau_U(x)>\operatorname{pos}_U(r),\, x\notin D_r.
\)
Such a job is exactly a violation of the later-low property at $r$.
The next position is the earliest later slot containing any low
exception. If several exceptions share that slot, the next position
is still unambiguous; any one of them can be used in the proof.

\begin{lemma}[Common-cutoff walk]\label{str:walk}
Fix a full schedule of $U$ normalized for a specified priority order.
Let $S$ and $S^+=S\cup\{j\}$ be consecutive prefixes of that order.
Suppose that $s$ separates $S$, where $s$ is an actual slot in $U$
or its exterior left boundary.
Starting at $r=s$, repeatedly apply the following rule:
if there is a low exception for $S^+$ at $r$, replace $r$ by the
earliest later slot containing such an exception; otherwise stop.
The process terminates at a position $t\geq s$ that separates
\emph{both} $S$ and $S^+$.

If the priority order is instead specified on a larger domain
$V\supseteq U$, use its restriction to $U$ and normalize for that
restriction. A promotion of a job outside $U$ leaves the restricted
cutoff and the separator position unchanged.
\end{lemma}

\begin{proof}
We first prove a fact about one advance for a fixed priority prefix
$L\subseteq U$. Assume that the current position $r$ has the
earlier-high property for $L$; we do not assume that it already
has the later-low property.
Let $t>r$ be the earliest slot containing a low exception $x$.

\textbf{1. All domain predecessors of $x$ occur strictly before $r$.}
If $p\in U$ and $p\prec x$, then $p\in L$, because $L$ is an ideal.
Feasibility puts $p$ before the slot $t$ of $x$.
It cannot be in $r$: that would make $x$ a descendant of $r$.
If it were strictly between $r$ and $t$, earliestness of the
exception slot would imply $p\in D_r$. Transitivity would again
give $x\in D_r$, contrary to the choice of $x$.
Thus every such predecessor is strictly before $r$.
If $r$ is exterior, this says that $x$ has no predecessor in $U$.

\textbf{2. Every high job from $r$ up to $t$ has a successor by $t$.}
Let $y\in U\setminus L$ satisfy
\[
 \operatorname{pos}_U(r)\leq\tau_U(y)
                        <\operatorname{pos}_U(t).
\]
Suppose, for a contradiction, that $y$ has no successor in $U$
scheduled at or before $t$. Exchange $x$ and $y$.
Moving $x$ to the old slot of $y$ crosses no predecessor of $x$,
by Step~1. Moving $y$ to $t$ crosses no successor of $y$, by the
supposition. The other precedence inequalities remain valid, and
each slot still contains exactly $m$ jobs. Exterior constraints
remain valid by Lemma~\ref{str:replacement}.
The exchange is therefore feasible.

Since $L$ is a priority prefix, the low job $x$ has better priority
than the high job $y$. The first changed slot is the old slot of
$y$; replacing $y$ by $x$ makes its priority-sorted list
lexicographically smaller. This contradicts normalization.
Consequently, $y$ has a successor $y_1\in U$ at or before $t$.
The successor is also high, because $U\setminus L$ is an upset.

\textbf{3. Such high jobs actually precede a job in $t$.}
If $y_1$ is in $t$, this is already established.
Otherwise apply Step~2 to $y_1$, obtaining a later high successor
$y_2$ still at or before $t$. Continuing gives a chain
\[
                         y\prec y_1\prec y_2\prec\cdots.
\]
The slot indices strictly increase and are bounded above by the
index of $t$. Thus the chain must reach a high job in $t$.
In particular, every high job at or after $r$ and before $t$
belongs to $P_t$.

\textbf{4. The earlier-high property survives the advance.}
Step~3 deals with the high jobs between $r$ and $t$, including
those in $r$. If $r$ is internal and a high job $y$ is earlier
than $r$, the assumed earlier-high property gives a job $w\in r$
with $y\prec w$. The job $w$ is high, by upward closure.
Step~3 puts $w$ in $P_t$, and transitivity puts $y$ in $P_t$.
At an exterior $r$ there are no earlier jobs of $U$ to consider.
Every high job before $t$ therefore belongs to $P_t$.

\textbf{5. Apply this fact to the new cutoff $S^+$.}
Initially $s$ has the earlier-high property for $S^+$, since
$U\setminus S^+\subseteq U\setminus S$.
Steps~1--4 preserve this property at every advance.
On termination there is no later low exception for $S^+$,
which is precisely its later-low property.
Thus the final position separates $S^+$.

\textbf{6. The final position also separates the old cutoff $S$.}
If there is no advance, $t=s$ already separates $S$.
Otherwise, the first exception must be $j$: every old low job
after $s$ lies in $D_s$, because $s$ separates $S$.
At the first target slot, $j$ is in the slot itself, so it is not
an earlier high job for the old cutoff.
All other old high jobs are high for $S^+$; Step~4 has already
established the required predecessor relation for them.
The first target therefore has the earlier-high property for $S$
as well as for $S^+$.

At every subsequent advance, $j$ is at or before the current
position. Hence any later exception belongs to the old set $S$.
The earliest exception slot for $S^+$ is now also the earliest
exception slot for $S$. We may apply Steps~1--4 with $L=S$,
which preserves the earlier-high property for the old cutoff.
Finally, the later-low property for $S^+$ implies that for
$S\subseteq S^+$. Both separator conditions for $S$ hold at the end.

\textbf{7. Termination and exterior promotions.}
Each advance strictly increases the slot index, and there are only
$q$ actual slots, so the process terminates.
For consecutive prefixes $A$ and $A\cup\{j\}$ of an order on
$V\supseteq U$, if $j\notin U$ then
\[
                  A\cap U=(A\cup\{j\})\cap U.
\]
The local low set has not changed, so its existing separator remains
valid and the position is left unchanged. If $j\in U$, the
restricted prefixes differ by exactly this one job, and the preceding
argument applies.
\end{proof}

\paragraph{Example: one promotion requiring two advances.}
Let $m=3$ and
\(
 U=J=\{a,b,c,d,e,x,j,h,k\}.
\)
The only precedence edges are
\(
                  a\prec x,\quad b\prec d,\quad
                  c\prec e,\quad j\prec h,
\)
and use the priority list $(a,b,c,d,e,x,j,h,k)$.
The normalized full schedule is
\(
 Z_1=\{a,b,c\},\qquad Z_2=\{d,e,j\},\qquad Z_3=\{x,h,k\}.
\)
To verify normalization, the lexicographically smallest possible
first slot is $\{a,b,c\}$, and the displayed schedule completes it.
For the second slot, the three smallest remaining jobs
$\{d,e,x\}$ would leave $\{j,h,k\}$ for the last slot, which is
infeasible because $j\prec h$. The next smallest choice
$\{d,e,j\}$ admits the displayed completion. Thus the schedule
is lexicographically smallest among feasible full schedules.

Start with the cutoff $S=\{a,b,c,d,e,x\}$ and position $s=Z_1$.
It is a separator: there are no earlier high jobs, and the later
low jobs $d,e,x$ all belong to
$D_{Z_1}=\{d,e,x\}$.
Promote $j$, so that $S^+=S\cup\{j\}$.
\begin{enumerate}
\item At $Z_1$, the newly low job $j$ is later and does not belong
to $D_{Z_1}$. The first exception is in $Z_2$, so move to $Z_2$.
\item At $Z_2$, the old low job $x$ is still later, but
$D_{Z_2}=\{h\}$ does not contain $x$. Thus move again, to $Z_3$.
\item At $Z_3$ there are no later jobs, so the walk stops.
For the old cutoff $S$, the only earlier high job is $j$, and
$j\prec h$ puts it in $P_{Z_3}$.
For the new cutoff $S^+$, the high jobs $h,k$ are in $Z_3$
itself. Both cutoffs are therefore separated at the final position.
\end{enumerate}
The changing descendant cones explain why reaching the promoted
job does not necessarily finish the walk: $x$ was covered by
$D_{Z_1}$ but is not covered by $D_{Z_2}$.
The proof's exchange argument is also visible here.
Swapping the low job $x$ with the old high job $j$ would improve
the second slot, but would put $j$ into the same slot as its
successor $h$. That successor is precisely the obstruction used
in the proof.

\paragraph{Why the common cutoff matters for the hierarchy.}
The starting position $s$ separates $S$, and the final position
$t$ separates both $S$ and $S^+$. Thus $s$ and $t$ share the
\emph{old} cutoff $S$. If $H\cap U=S$, Lemma~\ref{str:prefix}
describes both endpoint prefixes using the same global set $H$.
This is the information needed when the construction forms the
open child interval between the endpoints.
Intermediate positions of a walk need not separate either cutoff:
in the example, $Z_2$ fails the later-low condition because of $x$.
The construction therefore retains the final endpoint of a promotion,
rather than marking every intermediate advance.

\begin{corollary}[An actual separator]\label{str:actual}
Every priority prefix in a nonempty normalized full schedule has
an actual separator.
Every ideal of a feasible nonempty full-slot domain has an actual
separator in some feasible full schedule of that domain.
\end{corollary}

\begin{proof}
For the first assertion, begin at the first actual slot $Z_1$
with the empty cutoff $S_0=\varnothing$.
There are no earlier domain jobs and no low jobs, so both separator
conditions hold. Promote the jobs one by one in priority order.
Lemma~\ref{str:walk} supplies a separator for the new cutoff after
each promotion. The position either remains at an actual slot
or advances to another actual slot. Induction therefore gives
an actual separator for every cutoff, including the empty and
full cutoffs.

For the second assertion, let $I\subseteq U$ be an arbitrary
ideal. Choose a topological order of the induced order on $I$
and then a topological order on $U\setminus I$, and concatenate
them. No precedence can point from $U\setminus I$ into $I$,
by the ideal property, so this is a priority order on all of $U$
extending precedence. In that order $I$ is a prefix.
A feasible full schedule exists by hypothesis; normalize among
such schedules for the chosen priority and apply the first assertion.
The resulting separator is in this chosen normalized schedule.
The assertion does not require every feasible schedule to contain
a separator for $I$.
\end{proof}

\subsection{The construction at one node}

We now construct the marked slots and children of a nonempty node
$W=(a,b)$. The node comes with a complete feasible witness schedule,
a current frame, and a list $\mathcal K=(K_1,\ldots,K_\ell)$ of global
upsets satisfying (I1)--(I3). An empty node needs no construction.
All precedence relations, cones, ranks, and time comparisons below
use the current frame until a reversal is explicitly made.

A \emph{mark} is an actual full slot retained by the construction.
The \emph{center} $v$ is one distinguished marked slot: we first
construct the part before $v$ in the current frame, then process
the part after $v$ in the reversed frame. A \emph{child} is the job
set in a nonempty open gap between consecutive retained boundaries.
A child before the center inherits the current frame and is called
a \emph{same-direction child}.
\repairtext{A child on the other side inherits the reversed frame and is
an \emph{opposite-side~child}.}
These names describe how the child is processed; they do not assert
that all its jobs are precedence-related to the center.

\paragraph{The old qualifier priority order.}
Here \emph{old} means belonging to the parent list $\mathcal K$ before
we construct a child's list. Recall
\[
 Q(\mathcal K)=\bigcup_{i=1}^{\ell}K_i,\qquad
 O=W\cap Q(\mathcal K).
\]
Thus $Q(\mathcal K)$ is the set of global qualifiers, whereas $O$
contains only qualifiers inside the current node.
For $x\in Q(\mathcal K)$, put
\[
 i_{\mathcal K}(x)=\max\{i:x\in K_i\},\qquad
 \operatorname{key}_{\mathcal K}(x)
       =\bigl(i_{\mathcal K}(x),\operatorname{rank}(x)\bigr).
\]
The old qualifier priority order is the increasing lexicographic
order of these keys on the specified domain $Q(\mathcal K)$.
The first coordinate is compared first; the rank breaks ties.
If a job occurs in several entries, its \emph{last} membership
determines the first coordinate. Restricting this order to $O$
gives the order in which the node's qualifiers will be promoted.
This order compares jobs; it does not specify their time slots.

\paragraph{Suffixes and carries.}
For an ordered list $(x_1,\ldots,x_s)$, a \emph{suffix} is a final
segment $(x_k,\ldots,x_s)$, including its first retained element.
The entire list and the empty list are also allowed suffixes.
For example, $(x_3,x_4,x_5)$ is a suffix of
$(x_1,x_2,x_3,x_4,x_5)$.
We use a suffix of the qualifier order to keep the jobs whose
old keys are at least a specified threshold.

A \emph{carry} is the ordered list of global sets representing
the retained suffix. For an entry index $1\leq d\leq\ell$ and
a global rank threshold $1\leq r\leq N+1$, define
\[
 R_r=\{x\in J:\operatorname{rank}(x)\geq r\},\qquad
 \operatorname{carry}_{d,r}(\mathcal K)
       =(K_d\cap R_r,K_{d+1},\ldots,K_\ell).
\]
Here $r$ is a number in the global rank order, not a time slot.
This operation discards entries before $K_d$, restricts the first
retained entry by rank, and keeps every later entry unchanged.
Its qualifier union is exactly
\[
 \{x\in Q(\mathcal K):
       \operatorname{key}_{\mathcal K}(x)\geq_{\mathrm{lex}}(d,r)\}.
\]
The \emph{empty carry} is the empty list, whose union is
$\varnothing$. It is used when there are no jobs to retain.
Even when the threshold is selected using jobs of $W$, the carried
entries remain global subsets of $J$ and can contain exterior jobs.

\paragraph{Example of overlapping entries and a suffix carry.}
Take six independent jobs with ranks equal to their labels and
\[
 K_1=\{1,2,5\},\qquad K_2=\{2,3,6\},\qquad K_3=\{4,5,6\}.
\]
Every set here is an upset because there are no precedences.
The largest membership indices give the keys
\[
 (1,1),\ (2,2),\ (2,3),\ (3,4),\ (3,5),\ (3,6)
\]
for jobs $1,2,3,4,5,6$, respectively.
In particular, job $5$ uses its membership in $K_3$, although
it also belongs to $K_1$.
The threshold $(d,r)=(2,3)$ gives
\[
 \operatorname{carry}_{2,3}(\mathcal K)
       =(\{3,6\},\{4,5,6\}).
\]
Its qualifier union is the old suffix $\{3,4,5,6\}$.
The numerical entry indices have changed, but the relative
priority order of these four jobs is still $3,4,5,6$.

\paragraph{Pruning a list on a child.}
For any ordered list $\mathcal L=(L_1,\ldots,L_s)$ of global
sets and any child domain $V$, \emph{pruning on $V$} deletes
exactly the entries $L_i$ satisfying $L_i\cap V=\varnothing$.
The surviving entries keep their order and are then numbered
consecutively. We write the result as
$\operatorname{prune}_V(\mathcal L)$.
Every retained entry is the whole global set $L_i$.
In particular, pruning does not replace it by $L_i\cap V$.

For the carried list $(\{3,6\},\{4,5,6\})$ above, take
$V=\{1,2,3\}$. The first entry meets $V$ at job $3$ and is
retained; the second does not meet $V$ and is deleted. Thus
\[
 \operatorname{prune}_V((\{3,6\},\{4,5,6\}))=(\{3,6\}).
\]
Job $6$ remains a global qualifier even though it lies outside
$V$. In general, pruning preserves precisely the qualifiers
\emph{inside} $V$:
\[
 V\cap Q(\operatorname{prune}_V(\mathcal L))
                         =V\cap Q(\mathcal L).
\]
It can change global membership and the keys of exterior jobs.

\paragraph{An injection is a newly appended global upset.}
An \emph{injection} is a new entry added to the end of a carried
list. It is not a function or a movement of jobs.
For boundary sets $A,B$ in a fixed frame, define its job set by
\[
                         I(A,B)=D_A\setminus P_B.
\]
It contains the descendants of $A$ that do not precede any job
in $B$. This is a global upset: $D_A$ is an upset, $P_B$ is an
ideal, and
$I(A,B)=D_A\cap(J\setminus P_B)$ is an intersection of upsets.
Sentinel boundaries use the cone conventions from Section~7.1.

In the node construction the fresh entry is $I(a,b)$.
By (I1), $W\subseteq D_a\cup P_b$, and therefore
\[
                       W\cap I(a,b)=W\setminus P_b.
\]
Indeed, a job of $W$ outside $P_b$ must belong to $D_a$.
The injection supplies exactly this part of the parent domain.
The entry itself may contain additional jobs outside $W$.
Appending it last puts its jobs in the last priority block.
Jobs selected only by the carry come before that block.
A job in both the injection and a carried entry uses the
injection's later index.

For a concrete example, let
$A=\{a_1,b_1,c_1\}$, $W=\{a_2,b_2,c_2\}$, and
$B=\{a_3,b_3,c_3\}$ be three consecutive full slots.
Use the chains $a_1\prec a_2\prec a_3$ and
$b_1\prec b_2\prec b_3$, the relation $c_1\prec c_2$,
and no other precedence involving the $c$-jobs. Then
\[
 D_A=\{a_2,a_3,b_2,b_3,c_2\},\quad
 P_B=\{a_1,a_2,b_1,b_2\},\quad
 I(A,B)=\{a_3,b_3,c_2\}.
\]
Its restriction to $W$ is just $\{c_2\}=W\setminus P_B$.
The boundary jobs $a_3,b_3$ remain in the global entry.

\begin{lemma}[Carry and pruning comparisons]\label{str:carry}
A suffix carry retains exactly the global qualifiers whose old
keys are at least its threshold $(d,r)$, and preserves their
relative key order.
For pruning, let $\mathcal L'=\operatorname{prune}_V(\mathcal L)$.
If $x\in V\cap Q(\mathcal L')$ and $y\in Q(\mathcal L')$ satisfy
\[
       \operatorname{key}_{\mathcal L'}(x)
       <_{\mathrm{lex}}\operatorname{key}_{\mathcal L'}(y),
\]
then the same strict comparison holds for their keys in
$\mathcal L$, before pruning.
All entries obtained by carrying, appending an injection, and
pruning remain global upsets.
\end{lemma}

\begin{proof}
\textbf{Carry membership and order.}
Let $i$ be a job's largest old membership index.
If $i<d$, every entry containing it is discarded.
If $i=d$, it survives exactly when its rank is at least $r$.
If $i>d$, its membership in $K_i$ survives unchanged.
These are precisely the conditions for an old key at least
$(d,r)$. For surviving jobs, the largest index becomes $i-d+1$,
and ranks do not change. This increasing renumbering preserves
every relative key comparison.

\textbf{Pruning comparisons.}
Temporarily refer to retained entries by their original indices.
For a job $z$, write $M(z)$ for its largest index before pruning
and $M'(z)$ for its largest retained original index.
Every entry containing $x\in V$ meets $V$, so none of those
entries is removed. Hence $M'(x)=M(x)$.
For the possibly exterior job $y$, some memberships may disappear,
but $M'(y)\leq M(y)$.
Consecutive renumbering of retained indices preserves their order.
If the new comparison is decided by indices, then
$M(x)=M'(x)<M'(y)\leq M(y)$.
If retained indices tie, either $M(y)>M(x)$ already, or the
old indices also tie and the unchanged ranks decide the comparison.
In both cases the old key of $x$ is smaller.
The requirement $x\in V$ is essential to this argument.

\textbf{Closure.}
Rank suffixes are upsets because ranks increase along precedence.
Intersecting the first carried entry with a rank suffix preserves
upward closure. Injections are upsets by the argument above, and
pruning only deletes entries.
\end{proof}

\paragraph{Notation and order of the construction.}
We now work at a \emph{nonempty} node $W=(a,b)$ with its currently
chosen frame and its inherited list $\mathcal K=(K_1,\ldots,K_\ell)$.
All cones, ranks, and temporal comparisons refer to this frame unless
we explicitly reverse it.  Write
\[
 Q=Q(\mathcal K),\qquad O=W\cap Q.
\]
Thus $Q$ is a global upset, while $O$ is the set of qualifiers \emph{in}
$W$.  An \emph{associated cutoff} $S$ will always be a subset of $W$;
its complement $W\setminus S$ consists of the jobs called \emph{high}
for that cutoff.  The interval between two retained boundary positions
$p<q$ will be denoted $V=(p,q)$.  The letters $p,q$ refer to \emph{slots
or sentinels}, while $V$ is their open job set.  In Step~3 the symbols
$a,b$ are reused after reversing the frame; they then denote the
\emph{new} left and right boundaries, respectively.

The construction has four operations.  First it determines a
normalized schedule and a center slot $v$.  Next it creates the gaps
before $v$.  It then processes the opposite side in the reversed frame.
Finally it gives every nonempty gap a new list of global upsets.  The
cutoffs used to describe the gaps are \emph{witness data}; no cutoff
or normalized schedule has to be computed by the scheduling algorithm
in Sections~2--6.

\paragraph{Step 1: normalize the whole parent and choose a center.}
\emph{Purpose.}  We want separator positions whose associated cutoffs
have a particularly simple description in terms of the old list.  To
that end, arrange the jobs of $W$ in the following three \emph{priority
blocks} (a block is a consecutive portion of a priority list):
\begin{equation}\label{str:priorityblocks}
 \underbrace{(W\setminus O)\cap P_b}_{\text{first block}},\qquad
 \underbrace{O}_{\text{second block}},\qquad
 \underbrace{W\setminus P_b}_{\text{third block}}.
\end{equation}
Empty blocks are permitted.  Within the first and third blocks use
increasing global rank; within the second block use increasing old
qualifier key $\operatorname{key}_{\mathcal K}$.

\emph{Why is this a precedence extension?}  By (I1), $O\subseteq P_b$,
so the blocks are disjoint and cover exactly $W$.  Since $Q$ is a
global upset, $W\setminus O$ is an ideal of the order induced on $W$.
The predecessor cone $P_b$ is a global ideal.  Their intersection, the
first block, is therefore an ideal of $W$.  The third block is the
complement in $W$ of the ideal $W\cap P_b$, and hence an upset of $W$.
No precedence can consequently point from a later block to an earlier
one.  Increasing rank respects precedence in the first and third
blocks; the qualifier key order respects precedence in the second.
This proves that the concatenation in~\eqref{str:priorityblocks} is a
priority order on all of $W$.

Replace the current schedule of $W$, within its existing temporal
positions, by a feasible full schedule \emph{normalized} for this
priority.  Lemma~\ref{str:replacement} justifies this replacement
without moving any job outside $W$.  Normalization is chosen only for
the existence proof.

The initial low cutoff is the first priority block,
\[
 S_0=(W\setminus O)\cap P_b.
\]
Every job of $S_0$ is a nonqualifier, so invariant~(I1) implies
$S_0\subseteq D_a$. Hence the exterior left boundary $a$ separates
$S_0$ and serves as the starting position of the separator walk.

\paragraph{Retained separator positions and marked slots.}
We distinguish a \emph{separator position} from a \emph{marked slot}.
Initially, the boundary $a$ is the only \emph{retained separator
position}. It is an exterior boundary, not a marked slot of $W$.

Promote the jobs of $O$ one by one in increasing old key order.
Suppose that the next promoted job is $j$, and write $S$ for the
low cutoff immediately before its promotion. The new cutoff is
$S^+=S\cup\{j\}$. Starting at the current retained separator
position $p$, apply the walk of Lemma~\ref{str:walk} until it stops
at a position $q\ge p$. We call $q$ the \emph{final separator
position} of this promotion. The common-cutoff conclusion of
Lemma~\ref{str:walk} states that both $p$ and $q$ separate $S$,
and that $q$ also separates $S^+$.

If $q>p$, retain $q$ as the new separator position. Because $q$
is an actual full slot of $W$, we call that slot a \emph{marked
slot}, or simply a \emph{mark}. If $q=p$, update the cutoff to
$S^+$ but retain no additional position and create no new mark.
A single promotion may visit several slots before its walk stops;
these intermediate positions are \emph{not} marked. Thus only
distinct final separator positions of completed promotion walks
become marks.

After each promotion the low cutoff is $S_0$ together with the
jobs of $O$ promoted so far. Since the separator walks never
move backward in time, the retained positions are strictly
increasing. If a promotion advances from a retained position
$p$ to a new retained position $q$, the open interval $(p,q)$
is assigned the cutoff $S$ \emph{before} that promotion.
Both endpoints separate this common cutoff by
Lemma~\ref{str:walk}; this fact is used in Step~2 and
Lemma~\ref{str:labels}.

After all jobs of $O$ have been promoted, the \emph{exhausted cutoff}
is
\[
 S_{\mathrm{end}}=S_0\cup O=W\cap P_b.
\]
The term \emph{exhausted} refers only to the second block: the jobs of
$W\setminus P_b$ have not been promoted.  If the walk has reached an
actual position, call its last distinct actual endpoint $v$; repeated
promotions preserve the separator property, so $v$ separates
$S_{\mathrm{end}}$.  If all positions stayed at the exterior boundary
$a$, use Corollary~\ref{str:actual} to select an actual separator $v$
for $S_{\mathrm{end}}$.  This slot is the \emph{center}.  At the root,
$b$ is the right sentinel, $P_b=J$, and the last actual slot may be
chosen as $v$, leaving no opposite side to process.

\paragraph{Step 2: form the same-direction gaps.}
\emph{Purpose.}  The marks from Step~1 partition the portion of $W$
preceding $v$ into open gaps.  The crucial point is that the two
endpoints of each gap have a \emph{common} low cutoff.

List the distinct retained positions in increasing time order,
including the exterior starting boundary and the center:
\[
 a=p_0<p_1<\cdots<p_r=v.
\]
For consecutive positions $p<q$ put
\[
 V=(p,q)=\{x\in J:
     \operatorname{pos}(p)<\tau_f(x)<\operatorname{pos}(q)\}.
\]
Here $\tau_f$ is the slot number in the current frame, defined in
Section~7.1.  An empty gap has no child; a nonempty gap becomes a
\emph{same-direction child}, inheriting this frame.  Both boundary
slots themselves are excluded from $V$.

If the retained position moved from $p$ to $q$ during a promotion,
associate with $(p,q)$ the cutoff $S$ \emph{immediately before} that
promotion.  The shared-cutoff conclusion of Lemma~\ref{str:walk}
gives
\[
 p\text{ separates }S,\qquad q\text{ separates }S.
\]
Promotions which repeated at $p$ have already changed the cutoff, and
therefore must be included in this pre-promotion value of $S$.  For a
center appended after an entirely exterior walk, use
$S=S_{\mathrm{end}}$.  Every associated same-direction cutoff is
contained in $W\cap P_b$.

It is useful to deduce the two localization inclusions already here.
If $x\in V\cap S$, then $x$ is low and occurs after $p$; the
later-low separator condition at $p$ puts $x$ in $D_p$, and the
cutoff inclusion puts it in $P_b$.  If $x\in V\setminus S$, then it
is high and occurs before $q$; the earlier-high separator condition
at $q$ puts it in $P_q$.  Therefore
\[
 V\cap S\subseteq D_p\cap P_b,
 \qquad V\setminus S\subseteq P_q.
\]
These inclusions are the first assertion of Lemma~\ref{str:labels}.

\paragraph{Step 3: process the other side in reversed time.}
\emph{Purpose.}  Rather than prove a second separator-walk lemma
for right-to-left processing, apply the same lemma after reversing
time, precedence, and rank simultaneously.

The part of $W$ following the center in the current frame is processed
in the \emph{reversed frame} of Section~7.1.  In that frame, old
successor cones become predecessor cones, and the old right boundary
becomes the new left boundary.  From now until the end of the
opposite-side construction, rename the new boundaries $a,b$ and
interpret all cones, keys, ranks, and times in this reversed frame.
The job set $W$ and the actual center slot $v$ remain the same.

In the old frame $v$ separated $W\cap P_b^{\mathrm{old}}$.
Reversal of the separator property, Lemma~\ref{str:prefix}, says that
in the new frame $v$ separates the complement of this cutoff in $W$.
Since $P_b^{\mathrm{old}}$ becomes $D_a$, that complement is
\[
 X:=W\setminus D_a.
\]
The set $X$ is an ideal of $W$, because $D_a$ is an upset.  The
reversed form of the cone-cover invariant (I1) implies
$W\subseteq D_a\cup P_b$; hence $X\subseteq P_b$.

Let
\[
 E=(a,v)
\]
be the open side before $v$ in the \emph{new} frame (the side that
was after $v$ originally).  If $E=\varnothing$, nothing more is
required on this side.  Otherwise consider on $W$ the two-block
priority order
\[
 X,\qquad W\setminus X=W\cap D_a,
\]
with increasing \emph{new-frame} rank inside each block.  Since $X$
is an ideal, this is a precedence extension.  Normalize only the
contiguous interval $E$ for the restriction of this priority order,
keeping all other slots, including $v$, fixed.  Lemma~\ref{str:replacement}
ensures feasibility; in particular, this reordering cannot disturb
the marked slots already chosen before $v$ in the old frame.

Start at the new left exterior boundary $a$ with empty low set on $E$.
Promote the jobs of $X$ by increasing new-frame rank.  For a rank
threshold $h$, the corresponding global-on-$W$ low cutoff is
\begin{equation}\label{str:oppositecutoff}
 S_h=X\cap\{x\in J:\operatorname{rank}(x)<h\},
 \qquad 1\le h\le N+1.
\end{equation}
The low cutoff actually used by the walk on $E$ is $S_h\cap E$.
Promoting a job of $X\setminus E$ changes no local cutoff and leaves
the separator unchanged.  For a moving promotion retain only the
final actual endpoint and attach the pre-promotion cutoff $S_h$ to
its gap.  The endpoint-separation statements of these moves are on
$E$, not automatically on all of $W$.

After the final promotion set $h=N+1$ and hence $S_h=X$.  Add the
terminal open gap from the last retained position (or $a$ if none
exists) to $v$, with cutoff $X$.
    The left endpoint separates $X\cap E$ on $E$.
 The center separates $X$ on all of $W$, so its separator inequalities
apply to jobs of $E$.
 Every opposite-side gap therefore has a cutoff
$S=S_h\subseteq X\subseteq P_b$ with the required endpoint inequalities
on the processed domain.
 For the terminal right endpoint $v$, these inequalities come directly
from the center's separation on $W$.

\paragraph{The historical predicate required by Step 4.}
One more object is needed before the new lists can be stated.  For a
nonroot node $W$, its \emph{incoming edge} is the already constructed
parent-to-child link by which the gap $W$ was created.  It records
whether the link was same-direction or opposite-side, its associated
low cutoff in the historical parent, and a \emph{global ideal
extension} $H_*$ of that cutoff in the frame of that incoming edge.
An extension of a low set $S\subseteq U$ means precisely
$H_*\cap U=S$, where $U$ is the historical parent.  It need not
coincide with $S$ outside $U$.

When Step~3 reverses the frame, this already recorded global ideal
becomes a global upset on the \emph{same} set of jobs.  Write $u,d$
for the historical parent boundaries in the new frame, and write
$c,b$ for the current node boundaries in that frame, so $c=a$ in
the notation of Step~3.  If the historical incoming edge was
opposite-side, also denote its historical center by $w$.
Define the new global predicate by
\[
 G= D_u\cap(J\setminus\widehat P_c)\cap H_*
 \quad\text{(same-direction incoming edge)},
\]
and
\[
 G=D_u\cap(J\setminus\widehat P_c)
       \cap(J\setminus\widehat P_w)\cap H_*
 \quad\text{(opposite-side incoming edge)}.
\]
Lemma~\ref{str:history}, proved below using the \emph{already
constructed incoming edge}, establishes
\[
 W\setminus D_c\subseteq G,
 \qquad
 G\cap\{x\in J:\tau_f(x)\le\operatorname{pos}(c)\}=\varnothing.
\]
These are not additional properties assumed of an unknown child.
Their use in Lemma~\ref{str:labels} is part of an induction over
hierarchy depth: the incoming edge is available before the outgoing
opposite-side children are defined.  The root needs no such predicate,
because its opposite side is empty.

\paragraph{Step 4: assign a list to each nonempty child.}
\emph{Purpose.}  For a gap $V=(p,q)$ with associated low cutoff
$S\subseteq W$, we need a new ordered list $\mathcal L_V$ of global
upsets whose qualifier set inside $V$ is exactly its high part
$V\setminus S$.  We first establish the stronger identity on the
parent $W$, then prune the list on $V$.

\paragraph{Same-direction children.}
The cutoff $S$ contains the entire first priority block from
\eqref{str:priorityblocks} and an initial segment of the qualifiers
$O$ in their old key order.  The unpromoted qualifiers $O\setminus S$
therefore form a suffix of that order.  If this suffix is nonempty,
let $(d,r)$ be its least old key and take
\[
 C=\operatorname{carry}_{d,r}(\mathcal K)
   =(K_d\cap R_r,K_{d+1},\ldots,K_\ell).
\]
Otherwise take the empty carry $C=()$.  By Lemma~\ref{str:carry},
$W\cap Q(C)=O\setminus S$.  Recall that $C$ is a list of \emph{global}
upsets and may contain jobs outside $W$.

The remaining high parent jobs are the third priority block
$W\setminus P_b$.  They are selected by the \emph{fresh injection}
\[
 I(a,b):=D_a\setminus P_b,
 \qquad W\cap I(a,b)=W\setminus P_b,
\]
where the equality follows from (I1).  Append $I(a,b)$ as the
\emph{last entry} of the carry, writing
\[
 \mathcal L_V=(C,I(a,b)).
\]
This denotes list concatenation, not a two-element set whose first
element is $C$.  The qualifier set on $W$ is now
\[
\begin{aligned}
 W\cap Q(\mathcal L_V)
  &=(O\setminus S)\cup(W\setminus P_b)\\
  &=W\setminus S.
\end{aligned}
\]
The union on the first line is disjoint because $O\subseteq P_b$.
The carry preserves the relative key order of the unpromoted old qualifiers, while
placing the injection last gives every job of the third priority
block a later largest membership index.  The resulting key order on
$W\setminus S$ agrees with the parent priority order in Step~1.

\paragraph{Opposite-side children.}
For a nonempty gap $V=(p,q)$ produced in Step~3, let $S=S_h$
be its associated rank cutoff ($h=N+1$ for the final gap).
Form the unpruned list
\[
                       \mathcal L_V=(G\cap R_h,D_a)
\]
in the reversed frame, and prune it on $V$.  The historical
predicate $G$ has the coverage $W\setminus D_a\subseteq G$;
its construction and proof are deferred to
Lemma~\ref{str:history}.  Lemma~\ref{str:labels} now verifies
the child membership and cone properties.

Unlike the same-direction update, this operation discards the old
list and introduces a new initial group of two entries.  Both entries
are global upsets.  Recall that $X=W\setminus D_a\subseteq G$.
On $X$ the companion $D_a$ is absent, and the first entry selects
exactly
\[
 X\cap R_h=X\setminus S_h.
\]
On $W\cap D_a$ the second entry selects every job, whether or not it
also belongs to the first entry.  Hence
\[
 W\cap Q(\mathcal L_V)
   =(X\setminus S_h)\cup(W\cap D_a)=W\setminus S_h.
\]
Jobs selected only through the first entry have largest index~1;
jobs in the second entry have largest index~2, even on an overlap.
Rank breaks ties in each part, reproducing the two-block priority of
Step~3.  The \emph{past-exclusion} property of $G$, which is not
needed for this membership calculation, will be used for (I2).

\paragraph{Pruning in either case.}
Finally delete from $\mathcal L_V$ every entry with empty
intersection with $V$, and retain each surviving entry as its
original global set; denote the resulting list by
$\operatorname{prune}_V(\mathcal L_V)$.  Such pruning does not remove
membership of any job of $V$, nor change the relative key order of
the retained child qualifiers by Lemma~\ref{str:carry}.  In
particular,
\[
 V\cap Q(\operatorname{prune}_V(\mathcal L_V))=V\setminus S.
\]
The next lemma formalizes the cone and priority properties obtained
from these four steps.

\begin{lemma}[Child cone properties and list labels]\label{str:labels}
Every nonempty child $V=(p,q)$ has an actual right endpoint in its
own frame and an associated cutoff $S\subseteq W\cap P_b$ such that
\[
 V\cap S\subseteq D_p\cap P_b,\qquad
 V\setminus S\subseteq P_q.
\]
For the unpruned list $\mathcal L_V$ assigned in Step~4,
\[
 W\cap Q(\mathcal L_V)=W\setminus S.
\]
Its key order on $W\setminus S$ agrees with the priority order
specified in Step~1 or Step~3, respectively (in Step~3 the
normalization uses its restriction to $E$).  Pruning preserves the
qualifier set and relative key order inside $V$.  In particular, the
child satisfies (I1).
\end{lemma}

\noindent\emph{Proof idea.}
A shared pair of separators places low jobs of the gap in a successor
cone of its left endpoint, and high jobs in a predecessor cone of its
right endpoint.  The carry/injection and switch-pair identities say
that the new qualifiers are exactly the high jobs.  Pruning leaves
all memberships of child jobs unchanged.
\begin{proof}
\emph{Endpoints and cones.}
For a same-direction gap the right endpoint $q$ is a retained actual
position of Step~1 or its actual center.  For an opposite-side gap it
is a retained actual position of Step~3 or that same center.  Thus
$q$ is actual in the child's frame.  Take $x\in V\cap S$.  Since
$x$ occurs strictly after $p$ and $p$ separates $S$ on the relevant
walk domain, $x\in D_p$.  Furthermore, every associated cutoff was
chosen inside $W\cap P_b$, so $x\in P_b$.  This proves the first
inclusion.  If $x\in V\setminus S$, then $x$ is high and occurs
strictly before $q$.  Separation at $q$ gives $x\in P_q$.
  On an opposite-side gap with $q<v$, these conditions apply on $E$.
 For the terminal gap $q=v$, use the center's separation of $X$ on $W$
directly.
 This suffices because $V\subseteq E\subseteq W$.

\emph{Same-direction list.}
By Lemma~\ref{str:carry}, the suffix carry selects $O\setminus S$
inside $W$ and preserves their relative old keys.  By (I1), the
injection selects $W\setminus P_b$.  Since $O\subseteq P_b$ and
$S$ contains the entire first block,
\[
 W\setminus S=(O\setminus S)\mathbin{\dot\cup}(W\setminus P_b).
\]
Consequently the new qualifier union is exactly $W\setminus S$.
Appending the injection last places the third Step~1 priority block
last and uses rank within it; the carry supplies the correct earlier
key order.

\emph{Opposite-side list.}
Use $X=W\setminus D_a\subseteq G$ and $S=S_h$.
The first switched entry selects $X\setminus S_h$ on $X$, while the
second selects $W\cap D_a$.  Their union is $W\setminus S_h$.
Their respective largest indices are 1 and 2, with rank ordering
within each block.  This is the Step~3 priority on high jobs of the
parent cutoff.

\emph{Pruning and (I1).}
Pruning deletes no entry containing a job of $V$.  Thus it preserves
$V\cap Q(\mathcal L_V)=V\setminus S$, and Lemma~\ref{str:carry}
preserves the relative keys of these child jobs.  Together with the
cone inclusions proved above,
\[
\begin{aligned}
 V\cap Q(\operatorname{prune}_V(\mathcal L_V))
   &=V\setminus S\subseteq P_q,\\
 V\setminus Q(\operatorname{prune}_V(\mathcal L_V))
   &=V\cap S\subseteq D_p.
\end{aligned}
\]
These are exactly the two clauses of child (I1), using $p,q$ as
its new boundaries.  The global past condition (I2) and the
exchange condition (I3) are proved below.
\end{proof}

\paragraph{Example: a promotion skips an unmarked slot.}
We illustrate how the parent qualifiers, the low cutoff, the retained
separator positions, and the list assigned to a child fit together.
The important point is that a single promotion can move the separator
past an intermediate slot without marking that slot.

\emph{The instance and the parent list.}
Let $m=3$ and consider the nine jobs
\[
 W=\{a,b,c,d,e,x,j,h,k\},
\]
with generating precedence relations
\[
 a\prec x,\qquad b\prec d,\qquad c\prec e,\qquad j\prec h,
\]
and no other relations among these jobs.  Add three further jobs
$\alpha,\beta,\gamma$ that are independent of every job, and write
$A=\{\alpha,\beta,\gamma\}$ for the full exterior left boundary.
The global job set is $J=A\mathbin{\dot\cup}W$; the right boundary
is the sentinel $\top$.  Choose the global topological rank order
\[
 \alpha,\beta,\gamma,\ a,b,c,d,e,x,j,h,k.
\]
The complete witness schedule is
\[
 A,\qquad
 Z_1=\{a,b,c\},\qquad
 Z_2=\{d,e,j\},\qquad
 Z_3=\{x,h,k\}.
\]
Thus the current node is the open interval $W=(A,\top)$.
The three boundary jobs are outside $W$, even though the job
$a\in W$ has a similar name to the boundary $A$.

Take the inherited \emph{one-entry list}
\[
 \mathcal K=(W),\qquad K_1=W.
\]
The set $W$ is a global upset because all successors of its jobs
belong to $W$; the exterior jobs in $A$ are independent.  Consequently
\[
 Q(\mathcal K)=W,\qquad
 O:=W\cap Q(\mathcal K)=W.
\]
In particular, \emph{all nine jobs are parent qualifiers}.  Each has
key $(1,\operatorname{rank}(z))$, so the old qualifier order is
\[
 a,b,c,d,e,x,j,h,k.
\]
Since $P_{\top}=J$, the three Step~1 priority blocks are
\[
 \underbrace{(W\setminus O)\cap P_{\top}}_{\varnothing},
 \qquad \underbrace{O}_{W},\qquad
 \underbrace{W\setminus P_{\top}}_{\varnothing}.
\]
Thus the initial low cutoff is $S_0=\varnothing$, and all promotions
are taken from $O$ in the displayed key order.  The parent invariant
(I1) holds because $O=W\subseteq P_{\top}$.  Parent (I2) holds
because no qualifier occurs at or before $A$, and parent (I3) is
vacuous because no job of $A$ qualifies.

\emph{Why the displayed schedule is normalized.}
The first three priority jobs $a,b,c$ are incomparable and admit a
feasible completion, so the first slot of a normalized schedule is
$Z_1=\{a,b,c\}$.  Among the remaining jobs, the earliest-priority
triple would be $\{d,e,x\}$.  Choosing it as the second slot would
leave $\{j,h,k\}$ for the last slot, which is infeasible since
$j\prec h$.  The next possible triple in lexicographic priority
order is $\{d,e,j\}$, which admits the final slot
$\{x,h,k\}$.  Hence the displayed schedule is normalized.
Notice especially that $x$ has better priority than $j$ but is
scheduled later: the precedence $j\prec h$ prevents replacing $j$
in the second slot by $x$.

\emph{The separator walk and its marks.}
The successor cones relevant to the walk are
\[
 D_A=\varnothing,\qquad D_{Z_1}=\{d,e,x\},\qquad
 D_{Z_2}=\{h\},\qquad D_{Z_3}=\varnothing.
\]
Initially the separator is the exterior boundary $A$ and the low
cutoff is empty.  Promoting $a$ makes $a$ a low job scheduled after
$A$ but outside $D_A$; the earliest low exception is in $Z_1$.
The separator therefore moves from $A$ to $Z_1$.  Since $Z_1$ is
the \emph{final} endpoint of this promotion, it becomes the first
marked slot.  Promotions of $b,c$ do not move the separator because
both jobs already lie in $Z_1$.  Promotions of $d,e,x$ also leave the
separator at $Z_1$, since these three jobs belong to $D_{Z_1}$.

Immediately before promoting $j$, the low cutoff and current
retained separator are
\[
 S=\{a,b,c,d,e,x\},\qquad p=Z_1.
\]
Promoting $j$ changes the cutoff to $S^+=S\cup\{j\}$.
Since $j\in Z_2$ but $j\notin D_{Z_1}$, the walk first advances
from $Z_1$ to $Z_2$.  It cannot stop there: the previously promoted
job $x$ occurs later, in $Z_3$, but
\[
 x\notin D_{Z_2}=\{h\}.
\]
Thus $x$ is a new low exception at $Z_2$, and the walk must advance
a second time, from $Z_2$ to $Z_3$.  No job of $W$ occurs after
$Z_3$, so there is no later-low exception at $Z_3$.  The high jobs
for $S^+$ are $h,k$, both in $Z_3$; hence the earlier-high
condition also holds.  Therefore the complete walk is
\[
 Z_1\ \longrightarrow\ Z_2\ \longrightarrow\ Z_3.
\]
Only $Z_3$ is the \emph{final separator position} of this promotion
and hence a new mark; $Z_2$ is merely an intermediate position and
is not marked.

It is important that the new endpoint $Z_3$ also separates the
\emph{old} cutoff $S$.  Indeed, the high jobs for $S$ are
$\{j,h,k\}$, and the only one strictly before $Z_3$ is $j$.
Since $j\prec h$ and $h\in Z_3$, we have $j\in P_{Z_3}$.
The later-low property at $Z_3$ is automatic.  Thus both $Z_1$
and $Z_3$ separate the same cutoff $S$, as required by
Lemma~\ref{str:walk}.  The later promotions of $h$ and $k$ leave
the separator at $Z_3$.  Hence the distinct actual marks are
$Z_1,Z_3$, and the center is $v=Z_3$.

\emph{The child and its associated cutoff.}
The two retained marks enclose the nonempty same-direction gap
\[
 V=(Z_1,Z_3)=Z_2=\{d,e,j\}.
\]
The associated cutoff is the one \emph{before} the promotion
that moved the retained separator from $Z_1$ to $Z_3$, namely
\[
 S=\{a,b,c,d,e,x\}.
\]
It is \emph{not} the cutoff $S^+$ after promoting $j$.
Consequently the low and high parts of the child are respectively
\[
 V\cap S=\{d,e\},\qquad V\setminus S=\{j\}.
\]
In particular, $j$ is high relative to the child's associated
cutoff even though its promotion caused the separator advance.

\emph{The carry, the injection, and pruning.}
Recall that the \emph{parent} qualifier set is $O=W$.  At the
associated cutoff, the unpromoted parent qualifiers are
\[
 O\setminus S=\{j,h,k\}.
\]
Their least old key is $(1,\operatorname{rank}(j))=(1,10)$.
Since $\mathcal K=(K_1)$ with $K_1=W$, the suffix carry of
Step~4 is
\[
 C=\operatorname{carry}_{1,10}(\mathcal K)
   =(W\cap R_{10})=(\{j,h,k\}).
\]
The fresh injection is
\[
 I(A,\top)=D_A\setminus P_{\top}=\varnothing,
\]
so the unpruned list is
$\mathcal L_V=(\{j,h,k\},\varnothing)$.
In particular, the parent-wide identity from Lemma~\ref{str:labels}
is visible directly:
\[
 W\cap Q(\mathcal L_V)=\{j,h,k\}=W\setminus S.
\]
Pruning on $V$ deletes the empty second entry but retains
\emph{the complete global first entry} because
$\{j,h,k\}\cap V=\{j\}\ne\varnothing$.  Thus
\[
 \mathcal K_V=\operatorname{prune}_V(\mathcal L_V)
   =(\{j,h,k\}),\qquad
 O_V:=V\cap Q(\mathcal K_V)=\{j\}.
\]
The distinction between $O=W$, the nine \emph{parent}
qualifiers, and $O_V=\{j\}$, the only \emph{child} qualifier,
is crucial.  The jobs $h,k$ remain in the global list entry even
though they belong to the marked slot $Z_3$ outside $V$.
Pruning deletes whole entries; it does not restrict retained
global upsets to the child domain.

\emph{Verification of the child cone invariant.}
The left and right boundaries of the child are $p=Z_1$ and
$q=Z_3$.  We have
\[
 V\setminus S=\{j\}\subseteq P_{Z_3}
 \quad\text{because }j\prec h,\ h\in Z_3,
\]
and
\[
 V\cap S=\{d,e\}\subseteq D_{Z_1}
 \quad\text{because }b\prec d,\ c\prec e.
\]
Since $P_{\top}=J$, the additional inclusion
$V\cap S\subseteq P_{\top}$ is automatic.  Combining the
cone relations with $O_V=V\setminus S$ proves
\[
 V\cap Q(\mathcal K_V)=\{j\}\subseteq P_{Z_3},\qquad
 V\setminus Q(\mathcal K_V)=\{d,e\}\subseteq D_{Z_1},
\]
which is exactly invariant~(I1) for the child.  In this example
(I2) and (I3) also hold trivially: no global qualifier of the
child occurs at or before $Z_1$, and no job in $Z_1$ qualifies.

\paragraph{Feasibility and strict progress.}
The final marked slots and their intervening open gaps partition
$W$ in chronological order.  Every nonempty gap is a union of full
slots of the feasible witness schedule and is therefore feasible.
The opposite-side normalization changes only slots beyond the center
in the original frame; it does not move jobs across an already
selected same-direction mark, so its separator statements remain
valid.  Every child excludes the full center slot of $m$ jobs and
therefore contains at most $|W|-m$ jobs.  This is the size decrease
used in the later induction.

\subsection{Shortening stored information before a reversal}

\paragraph{Purpose and terminology.}
At a node $U$, the construction works with an \emph{exact ordered
list} $\mathcal K_U$ of global upsets.  The list determines the
qualifiers and their priority keys, including the memberships of
jobs outside $U$.  In Step~4 of Section~7.3 a same-direction child
inherits a suffix of this list, appends one \emph{injection}, and
prunes entries that do not meet the child.  An opposite-side child
instead replaces the list by a new \emph{switch pair}.  Hence a
switch pair may be inherited through several successive
same-direction steps, together with injections introduced later.
A subsequent switch replaces the entire inherited list.

There are two separate issues.  First, the exact entries inherited
from ancestors can have complicated formulas.  We shall describe
their \emph{union on the current descendant} by a simpler, or
\emph{flat}, expression (Lemma~\ref{str:flat}).  Second, when a later
switch reverses the frame, the construction needs a new \emph{global}
upset that also controls jobs outside the current node.  This is
obtained from a recorded historical cutoff (Lemma~\ref{str:history}).
The local replacement in the first step is not a global replacement
of the exact list.

\paragraph{Birth groups.}
Entries are grouped according to the operation that introduced
them.  The root's entry $J$ is one initial group.  A switch
introduces the ordered pair $(G_0\cap R_h,D_c)$ as one initial
group; its two entries may overlap.  Each subsequent injection
$I(A,B)=D_A\setminus P_B$ forms a separate group.  The \emph{birth
domain} of a group is the node on which it was introduced.  On a
chain of same-direction descendants, the exact list consists of
surviving entries of the initial group, followed by surviving
injections in their birth order.  A suffix carry restricts only
its first retained entry by a rank suffix; pruning deletes entire
entries but never replaces a retained global entry by its
intersection with a node.  In particular, the first entry of a
switch pair cannot be kept by a suffix carry while its later
companion is discarded by that same suffix operation.  The
companion may nevertheless have been deleted by pruning.

\paragraph{How an inherited switch pair occurs.}
For example, suppose a switch on $V_0$ creates the list
$(G_0\cap R_{h_0},D_c)$.  At a same-direction descendant $V_1$,
its suffix carry may retain both entries, possibly increasing the
first threshold to $t_1$, and append an injection $I_1$:
\[
 \mathcal K_{V_1}=(G_0\cap R_{t_1},D_c,I_1).
\]
A further same-direction descendant $V_2$ may carry these
entries and add $I_2$.  The switch-group assertion below concerns
\emph{only} the first two entries, not the injections $I_1,I_2$.
At a new switch, the inherited list is discarded and a new initial
pair is created.  The displayed sequence is schematic: any
particular carry or pruning step may delete entries.

\begin{lemma}[Flat unions on descendants]\label{str:flat}
The following identities concern restrictions to descendants of
the respective birth domains.

\emph{(a) Injection.} Suppose $I(A,B)=D_A\setminus P_B$ was
introduced on a node $V$ satisfying the birth cone cover
$V\subseteq D_A\cup P_B$.  For every descendant $U\subseteq V$,
\begin{equation}\label{str:flat-injection}
 U\cap I(A,B)=U\cap(J\setminus P_B).
\end{equation}
If a later suffix carry rank-restricts this injection by $R_t$,
then the corresponding restrictions also agree after intersecting
both sides of \eqref{str:flat-injection} with $R_t$.

\emph{(b) Switch group.} Suppose a switch introduces the two
entries
\[
                 (G_0\cap R_h,\ D_c)
\]
on a birth domain $V$, and its historical coverage property is
\begin{equation}\label{str:switch-birth-cover}
                    V\setminus D_c\subseteq G_0.
\end{equation}
Let $U\subseteq V$ be a descendant reached through subsequent
same-direction child constructions.  Assume that the \emph{first}
switch entry is still retained, with accumulated outer rank
restriction $R_t$; thus the retained first entry has the exact
form $G_0\cap R_t$.  Then the union of the \emph{surviving entries
of this switch group alone}, restricted to $U$, equals
\begin{equation}\label{str:switch-flat}
 U\cap(D_c\cup R_t).
\end{equation}
This is true both when the companion $D_c$ is still present and
when it was deleted by pruning at an intermediate ancestor of $U$.
Injections introduced after the switch are \emph{not} included
in \eqref{str:switch-flat}; they must be added separately.
If the first switch entry has disappeared, assertion~(b) is not
applied to the switch group.
\end{lemma}

\paragraph{Proof idea.}
Within a birth domain, a job not in the right predecessor cone
of an injection is already forced into its left successor cone.
For a switch, every job outside $D_c$ is already in $G_0$;
its membership in the first entry is therefore determined only
by the current rank suffix.  The companion handles all jobs
inside $D_c$, when it is present.

\begin{proof}
\emph{(a).} The inclusion from left to right in
\eqref{str:flat-injection} follows directly from
$I(A,B)=D_A\setminus P_B$.  Conversely, take
$x\in U\setminus P_B$.  Since $U\subseteq V$ and
$V\subseteq D_A\cup P_B$, we must have $x\in D_A$.
Hence $x\in I(A,B)$.  Intersecting an equality of sets with
$R_t$ preserves it.

\emph{(b).} Since $U\subseteq V$, the birth cover implies
\[
                     U\setminus D_c\subseteq G_0.
\]
First suppose both switch entries are present.  Partition
$U$ into $U\cap D_c$ and $U\setminus D_c$.  The companion
selects all jobs of the first part.  Every job of the second
part already belongs to $G_0$, so the first entry selects it
exactly when it belongs to $R_t$.  Formally,
\begin{align*}
 &U\cap\bigl((G_0\cap R_t)\cup D_c\bigr)\\
 &\quad=(U\cap D_c)\cup
       \bigl((U\setminus D_c)\cap G_0\cap R_t\bigr)\\
 &\quad=(U\cap D_c)\cup
       \bigl((U\setminus D_c)\cap R_t\bigr)\\
 &\quad=U\cap(D_c\cup R_t).
\end{align*}
If $D_c$ was pruned at an intermediate node $U'$ with
$U\subseteq U'\subseteq V$, then $U'\cap D_c=\varnothing$,
and hence $U\cap D_c=\varnothing$.  The birth cover now yields
$U\subseteq G_0$, so
\[
 U\cap(G_0\cap R_t)=U\cap R_t=U\cap(D_c\cup R_t).
\]
This proves the same expression for the surviving first entry.
No assertion is made about the entire current list or about
global equality outside $U$.
\end{proof}

\paragraph{Flat expressions for complete suffix carries.}
We now make the replacement used below explicit.  Fix a current
node $U$ and a suffix carry $\mathcal C=(E_1,\ldots,E_s)$ of
its \emph{exact} inherited list.  In the application to a
same-direction gap, this carry represents the unpromoted
qualifiers at the associated cutoff, \emph{before} the fresh
injection for that gap is appended.  A \emph{flat expression for
$\mathcal C$ on $U$} means a global Boolean set $F\subseteq J$
such that
\begin{equation}\label{str:flat-definition}
                  U\cap F=U\cap Q(\mathcal C).
\end{equation}
It is sufficient, and useful later, to choose $F$ to be a
global upset.  Equality in \eqref{str:flat-definition} is
\emph{only} an equality after intersection with $U$.

All surviving injections in the carry are denoted by
$I(A_i,B_i)=D_{A_i}\setminus P_{B_i}$, in their retained order.
The symbol $R_t$ refers to the \emph{current outer rank
restriction on the first entry of the carry}, with $R_1=J$
when no nontrivial restriction is present.  In every case,
there is at most one such current restriction.  The following
formulas exhaust the possible surviving initial groups.

\smallskip
\noindent
\emph{Case 1: the first switch entry survives.}
The first switch-group entry is $G_0\cap R_t$; its companion
$D_c$ is either present or was pruned earlier.  It is followed
by $s$ surviving injections.  By Lemma~\ref{str:flat}, the
switch group's contribution on $U$ is $D_c\cup R_t$, and
each injection's contribution is $J\setminus P_{B_i}$.
Thus we may take
\begin{equation}\label{str:flat-case-switch}
 F=D_c\cup R_t\cup
                    \bigcup_{i=1}^{s}(J\setminus P_{B_i}).
\end{equation}
If the companion was pruned, then $U\cap D_c=\varnothing$,
so retaining $D_c$ in this expression does not alter its
restriction to $U$.  The exact switch-group entries remain
separate in $\mathcal C$: the flat union does not record their
individual membership indices.

\smallskip
\noindent
\emph{Case 2: only the switch companion survives.}
If the first switch entry has disappeared, the companion
$D_c$ is the first retained entry of this group.  Its current
rank-restricted form is $D_c\cap R_t$, followed by surviving
injections.  Here we do \emph{not} use the switch identity
\eqref{str:switch-flat}; instead, directly take
\begin{equation}\label{str:flat-case-companion}
 F=(D_c\cap R_t)\cup
                    \bigcup_{i=1}^{s}(J\setminus P_{B_i}).
\end{equation}
The intersection $D_c\cap R_t$ in this case must not be
confused with the union $D_c\cup R_t$ in Case~1.

\smallskip
\noindent
\emph{Case 3: the root entry survives.}
If the initial group is the root entry $J$ and it is still
retained, its current form is $J\cap R_t=R_t$.  Thus
\begin{equation}\label{str:flat-case-root}
 F=R_t\cup\bigcup_{i=1}^{s}(J\setminus P_{B_i}).
\end{equation}

\smallskip
\noindent
\emph{Case 4: the initial group has disappeared.}
The carry then begins with an injection.  If it contains
$s\geq1$ injections, the first one may have the outer
restriction $R_t$, while the others are untrimmed.  Thus
\begin{equation}\label{str:flat-case-injections}
 F=\bigl((J\setminus P_{B_1})\cap R_t\bigr)
          \cup\bigcup_{i=2}^{s}(J\setminus P_{B_i}).
\end{equation}
For the empty carry, define $F=\varnothing$.

\smallskip
All five formulas consist of unions and intersections of
global upsets.  Thus each $F$ is a global upset, and
Lemma~\ref{str:flat} proves \eqref{str:flat-definition}.
The exact list is nevertheless retained unchanged for
priority keys, pruning, and the boundary invariants.  In
particular, a flat expression must \emph{not} replace its
exact entries outside $U$.  In Lemma~\ref{str:groups} we will
bound the number of surviving birth groups; together with
the formulas above, this will bound the number of atomic
occurrences in $F$, independently of the hierarchy depth.

\paragraph{The historical cutoff and its global ideal extension.}
We next explain the information needed for a \emph{later}
reversal.  Let $U=(a_U,b_U)$ be an already constructed node,
and suppose one of its gaps has become a child $W$.  The edge
$U\to W$ is the \emph{incoming historical edge} of $W$.
Let $S\subseteq U$ be the low cutoff associated with this
edge, in the frame used when the edge was constructed.
The construction records a set $H_*\subseteq J$ satisfying
\begin{equation}\label{str:historical-extension-def}
 H_*\text{ is a global ideal},\qquad H_*\cap U=S.
\end{equation}
We call such a set a \emph{global ideal extension of $S$ from
$U$}.  Being an ideal here means predecessor closure in the
\emph{entire} poset $J$.  The equality means that $H_*$
selects exactly the cutoff jobs among the jobs of $U$; it
may contain other jobs outside $U$.  An extension is not
unique.  We need a particular one with a short global
Boolean formula.

\paragraph{An elementary example of an extension.}
Suppose $J$ consists of $m$ independent chains, each with
three successive jobs, and let $B_1,B_2,B_3$ be their three
full chronological slots.  For $U=B_2\cup B_3$ and the
local ideal $S=B_2$, the set $S$ is not a global ideal:
its jobs have predecessors in $B_1$.  But
$H_*=B_1\cup B_2$ is a global ideal satisfying
$H_*\cap U=S$.  This example illustrates the term
\emph{extension}; the following construction additionally
ensures that the extension has a short formula in the
permitted global atoms.

For clarity, we derive that extension from the cutoffs of
Step~1 instead of assuming the formula.  Write
\[
 O_U=U\cap Q(\mathcal K_U)
\]
for the qualifiers at $U$.  If the incoming edge was
same-direction, its cutoff $S$ contains the entire first
priority block $(U\setminus O_U)\cap P_{b_U}$ and an
initial segment of $O_U$.  The remaining, unpromoted
qualifiers are $O_U\setminus S$.  Let $\mathcal C$ be the
suffix carry of those qualifiers, \emph{before} appending
the fresh injection for the edge.  By the suffix-carry
assertion of Lemma~\ref{str:carry},
\begin{equation}\label{str:cutoff-carry}
              U\cap Q(\mathcal C)=O_U\setminus S.
\end{equation}
Indeed, the unpromoted qualifiers form precisely the final
segment of the old key order, and the carry retains that
segment; if none remain, $\mathcal C$ is empty.
Since $O_U\subseteq P_{b_U}$ by parent (I1), we have
\[
 U\cap P_{b_U}=
   \bigl((U\setminus O_U)\cap P_{b_U}\bigr)
                   \mathbin{\dot\cup}O_U.
\]
Removing the unpromoted qualifiers leaves exactly the
low cutoff.  Equivalently,
\begin{equation}\label{str:cutoff-exact}
 S=U\cap\bigl(P_{b_U}\setminus Q(\mathcal C)\bigr).
\end{equation}
Choose the flat expression $F$ for $\mathcal C$ on $U$
from one of the explicit cases above.  Then
$U\cap F=U\cap Q(\mathcal C)$, and a set calculation gives
\begin{align}
 S&=U\cap\bigl(P_{b_U}\setminus Q(\mathcal C)\bigr)
       \notag\\
  &=(U\cap P_{b_U})\setminus
                          (U\cap Q(\mathcal C))\notag\\
  &=(U\cap P_{b_U})\setminus(U\cap F)\notag\\
  &=U\cap(P_{b_U}\setminus F).
                                  \label{str:cutoff-flat}
\end{align}
The substitution uses equality \emph{only on $U$}; it
does not require $F=Q(\mathcal C)$ globally.  Hence the
specific choice $H_*=P_{b_U}\setminus F$ satisfies the
extension property \eqref{str:historical-extension-def}.
Because $P_{b_U}$ is an ideal and $F$ an upset,
$H_*=P_{b_U}\cap(J\setminus F)$ is a global ideal.

For an opposite-side historical edge, Step~3 already
provides the low cutoff as a rank prefix of
$X=U\setminus D_{a_U}$.  Its extension therefore has a
simple form without a carry.  The next lemma records both
cases and explains how the recorded extension is used
after reversal.

\begin{lemma}[Historical extension and switch localization]
\label{str:history}
Let $U=(a_U,b_U)$ be the historical parent of an already
constructed edge $U\to W$, and let $S\subseteq U$ be its
associated low cutoff.  In the frame in which this edge
was constructed, define
\begin{equation}\label{str:history-ext-same}
 H_*=P_{b_U}\setminus F
 \quad\text{if the edge was same-direction},
\end{equation}
where $F$ is the flat expression for the suffix carry before
the fresh injection, and define
\begin{equation}\label{str:history-ext-opp}
 H_*=(J\setminus D_{a_U})\cap
                      \{x\in J:\operatorname{rank}(x)<h\}
 \quad\text{if the edge was opposite-side},
\end{equation}
where $h$ is the rank cutoff recorded on that edge.
In both cases $H_*$ is a global ideal satisfying
$H_*\cap U=S$.

Now suppose that the construction is about to switch
while processing $W$.  Reverse time, precedence, and
rank, but regard $H_*$ as the \emph{same subset of $J$}.
Write $U=(u,d)$ and $W=(c,b)$ for the historical parent
and current node in the \emph{new} frame.  In the
opposite-side historical case, let $w$ be the previously
recorded historical center, with temporal positions
$u<w\leq c<b\leq d$ (inequalities of boundary positions).
Define
\begin{equation}\label{str:history-G-same}
 G=D_u\cap(J\setminus\widehat P_c)\cap H_*
 \quad\text{for a same-direction historical edge},
\end{equation}
and
\begin{equation}\label{str:history-G-opp}
 G=D_u\cap(J\setminus\widehat P_c)
        \cap(J\setminus\widehat P_w)\cap H_*
 \quad\text{for an opposite-side historical edge}.
\end{equation}
Then $G$ is a global upset and
\begin{equation}\label{str:history-properties}
 W\setminus D_c\subseteq G,
 \qquad
 G\cap\{x\in J:\tau_f(x)\leq\operatorname{pos}(c)\}
             =\varnothing.
\end{equation}
These properties depend only on the already constructed
incoming historical edge; they do not assume an invariant
of a future child of $W$.
\end{lemma}

\paragraph{What has to be proved.}
There are two groups of assertions, and they refer to different frames.
\begin{enumerate}[label=\textup{(\Alph*)}]
\item In the frame in which the incoming edge $U\to W$ was constructed,
the prescribed set $H_*$ is a global ideal and satisfies $H_*\cap U=S$.
\item In the reversed frame, the prescribed set $G$ is a global upset,
contains $W\setminus D_c$, and contains no job at or before $c$.
The last two assertions are exactly \eqref{str:history-properties}.
\end{enumerate}
Here $H_*$ is always the same subset of $J$. Reversing the frame changes
which closure property it has: an old-frame ideal becomes a new-frame
upset. The identity $H_*\cap U=S$ may be used for a job only after that
job has been shown to belong to $U$.

\paragraph{Proof idea.}
First construct the global ideal representing the historical low cutoff.
Next reverse the frame. The historical separator properties imply
$W\setminus D_c\subseteq S\subseteq D_u$, which gives the required
coverage. To exclude a possible member of $G$ at or before $c$, first
use its time and its membership in $D_u$ to place it inside $U$.
It then belongs to the historical low cutoff $S$.
The old later-low separator property forces it into a weak predecessor
cone excluded from $G$. For an opposite-side historical edge, two
temporal ranges must be distinguished, because the right endpoint
separates the cutoff only on the processed side; the historical center
handles the other range.

\begin{proof}
\textbf{Step 1: distinguish the two frames and the boundary names.}
Use superscripts $\mathrm{old}$ and $\mathrm{new}$ on cones, and the
corresponding subscripts on ranks and times. The old frame is the frame
in which the incoming edge was actually constructed. For an opposite-side
historical edge, this is already the frame used in Step~3 of the
historical parent's construction.

In the old frame write
\[
 U=(a_U,b_U),\qquad W=(p,q).
\]
After reversal, the same boundary slots receive the names
\[
 u=b_U,\qquad d=a_U,\qquad c=q,\qquad b=p.
\]
Consequently $U=(u,d)$ and $W=(c,b)$ in the new frame. The sets
$U,W,S,H_*$ themselves do not change. If the complete schedule has
$T$ slots, then
\[
 x\prec_{\mathrm{new}}y\ \Longleftrightarrow\
 y\prec_{\mathrm{old}}x,
 \qquad
 \tau_{\mathrm{new}}(x)=T+1-\tau_{\mathrm{old}}(x).
\]
Boundary positions transform in the same way, including sentinel
positions $0$ and $T+1$. In particular,
\[
 \begin{aligned}
 P_{b_U}^{\mathrm{old}}&=D_u^{\mathrm{new}},&
 P_q^{\mathrm{old}}&=D_c^{\mathrm{new}},&
 D_q^{\mathrm{old}}&=P_c^{\mathrm{new}}.
 \end{aligned}
\]
In the opposite-side historical case, write $v$ for the historical
center in the old frame and $w=v$ for that same slot in the new frame.
The boundary orders are
\[
 \begin{aligned}
 &a_U\leq p<q\leq v<b_U &&\text{in the old frame},\\
 &u<w\leq c<b\leq d &&\text{in the new frame}.
 \end{aligned}
\]
All inequalities here concern temporal positions. The case $q=v$,
equivalently $c=w$, is the terminal opposite-side gap.

For reference, the boundary correspondence is summarized below.
Each row refers to the same boundary slot before and after reversal.

\begin{center}
\small
\begin{tabular}{@{}lll@{}}
\toprule
Boundary slot & Old name & New name \\
\midrule
Old left / new right boundary of $U$ & $a_U$ & $d$ \\
Old right / new left boundary of $U$ & $b_U$ & $u$ \\
Old left / new right boundary of $W$ & $p$ & $b$ \\
Old right / new left boundary of $W$ & $q$ & $c$ \\
Historical center & $v$ & $w$ \\
\bottomrule
\end{tabular}
\end{center}

The last row applies only to opposite-side history.
In its terminal case, $q=v$ and therefore $c=w$.

\medskip
\textbf{Step 2: prove (A) for a same-direction historical edge.}
All cones and ranks in this step are interpreted in the old frame.
Let $O=U\cap Q(\mathcal K_U)$ be the historical parent's qualifiers.
The construction orders $U$ in three blocks:
\[
 (U\setminus O)\cap P_{b_U}^{\mathrm{old}},
 \qquad O,\qquad U\setminus P_{b_U}^{\mathrm{old}}.
\]
Its associated cutoff $S$ consists of the whole first block and an
initial segment of $O$. The remaining qualifiers $O\setminus S$ form
the suffix represented by the carry $\mathcal C$, before the fresh
injection is appended. Hence
\[
 U\cap Q(\mathcal C)=O\setminus S.
\]
The parent cone invariant gives $O\subseteq P_{b_U}^{\mathrm{old}}$.
Therefore
\[
 U\cap P_{b_U}^{\mathrm{old}}
 =\bigl((U\setminus O)\cap P_{b_U}^{\mathrm{old}}\bigr)
       \mathbin{\dot\cup}O.
\]
Deleting precisely $O\setminus S$ from this disjoint union leaves $S$.
The flat expression satisfies
$U\cap F=U\cap Q(\mathcal C)$, so
\[
 \begin{aligned}
 S
 &=(U\cap P_{b_U}^{\mathrm{old}})\setminus(O\setminus S)\\
 &=(U\cap P_{b_U}^{\mathrm{old}})
                              \setminus(U\cap Q(\mathcal C))\\
 &=(U\cap P_{b_U}^{\mathrm{old}})\setminus(U\cap F)\\
 &=U\cap(P_{b_U}^{\mathrm{old}}\setminus F)
 =U\cap H_*.
 \end{aligned}
\]
This calculation uses only equality on $U$; it does not require the
global equality $F=Q(\mathcal C)$.

It remains to prove global predecessor closure. The set
$P_{b_U}^{\mathrm{old}}$ is a global ideal and $F$ is a global upset.
Thus $J\setminus F$ is a global ideal, and
\[
 H_*=P_{b_U}^{\mathrm{old}}\cap(J\setminus F)
\]
is an intersection of global ideals. Explicitly, if $y\in H_*$ and
$x\prec_{\mathrm{old}}y$ with $x\in J$, then
$x\in P_{b_U}^{\mathrm{old}}$. If $x$ belonged to $F$, upward closure
would put $y$ in $F$, a contradiction. Hence $x\in H_*$.
This proves both assertions in (A) for this case.

\medskip
\textbf{Step 3: prove (A) for an opposite-side historical edge.}
In the old frame put
\[
 X=U\setminus D_{a_U}^{\mathrm{old}},\qquad
 L_h=\{x\in J:\operatorname{rank}_{\mathrm{old}}(x)<h\}.
\]
Step~3 of the construction records the cutoff $S=X\cap L_h$,
including $h=N+1$ for the terminal gap. The prescribed set is
\[
 H_*=(J\setminus D_{a_U}^{\mathrm{old}})\cap L_h.
\]
Intersecting with $U$ gives exactly
\[
 U\cap H_*=(U\setminus D_{a_U}^{\mathrm{old}})\cap L_h
           =X\cap L_h=S.
\]
The first factor of $H_*$ is a global ideal, because
$D_{a_U}^{\mathrm{old}}$ is a global upset. The rank prefix $L_h$ is
also a global ideal: if $y\in L_h$ and $x\prec_{\mathrm{old}}y$, then
\[
 \operatorname{rank}_{\mathrm{old}}(x)
 <\operatorname{rank}_{\mathrm{old}}(y)<h.
\]
Thus $x\in L_h$. The intersection defining $H_*$ is consequently a
global ideal. This completes (A) in both historical cases.

\medskip
\textbf{Step 4: write down exactly which historical separator facts
will be used.}
A slot $r$ separates a cutoff $L$ on a domain $V$ when every high job
of $V$ strictly before $r$ lies in $P_r$, and every low job of $V$
strictly after $r$ lies in $D_r$. The latter condition applies only
to strict inequality. A job at $r$ belongs to the slot itself.
Thus, whenever separation applies on a domain containing $z$,
\[
 z\in L,\quad \tau(z)\geq\operatorname{pos}(r)
 \quad\Longrightarrow\quad z\in D_r\cup r=\widehat D_r.
\]
The equality case is supplied by $z\in r$, not by the later-low
condition.

For a same-direction historical edge, Step~2 of the construction
records a cutoff satisfying
\[
 S\subseteq U\cap P_{b_U}^{\mathrm{old}},\qquad
 q\text{ separates }S\text{ on all of }U.
\]
For a moving promotion, this is the pre-promotion cutoff, and the
common-cutoff walk proves separation at its final endpoint $q$.
For an appended center, it is the exhausted cutoff, which the center
separates. In either case every job of $W\setminus S$ is a high job
before $q$ and belongs to $P_q^{\mathrm{old}}$.

For an opposite-side historical edge, the recorded data are
\[
 \begin{gathered}
 X=U\setminus D_{a_U}^{\mathrm{old}},\qquad
 E=(a_U,v)\text{ in the old frame},\\
 S\subseteq X\subseteq P_{b_U}^{\mathrm{old}},\qquad
 v\text{ separates }X\text{ on all of }U.
 \end{gathered}
\]
The inclusion $X\subseteq P_{b_U}^{\mathrm{old}}$ follows from the
historical parent's cone cover
$U\subseteq D_{a_U}^{\mathrm{old}}\cup P_{b_U}^{\mathrm{old}}$.
If $q<v$, the common-cutoff walk on $E$ gives
\[
 q\text{ separates }S\cap E\text{ on }E.
\]
If $q=v$, the gap is terminal, its cutoff is $S=X$, and the center
separates that cutoff on all of $U$.
Since $W\subseteq E$, the earlier-high property again yields
$W\setminus S\subseteq P_q^{\mathrm{old}}$ in either case.

The center's separation remains valid when $E$ is normalized:
all moved jobs remain strictly before $v$, so neither the sets
before and after $v$ nor the relevant cones change. Similarly,
later reordering within $W$ changes no job's side of its boundary
$q$ or of the historical center.

We have therefore obtained, in both historical cases,
\[
 S\subseteq P_{b_U}^{\mathrm{old}},\qquad
 W\setminus S\subseteq P_q^{\mathrm{old}}.
\]
Rewriting these same sets in the new frame gives
\[
 S\subseteq D_u^{\mathrm{new}},\qquad
 W\setminus S\subseteq D_c^{\mathrm{new}}.
\]
In particular, if $x\in W\setminus D_c^{\mathrm{new}}$, then
$x\notin S$ would contradict the second inclusion. Hence
\[
 {\quad W\setminus D_c^{\mathrm{new}}
                    \subseteq S\subseteq D_u^{\mathrm{new}}.\quad}
\]
These facts come directly from the already constructed edge and its
separators. They do not use the list of a future child.

\medskip
\textbf{Step 5: prove that $G$ is a global upset in the new frame.}
First, $H_*$ is a global upset in that frame. Indeed, if $x\in H_*$
and $x\prec_{\mathrm{new}}y$, then $y\prec_{\mathrm{old}}x$.
The old-frame ideal property proved in Steps~2--3 therefore gives
$y\in H_*$.

The sets $D_u^{\mathrm{new}}$ and
$J\setminus\widehat P_c^{\mathrm{new}}$ are global upsets as well:
a successor cone is upward closed, and the complement of a weak
predecessor cone is upward closed. In the opposite-side historical
case, $J\setminus\widehat P_w^{\mathrm{new}}$ has the same property.
Every factor in the relevant definition of $G$ is thus a global
upset. Their intersection is a global upset. This proves the first
assertion of (B).

For the remainder of the proof, cones without a superscript are in
the new frame.

\medskip
\textbf{Step 6: prove the first condition in
\eqref{str:history-properties}, namely $W\setminus D_c\subseteq G$.}
Take any $x\in W\setminus D_c$. We check each factor defining $G$.
\begin{enumerate}[label=\textup{(\roman*)}]
\item Step~4 gives $x\in S\subseteq D_u$.
\item Since $x\in S=H_*\cap U$, we have $x\in H_*$.
\item Since $x\in W=(c,b)$, it occurs strictly after $c$.
It is not a job in $c$. It cannot precede a job in $c$ either,
because the schedule is feasible. Hence $x\notin\widehat P_c$.
\item For an opposite-side historical edge, $w\leq c$.
Thus $x$ also occurs strictly after $w$, and the same reasoning
gives $x\notin\widehat P_w$.
\end{enumerate}
For a same-direction historical edge, items (i)--(iii) put $x$ in
every factor of \eqref{str:history-G-same}. For an opposite-side
historical edge, all four items put $x$ in every factor of
\eqref{str:history-G-opp}. Therefore $x\in G$, as required.

\medskip
\textbf{Step 7: begin the proof of the second condition by placing
a hypothetical past member inside $U$.}
Suppose, for a contradiction, that
\[
 z\in G,\qquad
 \tau_{\mathrm{new}}(z)\leq\operatorname{pos}_{\mathrm{new}}(c).
\]
The factor $D_u$ gives $z\in D_u$. If $u$ is actual, feasibility
then places $z$ strictly after $u$. If $u$ is the left sentinel,
every real job is strictly after it, so the same conclusion holds.
As $c<d$, we obtain
\[
 \operatorname{pos}_{\mathrm{new}}(u)
 <\tau_{\mathrm{new}}(z)
 \leq\operatorname{pos}_{\mathrm{new}}(c)
 <\operatorname{pos}_{\mathrm{new}}(d).
\]
Thus $z\in U=(u,d)$. Membership in $G$ also gives $z\in H_*$,
and only now do we apply the extension identity:
\[
                          z\in H_*\cap U=S.
\]
Membership in $D_u$ alone would not have been sufficient to place
$z$ inside $U$; the upper bound on its time is essential.

\medskip
\textbf{Step 8: exclude that past member for a same-direction
historical edge.}
Reversing the inequality
$\tau_{\mathrm{new}}(z)\leq\operatorname{pos}_{\mathrm{new}}(c)$,
where $c=q$, gives
\[
 \tau_{\mathrm{old}}(z)\geq\operatorname{pos}_{\mathrm{old}}(q).
\]
We already know $z\in S\subseteq U$, and $q$ separates $S$ on
all of $U$.
If equality holds, $z$ lies in the slot $q=c$, so
$z\in\widehat P_c$.
If the inequality is strict, the old later-low property at $q$
gives
\[
                   z\in D_q^{\mathrm{old}}=P_c^{\mathrm{new}}
                     \subseteq\widehat P_c.
\]
Both cases contradict the factor $J\setminus\widehat P_c$ in $G$.
This proves the second condition of \eqref{str:history-properties}
for same-direction history.

\medskip
\textbf{Step 9: exclude that past member for an opposite-side
historical edge.}
Here $u<w\leq c<b\leq d$. Split the possible time of $z$ into
the following two ranges.

\emph{First range: $\tau_{\mathrm{new}}(z)
\leq\operatorname{pos}_{\mathrm{new}}(w)$.}
In the old frame this says that $z$ occurs at or after the
historical center $v=w$:
\[
 \tau_{\mathrm{old}}(z)\geq\operatorname{pos}_{\mathrm{old}}(v).
\]
We know $z\in S\subseteq X$ and $z\in U$. The center separates
$X$ on all of $U$, so its later-low condition is applicable.
If $z$ occurs strictly after $v$, it gives
\[
                  z\in D_v^{\mathrm{old}}=P_w^{\mathrm{new}}.
\]
If $z$ occurs at $v$, then $z\in v=w$.
In either case $z\in\widehat P_w$, contradicting the factor
$J\setminus\widehat P_w$ in $G$.

\emph{Second range:
$\operatorname{pos}_{\mathrm{new}}(w)
<\tau_{\mathrm{new}}(z)
\leq\operatorname{pos}_{\mathrm{new}}(c)$.}
Reversing these inequalities gives
\[
 \operatorname{pos}_{\mathrm{old}}(q)
 \leq\tau_{\mathrm{old}}(z)
 <\operatorname{pos}_{\mathrm{old}}(v).
\]
In particular $q<v$. Also $a_U<q$, so
$z\in E=(a_U,v)$ in the old frame. We now have exactly the
domain membership needed for the local separator assertion:
$z\in S\cap E$, and $q$ separates $S\cap E$ on $E$.
If $z$ occurs strictly after $q$, the old later-low property yields
\[
                  z\in D_q^{\mathrm{old}}=P_c^{\mathrm{new}}.
\]
If it occurs at $q$, then $z\in q=c$.
Thus $z\in\widehat P_c$ in both cases, contradicting its exclusion
from $G$.

The two ranges cover every possible time at or before $c$.
When $w=c$, the second range is empty and the first range handles
the terminal gap. Therefore
\[
 G\cap\{z\in J:\tau_{\mathrm{new}}(z)
                 \leq\operatorname{pos}_{\mathrm{new}}(c)\}
 =\varnothing.
\]
This proves the second condition of \eqref{str:history-properties}
also for opposite-side history, and completes (B).

\medskip
\textbf{Step 10: verify the stated dependence on the incoming edge.}
The proof used the recorded cutoff, its carry or rank threshold,
the already established separators of $U\to W$, and feasibility of
the complete witness schedule. None of these facts assumes an
invariant of a future child of $W$.
Reordering jobs inside $W=(c,b)$ preserves $W$ as a set, fixes its
boundary slots, and changes no job at or before $c$. It also leaves
all precedence cones and the recorded sets unchanged. Consequently
both conditions in \eqref{str:history-properties} remain valid
under the subsequent allowed reordering of $W$.
All assertions of the lemma have now been proved.
\end{proof}

\paragraph{Exact global forms after the reversal.}
The flat expression is used to define the historical
ideal $H_*$, but the new switch entry is still stored
\emph{exactly} as $G\cap R_t$; it is not replaced by
a flat local predicate.  We record its two possible
forms, which are used in the formula count later.

On a same-direction historical edge,
$H_*=P_{b_U}\setminus F_{\mathrm{hist}}$ in the old frame.
After reversal the cone $P_{b_U}$ becomes $D_u$, and
$F_{\mathrm{hist}}^{\mathrm{rev}}$ means the \emph{same subset
of $J$} expressed in reversed-frame atoms.  Intersecting
the resulting switch predicate with its current outer
rank suffix gives
\begin{equation}\label{str:complicatedsame}
 \repairmath{G\cap R_t}=D_u\cap(J\setminus\widehat P_c)
       \cap(J\setminus F_{\mathrm{hist}}^{\mathrm{rev}})
       \cap R_t.
\end{equation}
The frozen expression $F_{\mathrm{hist}}^{\mathrm{rev}}$ may
itself contain an old rank condition.  This condition is
part of the recorded history and is \emph{not} merged
with the new outer suffix $R_t$.

On an opposite-side historical edge, the old
complement $J\setminus D_{a_U}$ becomes
$J\setminus P_d$ in the new frame.  The old prefix
$\{\operatorname{rank}_{\mathrm{old}}<h\}$ becomes a new
suffix $R_k$, where $k=N+2-h$.  Therefore
\begin{equation}\label{str:complicatedopposite}
 G=D_u\cap(J\setminus\widehat P_c)
       \cap(J\setminus\widehat P_w)
       \cap(J\setminus P_d)\cap R_k.
\end{equation}
An additional outer suffix $R_t$ may be merged with
$R_k$, because $R_k\cap R_t=R_{\max\{k,t\}}$.
This merge does not change the local flat identity:
outside the switch companion, the birth cover already
forces membership in the exact initial predicate.
These two global forms, together with the explicit
flat carry cases above, are the inputs to the
leaf-count argument in Lemma~\ref{str:budget}.

\subsection{Why the invariants hold at every child}

\paragraph{Induction order and terminology.}
At a node $W=(a,b)$, the parent list $\mathcal K$ consists of
\emph{global} upsets, and $Q(\mathcal K)$ denotes their union.
All times, cones and priorities are taken in the frame in which the
child under consideration is constructed.  In particular, for an
opposite-side child this is the reversed frame of Step~3 in
Section~7.3.  A \emph{carried qualifier} is a job in the union of
the suffix carry $\mathcal C$ selected \emph{before} the new injection
is appended; it may lie outside $W$.  A \emph{strict-step child} is a
nonempty gap $(p,q)$ for which $q>p$ is the final actual separator
position of a promotion that advances from the previous retained
position $p$.  The associated cutoff $S$ is the cutoff \emph{before}
that promotion.  A \emph{birth group} is as defined in Section~7.4:
the root entry, the pair created by a switch (counted as one group),
or an individual subsequently appended injection.  We call a list
\emph{genuine} if it is a list actually obtained by the operations
of Section~7.3, including pruning; arbitrary lists of global upsets
are not covered by the counting assertion below.

We induct over the depth of the annotated hierarchy.  Assume
(I1)--(I3) for the current parent and its ancestors.  The
construction and Lemma~\ref{str:labels} establish child (I1).
For opposite-side children the historical predicate $G$ is already
available from Lemma~\ref{str:history}, using the incoming edge of
$W$ and no invariant of the child.  We next prove child (I2), then
child (I3), and only afterwards apply the list bound.  This order
prevents the use of any unproved child invariant.

\begin{lemma}[The past-boundary invariant]\label{str:past}
Assume that (I1)--(I3) hold for a parent $W$ and its ancestors.
For every nonempty child $V=(p,q)$, with unpruned list
$\mathcal L_V$ and pruned list
$\mathcal K_V=\operatorname{prune}_V(\mathcal L_V)$,
\[
 \{x\in Q(\mathcal L_V):\tau(x)\le\operatorname{pos}(p)\}
     \subseteq\widehat P_p.
\]
Consequently $\mathcal K_V$ satisfies (I2).
\end{lemma}
\noindent\emph{Proof idea.}
A new qualifier lying in the parent interval is high at the
associated cutoff and is handled by the left separator.  A
qualifier occurring before the parent's left boundary can occur
only in an inherited same-direction carry.  Parent (I2) connects
it to a qualifier in the left boundary slot.  The first strict
separator advance, together with parent (I3), forces this boundary
qualifier to precede a job in the first target slot; the normalized
walk propagates the chain to the child's left boundary.
\begin{proof}
Write $W=(a,b)$ in the frame used for the child, and let
$x\in Q(\mathcal L_V)$ satisfy
$\tau(x)\le\operatorname{pos}(p)$.  We prove
$x\in\widehat P_p=P_p\cup\operatorname{jobs}(p)$.
If $p$ is the left sentinel, no real job satisfies the stated
time inequality, so assume otherwise.

\emph{Jobs strictly after $a$.}
Suppose $\tau(x)>\operatorname{pos}(a)$.  If $x$ occurs at $p$,
the conclusion is immediate.  Otherwise $a<x<p$ in temporal
position.  Thus $x$ belongs to the parent domain $W$ (and,
for an opposite-side child, to the processed domain $E$).
By Lemma~\ref{str:labels},
\[
 W\cap Q(\mathcal L_V)=W\setminus S.
\]
Therefore $x$ is high for the associated cutoff on the relevant
walk domain.  Its left endpoint $p$ separates that cutoff there;
the earlier-high separator condition gives $x\in P_p$.
This covers both child directions, including a terminal gap in
the reversed processed domain.

\emph{Jobs at or before $a$.}
Suppose $\tau(x)\le\operatorname{pos}(a)$.  For an opposite-side
child the two unpruned entries are $G\cap R_h$ and $D_a$.
Lemma~\ref{str:history} gives
$G\cap\{z:\tau(z)\le\operatorname{pos}(a)\}=\varnothing$;
the same exclusion holds for $D_a$ by feasibility (or the left
sentinel convention).  Hence this case cannot occur for an
opposite-side child.  For a same-direction child the fresh
injection $D_a\setminus P_b$ also contains no such job.
Consequently $x$ lies in the suffix carry $\mathcal C$ of the
old list $\mathcal K$.
In particular $x\in Q(\mathcal K)$; parent (I2) implies
$x\in\widehat P_a$.  If $a$ is actual, choose
$y\in\operatorname{jobs}(a)$ with $x=y$ or $x\prec y$.
Since $Q(\mathcal C)$ is a global upset,
$y\in Q(\mathcal C)\subseteq Q(\mathcal K)$.
The sentinel case cannot arise here, because there is no real
job at or before the left sentinel.
If $p=a$, then $x\in\widehat P_p$, as required.

Now assume $p>a$. Remember the assumptions here: $y \in \operatorname{jobs}(a) \cap
Q(\mathcal C)$ and $x = y$ or $x\prec y$.
Let $S$ be the cutoff assigned to this child;
the fixed carry $\mathcal C$ consists of the parent qualifiers
unpromoted at $S$.  Therefore every job of $Q(\mathcal C)$ that
belongs to $W$ is high for \emph{every} cutoff occurring in the
separator walk from $a$ to $p$, including its intermediate
advances.  Moreover, each promotion preceding this child has
old key smaller than the key of $y$: membership of $y$ in the
final suffix carry gives precisely this comparison.

\emph{The first strict advance from $a$.}
Let $t$ be its first target slot.  The current separator is $a$,
and the cutoff immediately before the promotion is an ideal $L$
separated by $a$.  If $j$ is the job just promoted, then every
old-low job scheduled after $a$ lies in $D_a$.  Hence the first
low exception causing the advance must be $j$ itself, with
$j\notin D_a$ and $j\in t$.  Since $j$ was promoted before $S$,
$\operatorname{key}_{\mathcal K}(j)<
\operatorname{key}_{\mathcal K}(y)$.
Also $j\in W\setminus D_a$, so parent (I1) makes it a qualifier.

We check the feasibility of the boundary exchange rather than
assuming it.  Every predecessor $z\prec j$ in $W$ is in $L$,
because the parent priority order extends precedence and $j$ is
the next promoted job.  As $a$ separates $L$, each such $z$
occurring after $a$ belongs to $D_a$, and thus $j\in D_a$ by
transitivity, a contradiction.  A predecessor of $j$ in the
boundary slot $a$ gives the same contradiction; a predecessor
outside $W$ cannot occur strictly between $a$ and $t$.
Therefore \emph{every} predecessor of $j$ in the complete
witness occurs strictly before $a$.

Suppose that $y$ had no successor scheduled at or before $t$.
Swapping $j$ (in $t$) with $y$ (in $a$) would then be feasible:
$j$ crosses none of its predecessors, while $y$ crosses none of
its successors.  The two slots remain full and all other jobs
keep their times.  This is forbidden by parent (I3), which
applies after the present feasible reordering of $W$.
Thus $y$ has a strict successor $z$ scheduled at or before $t$.
It lies after $a$, and it belongs to $Q(\mathcal C)$ because this
union is a global upset.  If $z$ lies in $t$, we already have
a carry-successor of $y$ in $t$.  Otherwise $z$ is a high job in
$W$ strictly between $a$ and $t$.  The high-job propagation
proved in Steps~2--3 of Lemma~\ref{str:walk} gives a strict
high successor of $z$ in $t$.  Upward closure again keeps that
successor in $Q(\mathcal C)$.  Hence in either case there exists
$z_t\in t\cap Q(\mathcal C)$ with $y\prec z_t$.

\emph{Propagation through the remaining advances.}
Consider any following advance $r\to t'$ before the final
retained position $p$, whether it is an intermediate advance
within one promotion or part of a later promotion.  The job
previously reached in $r$ belongs to $Q(\mathcal C)\cap W$, and
is therefore high for the current cutoff.  Steps~2--3 of
Lemma~\ref{str:walk}, applied to that advance in the normalized
parent schedule, supply a strict high successor in $t'$.
Iterating along the finitely many advances produces
$z_p\in p$ such that $y\prec z_p$.
Because $x=y$ or $x\prec y$, we conclude
$x\in\widehat P_p$.  The opposite-side normalization performed
after the same-direction marks are chosen does not move jobs
across $p$, so this conclusion persists.

We have proved (I2) for the unpruned list.  Pruning only deletes
entries and cannot introduce a new qualifier; hence
$Q(\mathcal K_V)\subseteq Q(\mathcal L_V)$ and the assertion
holds for the child's actual list.
\end{proof}

\begin{lemma}[The exchange invariant for all child reorderings]
\label{str:exchange}
Under the same induction hypotheses, every nonempty child with
its pruned list satisfies (I3), for \emph{every} feasible
reordering of the child's interior slots.
\end{lemma}
\noindent\emph{Proof idea.}
When the child's left boundary is a slot of the normalized
parent (or of the normalized reversed processed side), a
feasible exchange would improve that normalized schedule at its
first changed slot.  When the left boundary is inherited from
the parent, comparison of carried keys reduces the exchange to
the parent's (I3).
\begin{proof}
Let $V=(p,q)$ and fix an \emph{arbitrary} feasible full
reordering of its interior slots while all exterior slots stay
fixed.  Put
$\mathcal K'=\operatorname{prune}_V(\mathcal L_V)$.
Suppose, seeking a contradiction, that the exchange described in
(I3) is feasible for some
\[
 x\in V\setminus D_p,
 \qquad y\in\operatorname{jobs}(p)\cap Q(\mathcal K'),
 \qquad
 \operatorname{key}_{\mathcal K'}(x)
   <_{\mathrm{lex}}\operatorname{key}_{\mathcal K'}(y).
\]
Child (I1), already proved, gives
$x\in V\cap Q(\mathcal K')$, so both keys are defined.
The pruning comparison of Lemma~\ref{str:carry} lifts this
strict inequality to the unpruned list.  Write $S$ for the
cutoff associated with $V$.

\emph{An internal left endpoint $p$.}
By ``internal'' we mean that $p$ is an actual slot belonging to
the domain normalized when these marks were selected:
$W$ for a same-direction child, or $E$ for an opposite-side
child.  Both $x$ and $y$ then lie in that domain.  Since they
qualify for $\mathcal L_V$ on $W$, Lemma~\ref{str:labels}
gives $x,y\in W\setminus S$.  Its key-order assertion turns
the lifted inequality into $x\triangleleft y$ in the \emph{fixed}
priority order used for that normalization.  For an
opposite-side child we restrict the Step~3 priority to $E$.

The current rearrangement of $V$ affects only slots strictly
after $p$.  All other normalizations and later subtree
reorderings relevant to the present construction leave the
normalized prefix ending at $p$ unchanged.  If the proposed
exchange were feasible, the resulting feasible local schedule
of the normalized domain would agree with its normalized
witness in every slot before $p$, but would replace $y$ by
higher-priority $x$ in slot $p$.  This contradicts the prefix
minimality conclusion of Lemma~\ref{str:replacement}, which
explicitly allows arbitrary feasible later reorderings.

\emph{An inherited left endpoint $p=a$.}
If $a$ is a sentinel there is no boundary job $y$.
For an opposite-side child, the first entry
$G\cap R_h$ has no job at or before $a$ by
Lemma~\ref{str:history}, and neither does $D_a$; hence no $y$
qualifies.  For a same-direction child, the appended injection
$D_a\setminus P_b$ contains neither $x$ (since $x\notin D_a$)
nor $y$ (since an actual slot is an antichain).  Both jobs must
therefore qualify through the suffix carry.
Carrying preserves their relative old-key order, and pruning
lifts their comparison by Lemma~\ref{str:carry}, because
$x\in V$ is an interior job.  Thus
$\operatorname{key}_{\mathcal K}(x)
 <_{\mathrm{lex}}\operatorname{key}_{\mathcal K}(y)$.
As $V\subseteq W$, and $p=a$, the proposed exchange is exactly
one of the exchanges forbidden by parent (I3).  In particular,
the feasible normalizations of the parent's interior followed
by the arbitrary reordering of $V$ form a feasible interior
reordering to which that invariant applies (by
Lemma~\ref{str:replacement}).  We obtain a contradiction in
this case as well.
\end{proof}

\paragraph{Counting groups at a marked endpoint.}
A switch pair is treated as \emph{one} birth group even if both
of its entries survive.  We count a birth group as nonempty at a
node precisely when at least one of its retained entries meets
the current node.  For a strict-step child $(p,q)$ the final
separator advance provides a job in $q$ that cannot be reached
from the old-high jobs.  We first prove that separate surviving
groups cannot be charged to the same job in $q$, then handle
children not coming from a strict advance.

\begin{lemma}[A spare endpoint and at most $m$ entries]
\label{str:groups}
Every genuine pruned list has at most $m-1$ nonempty birth
groups and at most $m$ entries.  Among its entries only the
first can have a \emph{current outer} rank-suffix restriction;
rank conditions already frozen inside the historical first
entry are not counted as current outer restrictions.
\end{lemma}
\noindent\emph{Proof idea.}
For a strict separator move, charge each surviving group to a
distinct job in its final endpoint slot.  A low exception causing
the last advance cannot receive a charge, leaving at most $m-1$
slots for charges.  Other child types start from one group or
from a fresh injection alone.
\begin{proof}
\emph{The spare job for a strict advance.}
Consider a strict-step child $V=(p,q)$ of a parent $W$.
The endpoint $q$ was retained after a promotion from cutoff
$S$ to $S^+=S\cup\{j\}$.  Let $r\to q$ be the last individual
advance of that promotion, and choose a low exception
$e\in\operatorname{jobs}(q)$ causing this advance.
If it is the first advance, the only newly low job is $j$;
previously low jobs were already covered by $D_p$.  Hence
$e=j$.  Because $S$ is a priority prefix, every predecessor
of $j$ in the parent belongs to $S$.  If an earlier advance
occurred, then $j$ lies at or before $r$.  Thus the exception
$e$ in the later slot $q$ belongs to the old cutoff $S$;
all of its parent-domain predecessors lie in $S$, since $S$
is an ideal.  In either case we obtain
\[
 \text{there is no }z\in W\setminus S
       \text{ with }z\prec e.
\]
For an opposite-side strict step the walk was performed on
$E\subseteq W$.  The old-high jobs in $W\setminus E$ occur
at or after the center, hence cannot precede $e\in q\subseteq E$.
Thus the spare-job property holds for the entire parent also in
this case.

\emph{Disjointness of different birth groups on current nodes.}
We establish by induction along a chain of genuine list
updates that distinct groups have disjoint qualifier sets on
the current node.  It holds for the root's single group and
immediately after a switch, which discards all older groups
and creates one new group.  At a same-direction update from a
parent $W=(a,b)$, every old group is a subset of the union of
the parent's old qualifiers.  By parent (I1), those qualifiers
that belong to $W$ lie in $P_b$.  The new injection
$I(a,b)=D_a\setminus P_b$ is therefore disjoint from all old
groups \emph{on $W$}.  Suffix carrying and pruning only remove
memberships or entire entries; in particular, they cannot
create an overlap on a descendant $V\subseteq W$ between
different old groups.  This proves the induction step.
Note that the entries of the \emph{same} switch group may
overlap, and different groups may overlap outside the current
node; neither phenomenon affects the argument.

\emph{Charging at a strict-step child.}
Let $\mathcal B$ be a group surviving pruning on $V$.
There is a job $x\in V$ belonging to this group.  By
Lemma~\ref{str:labels}, $x$ is high for the associated cutoff
$S$, so separation at the right endpoint $q$ gives
$x\prec y$ for some $y\in\operatorname{jobs}(q)$.
Since every entry in $\mathcal B$ is a global upset, the
successor $y$ belongs to $\mathcal B$.

\ifshowrevisions
\begin{revisionold}
\renewcommand{\theHequation}{deleted.0001.\arabic{equation}}
    The endpoint slot $q$ is contained in $W$, and the disjointness just
proved therefore ensures that distinct groups receive distinct jobs
$y$ of its full $m$-job slot.
\end{revisionold}
\begin{revisionnew}
Before restriction to the child $V$, distinct updated groups are already
disjoint on the whole parent $W$: carried groups retain the parent's
disjointness, and the fresh injection is disjoint from every carried
group on $W$. An opposite-side update has only one switch group.
Since the endpoint slot $q$ lies in $W$, this parent-domain disjointness
applies to the charged jobs $y\in\operatorname{jobs}(q)$.
Thus distinct surviving groups receive distinct jobs of the full
$m$-job endpoint slot.
\end{revisionnew}
\else
Before restriction to the child $V$, distinct updated groups are already
disjoint on the whole parent $W$: carried groups retain the parent's
disjointness, and the fresh injection is disjoint from every carried
group on $W$. An opposite-side update has only one switch group.
Since the endpoint slot $q$ lies in $W$, this parent-domain disjointness
applies to the charged jobs $y\in\operatorname{jobs}(q)$.
Thus distinct surviving groups receive distinct jobs of the full
$m$-job endpoint slot.
\fi
  No group can be charged to the spare job $e$,
by the spare-job property.  There are consequently at most
$m-1$ surviving groups.

\emph{Other child types and the number of entries.}
For a same-direction terminal gap associated with the exhausted
cutoff $S_{\mathrm{end}}=W\cap P_b$, the carry of unpromoted
qualifiers is empty; only the fresh injection can survive.
Every opposite-side child is created by a switch and thus has
just one birth group (its two-entry pair).  At the root there
is one group.  These are the remaining cases: every other
same-direction nonempty child arises from a strict retained
advance.  Since $m\ge3$, one group also satisfies the bound
$m-1$.

The root group has one entry, a switch group at most two, and
every injection group one.  There is at most one initial
switch group in a genuine list, so $m-1$ groups yield at most
$(m-1)+1=m$ entries.  Each suffix carry can rank-restrict
only its first retained entry; multiple such restrictions
merge into a single threshold.  New injections are appended
without an outer rank restriction, and pruning only deletes
entries.  A historical rank restriction frozen \emph{inside}
a switched first entry is part of that entry's predicate,
not an additional outer restriction on a later entry.
This proves all assertions.
\end{proof}

\begin{proposition}[Existence of the annotated hierarchy]
\label{str:annotated}
Every feasible full instance admits a finite hierarchy whose
nodes carry the frames, complete witness schedules, boundary
slots, genuine lists and recorded incoming-edge information of
Sections~7.1--7.4.  Each node satisfies (I1)--(I3), its list
has at most $m$ entries, every nonempty node has at least one
marked full slot, and each of its children has at most
$|W|-m$ jobs.
\end{proposition}
\noindent\emph{Proof idea.}
Construct the children only from verified parent information;
prove their three invariants before recursing.  The marked
center supplies a uniform decrease of at least $m$ jobs.
\begin{proof}
Use the forward frame, the two sentinel boundaries,
$W=J$ and list $(J)$ at the root.  The sentinel cone
conventions give (I1); invariants (I2) and (I3) are vacuous.
The root has one group and one entry.

Suppose the parent and its ancestors have been constructed
and satisfy the asserted properties.  Lemma~\ref{str:replacement}
allows each required normalization while keeping the exterior
slots fixed.  Lemma~\ref{str:walk} and
Corollary~\ref{str:actual} supply the separator walks and the
actual center.  The gaps are unions of full slots and hence
have feasible full schedules.  The lists for same-direction
children are obtained by carrying, injecting and pruning.
For opposite-side children the parent has an incoming edge;
Lemmas~\ref{str:flat} and~\ref{str:history} construct the
switch predicate from \emph{previously recorded} information.
At the root no such predicate is needed, because its
opposite side is empty.

Lemma~\ref{str:labels} now establishes (I1) for every new
child.  Lemma~\ref{str:past} establishes (I2), and
Lemma~\ref{str:exchange} establishes (I3).  The counting
Lemma~\ref{str:groups}, applied to the resulting genuine
pruned lists, bounds their groups and entries.  Thus every
hypothesis needed to use a child as the parent of the next
level has been verified, without invoking any property of a
future node.

Finally the center is a marked slot of exactly $m$ jobs, and
no child contains any of its jobs.  Hence each child's set has
cardinality at most $|W|-m$.  This strict decrease makes the
hierarchy finite, completing the induction.
\end{proof}

\subsection{Exact formulas and their sizes}

\paragraph{Purpose, domains, and interpretation of a formula.}
Proposition~\ref{str:annotated} constructs a finite annotated hierarchy.
To prove Theorem~\ref{af:structure}, we must also describe the job set
of every node and the temporal prefix before every marked slot by
Boolean formulas from Definition~2.2.  All such formulas are evaluated
on the \emph{fixed global ground set} $J$; their value is not an
expression whose universe changes when we pass to a child.

We distinguish three objects throughout this subsection.
The \emph{parent domain} $W=(a,b)$ consists of the jobs strictly
between its two boundary slots in the complete feasible witness.
A \emph{child domain} $V=(p,q)$ is an open temporal gap within $W$.
The associated \emph{low cutoff} $S\subseteq W$ is a priority prefix
in the frame in which that gap was constructed.  The child jobs
in $V\cap S$ are called \emph{low}, and those in $V\setminus S$
are called \emph{high}.  These classifications refer to priority,
not to execution time.  We write $Q'=Q(\mathcal K_V)$ for the union
of the entries of the child's genuine \emph{pruned} list
$\mathcal K_V$.  Its entries, and hence $Q'$, are global sets;
in particular, $Q'$ may contain jobs outside $V$.

Unless stated otherwise, each assertion below is made in the
\emph{working frame of the particular child}.  For a same-direction
child this is the parent's frame; for an opposite-side child it is
the reversed frame of Step~3 of Section~7.3.  In either frame,
$\tau(x)$ denotes the slot of $x$ in the \emph{complete} witness
and $\operatorname{pos}(r)$ the position of boundary slot $r$.
For an actual boundary $r$ we use
\[
 P_r=\{x\in J:\exists u\in\operatorname{jobs}(r),\ x\prec u\},
 \qquad
 D_r=\{x\in J:\exists u\in\operatorname{jobs}(r),\ u\prec x\},
\]
and $\widehat P_r=P_r\cup\operatorname{jobs}(r)$,
$\widehat D_r=D_r\cup\operatorname{jobs}(r)$.
We retain the sentinel conventions of Section~7.1.  Feasibility
implies the following elementary temporal consequences:
\begin{equation*}
 x\in P_r\ \Longrightarrow\ \tau(x)<\operatorname{pos}(r),
 \qquad
 x\in D_r\ \Longrightarrow\ \tau(x)>\operatorname{pos}(r).
\end{equation*}
Conversely, the corresponding time inequalities do \emph{not}
by themselves imply cone membership.  The separator conditions
are needed whenever a proof makes such an inference.

\paragraph{Two previously established identities.}
For every genuine child, Lemmas~\ref{str:labels},
\ref{str:past} and the pruning rule give
\begin{equation}
 V\cap Q'=V\setminus S,
 \qquad V\cap S\subseteq D_p\cap P_b,
 \qquad V\setminus S\subseteq P_q,
 \label{str:local-ingredients}
\end{equation}
and the child's past-boundary invariant
\begin{equation}
 x\in Q',\ \tau(x)\le\operatorname{pos}(p)
 \quad\Longrightarrow\quad x\in\widehat P_p.
 \label{str:local-past}
\end{equation}
These statements have different logical roles: the first identifies
\emph{which child jobs} qualify; the second rules out false
qualifiers \emph{outside} the child.  In particular, it would be
incorrect to substitute $Q'=V\setminus S$ as a global set identity.
For same-direction children, if $\mathcal C$ denotes the suffix
carry of unpromoted parent qualifiers before the new injection,
Section~7.4 supplies a \emph{global upset formula} $F$ satisfying
\begin{equation}
 W\cap F=W\cap Q(\mathcal C)=O\setminus S,
 \qquad O:=W\cap Q(\mathcal K_W).
 \label{str:local-flat}
\end{equation}
The same construction gives
\begin{equation}
 S=W\cap(P_b\setminus F).
 \label{str:local-cutoff}
\end{equation}
Indeed $S$ contains the first priority block
$(W\setminus O)\cap P_b$ and the promoted qualifiers, whereas
$O\setminus S$ is precisely the unpromoted carry.  Neither
\eqref{str:local-flat} nor \eqref{str:local-cutoff} is asserted
outside $W$.
For opposite-side children let
\[
 X:=W\setminus D_a,\qquad
 L_h:=\{x\in J:\operatorname{rank}(x)<h\}=J\setminus R_h,
 \qquad H_h:=(J\setminus D_a)\cap L_h.
\]
Here $h$ is the rank threshold attached to the child, so that
\begin{equation}
 S=S_h=X\cap L_h=W\cap H_h.
 \label{str:local-opcut}
\end{equation}
The processed side is $E=(a,v)$, where $v$ is the center in the
reversed working frame.
    The center separates $X$ on $W$.
 If $q<v$, then $q$ separates $S_h\cap E$ on $E$.
 If $q=v$, the gap is terminal and the center's separator properties
on $W$ are used directly.

\begin{lemma}[Exact global localization of a child]\label{str:localization}
Let $V=(p,q)$ be a nonempty genuine child of $W=(a,b)$,
let $S\subseteq W$ be its associated cutoff, and let
$Q'=Q(\mathcal K_V)$ be the global union of its pruned list.
In the child's working frame, the high part has the \emph{global}
set identity
\begin{equation}
 V\setminus S=(Q'\cap P_q)\setminus\widehat P_p.
 \label{str:high-exact}
\end{equation}
If $V$ is same-direction and $F$ is the flat formula of
\eqref{str:local-flat}, then
\begin{equation}
 V\cap S=(D_p\cap P_b)\setminus(\widehat D_q\cup F).
 \label{str:low-same-exact}
\end{equation}
If $V$ is opposite-side, write $E=(a,v)$ and use the threshold
$h$ of \eqref{str:local-opcut}.  Then
\begin{equation}
 V\cap S=
 \bigl(D_p\cap P_b\cap(J\setminus D_a)\cap L_h\bigr)
 \setminus(\widehat D_q\cup\widehat D_v).
 \label{str:low-op-exact}
\end{equation}
All right-hand sides are formulas evaluated on $J$.  In each
case their high and low parts are disjoint and have union
exactly $V$ as a subset of $J$.
\end{lemma}
\noindent\emph{Proof idea and purpose.}
The locally known identity $V\cap Q'=V\setminus S$ must be
made global.  The cone $P_q$ places high qualifiers before the
right endpoint, and (I2) eliminates qualifiers at or before the
left endpoint.  For low jobs, $D_p\cap P_b$ first establishes
membership in the \emph{parent}, so that a local cutoff formula
may legitimately be used.  The exclusions of weak successor
cones then enforce the right temporal boundary.  On the
opposite side, a second separator at the center is needed to
localize the job to the processed domain $E$.
\begin{proof}
\emph{High part: inclusion from left to right.}
Take $x\in V\setminus S$.
By \eqref{str:local-ingredients}, $x\in Q'$ and $x\in P_q$.
Also $\tau(x)>\operatorname{pos}(p)$ because $x\in(p,q)$.
If $x\in\widehat P_p$, either $x$ would be in the slot $p$,
or some job of $p$ would be scheduled after $x$; both are
incompatible with $\tau(x)>\operatorname{pos}(p)$ and
feasibility.  Hence $x\notin\widehat P_p$, proving the first
inclusion in \eqref{str:high-exact}.

\emph{High part: inclusion from right to left.}
Take $x\in(Q'\cap P_q)\setminus\widehat P_p$.
Since $x\in P_q$, feasibility gives
$\tau(x)<\operatorname{pos}(q)$.  If
$\tau(x)\le\operatorname{pos}(p)$, then $x\in Q'$ and
\eqref{str:local-past} imply $x\in\widehat P_p$, a
contradiction.  Therefore
\[
 \operatorname{pos}(p)<\tau(x)<\operatorname{pos}(q).
\]
This is exactly $x\in V$.  Now \emph{and only now} we apply
the first identity of \eqref{str:local-ingredients}:
$x\in V\cap Q'=V\setminus S$.  This proves
\eqref{str:high-exact} globally, without any assumption about
qualifiers elsewhere in $J$.

\emph{Same-direction low part: inclusion from left to right.}
Take $x\in V\cap S$.  Lemma~\ref{str:labels} gives
$x\in D_p\cap P_b$.  Since $x$ occurs strictly before $q$,
it belongs neither to the slot $q$ nor to $D_q$; thus
$x\notin\widehat D_q$.  Also $x\in W\cap S$, while
\eqref{str:local-flat} says that $W\cap F=O\setminus S$.
As $S\cap(O\setminus S)=\varnothing$, we have $x\notin F$.
This proves the first inclusion in \eqref{str:low-same-exact}.

\emph{Same-direction low part: inclusion from right to left.}
Take
\[
 x\in(D_p\cap P_b)\setminus(\widehat D_q\cup F).
\]
The membership $x\in D_p$ implies
$\operatorname{pos}(p)<\tau(x)$, also when $p$ is the left
sentinel by its cone convention.  Membership $x\in P_b$
implies $\tau(x)<\operatorname{pos}(b)$, also at the right
sentinel.  Because the two child endpoints satisfy
$\operatorname{pos}(a)\le\operatorname{pos}(p)<
\operatorname{pos}(q)\le\operatorname{pos}(b)$, it follows that
\[
 \operatorname{pos}(a)<\tau(x)<\operatorname{pos}(b),
 \quad\text{hence}\quad x\in W.
\]
Only with this domain membership can we use
\eqref{str:local-cutoff}.  From $x\in W\cap P_b$ and
$x\notin F$ it follows that $x\in S$.
The right endpoint $q$ separates $S$ on $W$.
If $\tau(x)=\operatorname{pos}(q)$, then $x\in
\operatorname{jobs}(q)\subseteq\widehat D_q$; if
$\tau(x)>\operatorname{pos}(q)$, the \emph{later-low}
separator condition gives $x\in D_q\subseteq\widehat D_q$.
Both possibilities contradict the assumption.
Consequently $\tau(x)<\operatorname{pos}(q)$, and with
$\tau(x)>\operatorname{pos}(p)$ this yields $x\in V\cap S$.
This proves \eqref{str:low-same-exact}.

\emph{Opposite-side low part: inclusion from left to right.}
We work throughout in the reversed frame used to construct the
opposite-side child.  Take $x\in V\cap S$.
The cone inclusions of Lemma~\ref{str:labels} give
$x\in D_p\cap P_b$.  By \eqref{str:local-opcut},
$x\in S\subseteq W\cap H_h$, so
$x\in (J\setminus D_a)\cap L_h$.
Furthermore $V=(p,q)\subseteq E=(a,v)$ and
$\tau(x)<\operatorname{pos}(q)\le\operatorname{pos}(v)$.
Feasibility implies that $x$ is in neither
$\widehat D_q$ nor $\widehat D_v$.  Thus it belongs to the
right-hand side of \eqref{str:low-op-exact}.

\emph{Opposite-side low part: inclusion from right to left.}
Take $x$ in the right-hand side of
\eqref{str:low-op-exact}.  As in the same-direction case,
$x\in D_p\cap P_b$ and $p\ge a$ place it in the \emph{parent}
$W=(a,b)$: namely
$\operatorname{pos}(a)<\tau(x)<\operatorname{pos}(b)$.
Since also $x\in(J\setminus D_a)\cap L_h$, we may use
\eqref{str:local-opcut} to conclude
\[
 x\in W\cap H_h=S_h=S\subseteq X.
\]
We next show explicitly why $x\in E$.  The center $v$
separates $X$ on the \emph{whole} parent $W$.
If $\tau(x)=\operatorname{pos}(v)$, then
$x\in\operatorname{jobs}(v)\subseteq\widehat D_v$.
If $\tau(x)>\operatorname{pos}(v)$, the later-low property
for $X$ at $v$, applied to $x\in X\cap W$, yields
$x\in D_v\subseteq\widehat D_v$.  Both contradict the
exclusion of $\widehat D_v$.  Hence
$\tau(x)<\operatorname{pos}(v)$.  Since
$x\in D_p$ and $p\ge a$, it follows that
\[
 \operatorname{pos}(a)<\tau(x)<\operatorname{pos}(v),
 \quad\text{so}\quad x\in E.
\]
Now $x\in S_h\cap E$, exactly the local low cutoff on which
the separator assertion for $q$ is available.  If
$\tau(x)=\operatorname{pos}(q)$, then $x\in\widehat D_q$;
if $\tau(x)>\operatorname{pos}(q)$, the later-low property
at $q$ on $E$ implies $x\in D_q$.  When $q=v$, the same
conclusion follows from the center's separator property on
$W$.  Both contradict $x\notin\widehat D_q$.
Thus $\tau(x)<\operatorname{pos}(q)$, and the membership
$x\in D_p$ places it after $p$.  We obtain $x\in V\cap S$.
This proves \eqref{str:low-op-exact} and the lemma.
\end{proof}

\paragraph{The exact counting convention.}
The atomic language is precisely the one in Definition~2.2.
A \emph{leaf} is one occurrence of a permitted atom:
$J$, $\varnothing$, a rank suffix $R_h$, or a boundary set
$A$ of size $m$ and one of $P_A,D_A,\widehat P_A,\widehat D_A$.
The chosen boundary slot has $m$ jobs, so its set and cones
are permitted atoms.  A sentinel cone denotes the appropriate
constant $J$ or $\varnothing$ by the conventions of
Section~7.1.  A rank prefix $L_h=J\setminus R_h$ costs one
leaf (the complement operation does not add a leaf).
Likewise, a set difference $B\setminus C$ is shorthand for
$B\cap(J\setminus C)$, where the $J$ in the complement
notation does \emph{not} introduce another atomic occurrence.
Union, intersection, and complement themselves cost no leaves;
repeated appearances of an atom are counted repeatedly.
A standalone constant such as $J$ or $\varnothing$ has one leaf.
The leaf count of a union is at most the sum of the leaf
counts of the displayed subformulas; no Boolean simplification
is being assumed unless explicitly described.

We also distinguish the \emph{exact entries} of a list from a
\emph{flat carry}.  A flat formula $F$ need only satisfy
$U\cap F=U\cap Q(\mathcal C)$ for the specified current
domain $U$; it cannot replace the global entries when defining
$Q'$ or applying (I2).  By contrast, the formulas counted for
$Q'$ below are exact on all of $J$.

\begin{lemma}[A linear formula budget]\label{str:budget}
For $m\ge3$, the hierarchy of Proposition~\ref{str:annotated}
has the following properties, with leaf counts in the atomic
language of Definition~2.2:
\begin{enumerate}[label=\textup{(\arabic*)}]
\item Every suffix carry from a genuine pruned list has a flat
      global upset formula on its parent domain with at most
      $\revchange{m+2}{m}$ leaves.
\item The possible complicated first exact entry of a genuine
      list has at most $\revchange{m+5}{m+3}$ leaves.
\item The exact global qualifier union of every genuine pruned
      list has a formula with at most $\revchange{3m+4}{3m}$ leaves.
\item Every genuine node set, evaluated as a subset of $J$, has
      a formula with at most $\revchange{4m+12}{4m+5}$ leaves.
\item For every marked slot $z$ at every node $W$, there is a
      global formula $C_z$ with at most $\revchange{m+6}{m+3}$ leaves such that
      $W\cap C_z$ is the set of jobs occurring strictly before
      $z$ in the \emph{original input frame}.
\end{enumerate}
Frame changes do not enlarge the leaf count of any set formula:
they merely exchange the permitted predecessor and successor
cone atoms and replace rank suffixes by complemented rank
suffixes.  
\ifshowrevisions
\begin{revisionold}
\renewcommand{\theHequation}{deleted.0008.\arabic{equation}}
The possible extra boundary-set leaf needed to
convert a \emph{temporal prefix} is included in~(5).
\end{revisionold}
\begin{revisionnew}
In~(5), the input-time prefix is represented directly by a working-time
prefix or suffix formula. Thus frame conversion requires no additional
boundary-set leaf.
\end{revisionnew}
\else
In~(5), the input-time prefix is represented directly by a working-time
prefix or suffix formula. Thus frame conversion requires no additional
boundary-set leaf.
\fi

\end{lemma}
\noindent\emph{Proof idea and purpose.}
Use Lemma~\ref{str:groups} to bound surviving birth groups,
then inspect the exhaustive flat-union cases of Section~7.4.
An exact historical entry can be complicated, but there is at
most one such entry in any current list, and it is the initial
one.  Count these exact entries before using the two global
localization formulas of Lemma~\ref{str:localization}.

\ifshowrevisions
\begin{revisionold}
\renewcommand{\theHequation}{deleted.0009.\arabic{equation}}
Finally, construct the global prefix formula first in the
frame in which its separator was proved, and explicitly
convert it to input time.  This avoids treating a local
reversed prefix as if it were already an input-time prefix.
\end{revisionold}
\begin{revisionnew}
Finally, construct both temporal sides of each separator directly.
Choose its working-time prefix or suffix according to its orientation
relative to the input frame. This preserves the distinction between
working time and input time without adding a boundary-set leaf.
\end{revisionnew}
\else
Finally, construct both temporal sides of each separator directly.
Choose its working-time prefix or suffix according to its orientation
relative to the input frame. This preserves the distinction between
working time and input time without adding a boundary-set leaf.
\fi

\begin{proof}
\emph{(1) Flat carries.}
Fix a parent node $U$ and a suffix carry $\mathcal C$ selected
from its genuine pruned list.  A birth group surviving in the
carry also survived in the parent list, so Lemma~\ref{str:groups}
provides at most $m-1$ groups.
    The flat formulas
\eqref{str:flat-case-switch}--\eqref{str:flat-case-injections}
of Section~7.4, together with the empty-carry formula, cover
all possible surviving initial groups.  We count them explicitly.
\begin{itemize}
\item If the first switch entry $G_0\cap R_t$ survives, its
      switch group contributes the two atoms $D_c$ and $R_t$
      through the flat expression $D_c\cup R_t$.  If the
      companion was pruned, the same expression still agrees
      with the group on $U$, as proved in Lemma~\ref{str:flat}.
\item If only the switch companion survives as the initial
      entry, its contribution is $D_c\cap R_t$, at most two
      leaves.  (With no nontrivial rank restriction one may
      omit the factor $R_1=J$.)
\item If the root entry survives, its contribution is $R_t$,
      one leaf.
\item If the initial group has disappeared, the carry begins
      with an injection.  The first injection contributes
      $(J\setminus P_{B_1})\cap R_t$ on $U$, at most two
      leaves; later injections contribute $J\setminus P_{B_i}$.
      The identity is only after intersection with $U$, as
      required for a flat expression.
\item For an empty carry, the constant $\varnothing$ is a
      one-leaf flat formula.
\end{itemize}
Every later injection group contributes just one additional
atom $P_{B_i}$ to the flat union.  If a switch group survives,
there are at most $(m-1)-1=m-2$ injection groups, hence at
most $2+(m-2)=m$ leaves.  If no switch group survives, the
first remaining group uses at most two leaves and the other
at most $m-2$ groups contribute one each, again at most $m$.

\ifshowrevisions
\begin{revisionold}
\renewcommand{\theHequation}{deleted.0006.\arabic{equation}}
The empty case uses one leaf.  Thus the stated, slightly
relaxed bound $m+2$ holds in every case.
\end{revisionold}
\begin{revisionnew}
The empty case uses one leaf. Thus the bound $m$ holds in every case.
\end{revisionnew}
\else
The empty case uses one leaf. Thus the bound $m$ holds in every case.
\fi
  Each displayed
flat formula is a global upset: rank suffixes, complements of
predecessor cones, and successor cones are upsets, and
unions/intersections preserve that property.

\emph{(2) The complicated exact entry.}
The operations of Section~7.3 imply that an inherited list
starts with a surviving part of one root/switch group or with
an injection; a later switch discards the entire earlier list.
An injection has the exact two-leaf predicate $D_A\setminus P_B$.
Only the \emph{first} entry of a switch pair can have the
historically complicated form.  Carrying can only intersect
the first retained entry with a current outer rank suffix,
and pruning can only remove entries.  Therefore no later entry
can become a new complicated historical entry.

For a same-direction historical incoming edge, the exact form
\eqref{str:complicatedsame} of Section~7.4, interpreted in the new
frame, is
\[
 G\cap R_t
 =D_u\cap(J\setminus\widehat P_c)
   \cap(J\setminus F_{\mathrm{hist}}^{\mathrm{rev}})
   \cap R_t.
\]
The frozen flat formula $F_{\mathrm{hist}}$ was formed from
an earlier genuine pruned list; by~(1), it has at most $\revchange{m+2}{m}$
leaves.  The remaining factors contribute exactly three
additional atomic occurrences: $D_u$, $\widehat P_c$, and
$R_t$.  Hence this exact entry has at most $\revchange{m+5}{m+3}$ leaves.
Here $F_{\mathrm{hist}}^{\mathrm{rev}}$ denotes the \emph{same}
subset of $J$ written in the reversed atomic language; it
is not the set complement of $F_{\mathrm{hist}}$.

For an opposite-side historical incoming edge, the exact formula
\eqref{str:complicatedopposite} is
\[
 G=D_u\cap(J\setminus\widehat P_c)
       \cap(J\setminus\widehat P_w)
       \cap(J\setminus P_d)\cap R_k.
\]
It uses five leaves.  An additional current outer suffix
$R_t$ can be merged with $R_k$ because
$R_k\cap R_t=R_{\max\{k,t\}}$ in the same frame.
Thus the bound $\revchange{m+5}{m+3}$ holds for either historical type
($m\ge3$).  Rank conditions already frozen inside
$F_{\mathrm{hist}}^{\mathrm{rev}}$ are not treated as another
current outer suffix.

\emph{(3) Exact global qualifier unions.}
Let $\mathcal K$ be a genuine pruned list, and write
$Q(\mathcal K)$ as the union of its \emph{exact} retained
entries.  By Lemma~\ref{str:groups}, it has at most $m-1$
nonempty birth groups.  The exceptional case is a surviving
complicated first entry of a switch group.  This group has
at most one additional companion entry, costing one cone
leaf, and there are at most $m-2$ other groups, each an
injection with an exact two-leaf formula.  Thus
\[
 \#\operatorname{leaves}(Q(\mathcal K))
 \le(m+3)+1+2(m-2)=3m.
\]

If no complicated switch entry survives, a remaining root entry
or simple switch companion has at most two leaves.
 A first injection may instead have the three-leaf formula
$(D_A\setminus P_B)\cap R_t$.
 Each later injection costs two leaves.
 Thus, with $1\le g\le m-1$ nonempty groups, the total is at most
$3+2(g-1)\le2m-1$;
the empty list is represented by the one-leaf constant
$\varnothing$.  These counts are no greater than $\revchange{3m+4}{3m}$
for $m\ge3$.  Hence there is an \emph{exact global} formula
for $Q(\mathcal K)$ using at most $\revchange{3m+4}{3m}$ leaves.

\emph{(4) Exact node sets.}
At any nonroot node $V$, Lemma~\ref{str:localization}
expresses $V$ as the disjoint union of its high and low parts.
The high-part formula \eqref{str:high-exact} uses the exact
qualifier union $Q'$ of the child's \emph{pruned} list and the
two weak/strict predecessor-cone atoms $P_q$ and
$\widehat P_p$.  Its leaf count is therefore at most
\[
  3m+2.
\]

For a same-direction child, \eqref{str:low-same-exact} adds
the $m$-leaf flat formula of its \emph{parent} carry and
the three cone occurrences $D_p,P_b,\widehat D_q$.
Thus its low part has at most $m+3$ leaves, and the
complete set $V$ has at most
\[
  (3m+2)+(m+3)=4m+5
\]
leaves.  For an opposite-side child, formula
\eqref{str:low-op-exact} has six atomic occurrences:
$D_p,P_b,D_a,R_h,\widehat D_q,\widehat D_v$.
The complete set consequently has at most
\[
  (3m+2)+6=3m+8\le4m+5
  \qquad(m\ge3)
\]
leaves.  The root set $J$ is already a single atom.
These formulas are \emph{exact on all of $J$}; no uncounted
occurrence of the parent formula is used to localize the child.

\ifshowrevisions
\begin{revisionold}
\renewcommand{\theHequation}{deleted.0000.\arabic{equation}}
\emph{(5) Global cuts for marked prefixes.}
We first formulate the frame-conversion rule once.  Let $z$
be an actual marked slot of a node $W$ and let $A_z\subseteq J$
be a global formula satisfying
\begin{equation}
 W\cap A_z
 =\{x\in W:\tau_{\mathrm{work}}(x)
          <\operatorname{pos}_{\mathrm{work}}(z)\}.
 \label{deleted:str:working-prefix}
\end{equation}
If the working frame is the original input frame, take
$C_z=A_z$.  If the working frame is reversed relative to
the input, then within $W$ the jobs strictly before $z$
in input time are precisely the jobs \emph{strictly after}
$z$ in working time.  Thus take
\begin{equation}
 C_z=J\setminus(A_z\cup\operatorname{jobs}(z)).
 \label{deleted:str:frame-prefix-conversion}
\end{equation}
In either case $W\cap C_z$ is exactly the required input-time
prefix.  Indeed, within $W$ the working-time prefix,
the marked slot, and the working-time suffix partition the
jobs of $W$.  In the second case the conversion adds just
one boundary-set leaf; the complement operation adds none.
The reversal of cone and rank atoms itself costs no leaves:
$P_A$ and $D_A$ (and the weak versions) are exchanged,
while a rank suffix becomes the complement of one rank
suffix in the opposite frame.

\emph{Same-direction marked slots.}
Let $z$ be a retained actual separator from Step~1 of
Section~7.3, including the center.  Choose a low cutoff
$S_z$ that $z$ separates on the \emph{whole parent} $W$:
for a strict retained advance one may use its pre-promotion
cutoff, and for the center one may use the exhausted cutoff
$W\cap P_b$.  Let $F_z$ be the flat formula for the
unpromoted carry at this cutoff.  By
\eqref{str:local-cutoff}, the global ideal
\[
 H_z:=P_b\setminus F_z
\]
satisfies $W\cap H_z=S_z$.  Lemma~\ref{str:prefix}
then gives \eqref{deleted:str:working-prefix} with
\begin{equation}
 A_z=B_z(H_z)
    =P_z\cup\bigl((P_b\setminus F_z)
                         \setminus\widehat D_z\bigr).
 \label{deleted:str:same-mark-global}
\end{equation}
The displayed formula uses at most $(m+2)+3=m+5$
leaves: the flat carry, $P_b$, $P_z$, and $\widehat D_z$.
If the node's frame is reversed relative to the input,
\eqref{deleted:str:frame-prefix-conversion} increases the count to
at most $m+6$.  The opposite-side normalization occurs on
the other side of the center, so it leaves the temporal
prefix of every already retained same-direction mark fixed.

\emph{Opposite-side marked slots.}
We work in the reversed frame used in Step~3 of Section~7.3;
all precedence cones, time positions, and ranks in this paragraph
refer to that frame.  Let $v$ be the center of the parent
$W=(a,b)$, and put
\[
 X:=W\setminus D_a,\qquad
 H_0:=J\setminus D_a,\qquad
 E:=(a,v).
\]
The set $H_0$ is a global ideal and $W\cap H_0=X$.
By Step~3, the center $v$ separates $X$ on the entire
parent $W$.  Lemma~\ref{str:prefix} therefore gives the
\emph{working-frame} prefix identity
\begin{equation}
 E=\{x\in W:\tau(x)<\operatorname{pos}(v)\}
   =W\cap B_v(H_0),\qquad
 B_v(H_0)=P_v\cup(H_0\setminus\widehat D_v).
 \label{deleted:str:side-by-center}
\end{equation}
For an opposite-side retained separator $q\le v$, let $h$
be the rank threshold recorded with its associated cutoff.
As in Step~3, write
\[
 L_h:=\{x\in J:\operatorname{rank}(x)<h\}
      =J\setminus R_h,\qquad
 S_h:=X\cap L_h,\qquad
 H_h:=H_0\cap L_h.
\]
In particular, $H_h$ is a global ideal and
$H_h\cap E=S_h\cap E$.  We distinguish the two possible
locations of $q$.

If $q<v$, then $q$ is an actual slot \emph{inside} the
processed domain $E$.  The separator walk on $E$ ensures
that $q$ separates $S_h\cap E$ on $E$.  Applying
Lemma~\ref{str:prefix} with domain $E$ and global ideal
extension $H_h$, we obtain
\[
 \{x\in E:\tau(x)<\operatorname{pos}(q)\}
       =E\cap B_q(H_h).
\]
Because $q<v$, every job of $W$ occurring strictly before
$q$ lies in $E$.  Inserting the identity
$E=W\cap B_v(H_0)$ therefore yields
\[
 \{x\in W:\tau(x)<\operatorname{pos}(q)\}
       =W\cap B_v(H_0)\cap B_q(H_h).
\]

If $q=v$, this is the terminal endpoint, which does not
belong to the open domain $E$.  Thus we do \emph{not} apply
the separator lemma to $q$ on $E$.  The recorded terminal
threshold is $h=N+1$, so $L_h=J$ and $H_h=H_0$.
The center's separator identity above directly gives
\[
 \{x\in W:\tau(x)<\operatorname{pos}(q)\}
    =E=W\cap B_v(H_0)
     =W\cap B_v(H_0)\cap B_q(H_h).
\]
Here $q=v$, so the two factors are the same set.

Consequently, in \emph{both} cases the global formula
\begin{equation}
 {\displaystyle
 A_q:=B_v(H_0)\cap B_q(H_h)
      =B_v(J\setminus D_a)\cap
        B_q\bigl((J\setminus D_a)\cap L_h\bigr),}
 \qquad B_r(H):=P_r\cup(H\setminus\widehat D_r)
 \label{deleted:str:op-mark-global}
\end{equation}
satisfies the working-frame prefix identity
\eqref{deleted:str:working-prefix}:
\[
 W\cap A_q
  =\{x\in W:\tau_{\mathrm{work}}(x)
                    <\operatorname{pos}_{\mathrm{work}}(q)\}.
\]
The rank cutoff must be placed \emph{inside the ideal
argument} of $B_q$.  Indeed,
\[
 B_q(H_h)=P_q\cup\bigl((H_0\cap L_h)\setminus\widehat D_q\bigr),
\]
whereas intersecting $B_q(H_0)$ with $L_h$ would also
intersect the $P_q$ term with $L_h$.  The separator property
does not require every earlier high job in $P_q$ to have
rank below $h$, so that alternative expression is not
justified.

We count atomic occurrences, including repeated ones.
The factor $B_v(H_0)$ has three leaves
$P_v,D_a,\widehat D_v$.  The factor $B_q(H_h)$ has four
leaves $P_q,D_a,R_h,\widehat D_q$; recall that
$L_h=J\setminus R_h$ costs just one leaf.  Hence $A_q$
uses at most seven leaves.  If the working frame is the
original input frame, set $C_q=A_q$.  If it is reversed
relative to the input frame, the input-time prefix is the
\emph{working-time suffix}, with the marked slot removed.
Thus, by \eqref{deleted:str:frame-prefix-conversion}, the correct
global cut is
\[
 {\displaystyle
 C_q:=J\setminus\Bigl(
       \bigl[B_v(H_0)\cap B_q(H_h)\bigr]
       \cup\operatorname{jobs}(q)\Bigr).}
\]
The parentheses are essential: the marked slot must be
\emph{inside} the complemented union.  Indeed, the
working-time prefix, the marked slot, and the working-time
suffix partition $W$; the last of these is exactly
$W\cap C_q$.  The extra boundary-set atom
$\operatorname{jobs}(q)$ adds one leaf, so in either
orientation the global cut has at most eight leaves.

Combining the two types of marks gives the uniform bound
\[
 \max\{m+6,8\}=m+6\qquad(m\ge3).
\]
The cut formula need not itself be an ideal of $J$;
Definition~\ref{af:def-hierarchy} only requires that its
intersection with $W$ be the indicated input-time prefix.
This completes all five assertions.

\end{revisionold}
\begin{revisionnew}
\emph{(5) Global cuts for marked prefixes.}
We describe both temporal sides of a separator directly. This avoids an
additional boundary-set leaf when its working frame is reversed relative
to the original input frame.

Suppose that an actual slot $z$ separates a cutoff $S$ on a domain $U$,
and let $H\subseteq J$ satisfy $U\cap H=S$. Define
\[
 B_z(H)=P_z\cup(H\setminus\widehat D_z),\qquad
 T_z(H)=D_z\cup\bigl((J\setminus H)\setminus\widehat P_z\bigr).
\]
Then $U\cap B_z(H)$ is the strict working-time prefix of $z$ by
Lemma~\ref{str:prefix}, and $U\cap T_z(H)$ is its strict working-time
suffix. To see the latter identity, every low job after $z$ belongs to
$D_z$, and every high job after $z$ belongs to $J\setminus H$ but not to
$\widehat P_z$. Conversely, a job of $D_z$ is after $z$, while a selected
high job can be neither in $z$ nor before it: in the latter case the
earlier-high separator property would put it in $P_z$. Thus both formulas
are exact on the domain. If $H$ uses $k$ leaves, each formula uses at
most $k+2$ leaves.

\emph{Same-direction marks.}
Let $z$ be a retained actual separator from Step~1, including the center.
For a strict retained advance, choose its pre-promotion cutoff $S_z$;
for the center, choose the exhausted cutoff $W\cap P_b$.
In each case $z$ separates $S_z$ on the whole parent $W$.
Let $F_z$ be the flat formula for the unpromoted carry at that cutoff. The global ideal $H_z=P_b\setminus F_z$ satisfies
$W\cap H_z=S_z$ and uses at most $m+1$ leaves. Hence the prefix and suffix
formulas $B_z(H_z)$ and $T_z(H_z)$ both use at most $m+3$ leaves. Choose
$C_z=B_z(H_z)$ if the working frame is the input frame, and
$C_z=T_z(H_z)$ otherwise. In both cases $W\cap C_z$ is exactly the
strict input-time prefix. The later opposite-side normalization changes
no job's side of a same-direction mark.

\emph{Opposite-side marks.}
Work in the reversed frame used in Step~3, and put
\[
 W=(a,b),\quad E=(a,v),\quad X=W\setminus D_a,\quad
 H_0=J\setminus D_a,
\]
where $v$ is the center and separates $X$ on all of $W$.
For a retained mark $q<v$, let
\[
 L_h=J\setminus R_h,\qquad H_h=H_0\cap L_h,
 \qquad S_h=W\cap H_h.
\]
The slot $q$ separates $S_h\cap E$ on $E$. Define
\begin{align*}
 A_q&=P_q\cup\Bigl(H_h\setminus
                    (\widehat D_q\cup\widehat D_v)\Bigr),\\
 Z_q&=D_q\cup\widehat D_v\cup
                    \Bigl((J\setminus H_h)\setminus\widehat P_q\Bigr).
\end{align*}
We claim that $W\cap A_q$ is the strict working-time prefix of $q$ and
$W\cap Z_q$ its strict working-time suffix.

For the prefix, every earlier high job lies in $E$ and hence in $P_q$;
every earlier low job belongs to $H_h$ and avoids both weak successor
cones. Conversely, a selected job of $W\cap H_h=S_h\subseteq X$ cannot
be at or after $v$, because the center's separator property would put
it in $\widehat D_v$. It therefore lies in $E$, where separation at $q$
excludes its being at or after $q$. The term $P_q$ contains only jobs
before $q$, by feasibility.

For the suffix, $D_q$ and $\widehat D_v$ contain only jobs strictly after
$q$, using $q<v$. A selected high job cannot be in $q$, and if it were
before $q$ the separator on $E$ would put it in $P_q$. Conversely, a
high job after $q$ is selected by the last term. A low job after $q$
but before $v$ belongs to $D_q$ by the local separator, and a low job
at or after $v$ belongs to $\widehat D_v$ by the center separator.
These arguments prove both identities. Notice that $L_h$ restricts
only the low predicate $H_h$, not the term $P_q$: an earlier high job
need not have rank below $h$.

The formula for $A_q$ uses the five occurrences
$P_q,D_a,R_h,\widehat D_q,\widehat D_v$; that for $Z_q$ uses
$D_q,\widehat D_v,D_a,R_h,\widehat P_q$.
If $q=v$, use the center directly: the terminal cutoff is $X$, so
$B_v(H_0)$ and $T_v(H_0)$ give the two sides with three leaves each.
In particular, the displayed formula for $Z_q$ for $q<v$ is not used
at $q=v$, where its weak-center term would include the marked slot.
Choose the prefix formula when this working frame equals the input
frame, and the suffix formula otherwise.

All these are global Boolean formulas whose restrictions to $W$ are
the required temporal prefixes; the formulas need not be global ideals.
Rewriting their cones and rank conditions in the original atomic
language preserves leaf counts. Combining the two types of marks gives
\[
 K_{\mathrm{cut}}\le\max\{m+3,5\}=m+3\qquad(m\ge3).
\]
This completes all five assertions.

\end{revisionnew}
\else
\emph{(5) Global cuts for marked prefixes.}
We describe both temporal sides of a separator directly. This avoids an
additional boundary-set leaf when its working frame is reversed relative
to the original input frame.

Suppose that an actual slot $z$ separates a cutoff $S$ on a domain $U$,
and let $H\subseteq J$ satisfy $U\cap H=S$. Define
\[
 B_z(H)=P_z\cup(H\setminus\widehat D_z),\qquad
 T_z(H)=D_z\cup\bigl((J\setminus H)\setminus\widehat P_z\bigr).
\]
Then $U\cap B_z(H)$ is the strict working-time prefix of $z$ by
Lemma~\ref{str:prefix}, and $U\cap T_z(H)$ is its strict working-time
suffix. To see the latter identity, every low job after $z$ belongs to
$D_z$, and every high job after $z$ belongs to $J\setminus H$ but not to
$\widehat P_z$. Conversely, a job of $D_z$ is after $z$, while a selected
high job can be neither in $z$ nor before it: in the latter case the
earlier-high separator property would put it in $P_z$. Thus both formulas
are exact on the domain. If $H$ uses $k$ leaves, each formula uses at
most $k+2$ leaves.

\emph{Same-direction marks.}
Let $z$ be a retained actual separator from Step~1, including the center.
For a strict retained advance, choose its pre-promotion cutoff $S_z$;
for the center, choose the exhausted cutoff $W\cap P_b$.
In each case $z$ separates $S_z$ on the whole parent $W$.
Let $F_z$ be the flat formula for the unpromoted carry at that cutoff. The global ideal $H_z=P_b\setminus F_z$ satisfies
$W\cap H_z=S_z$ and uses at most $m+1$ leaves. Hence the prefix and suffix
formulas $B_z(H_z)$ and $T_z(H_z)$ both use at most $m+3$ leaves. Choose
$C_z=B_z(H_z)$ if the working frame is the input frame, and
$C_z=T_z(H_z)$ otherwise. In both cases $W\cap C_z$ is exactly the
strict input-time prefix. The later opposite-side normalization changes
no job's side of a same-direction mark.

\emph{Opposite-side marks.}
Work in the reversed frame used in Step~3, and put
\[
 W=(a,b),\quad E=(a,v),\quad X=W\setminus D_a,\quad
 H_0=J\setminus D_a,
\]
where $v$ is the center and separates $X$ on all of $W$.
For a retained mark $q<v$, let
\[
 L_h=J\setminus R_h,\qquad H_h=H_0\cap L_h,
 \qquad S_h=W\cap H_h.
\]
The slot $q$ separates $S_h\cap E$ on $E$. Define
\begin{align*}
 A_q&=P_q\cup\Bigl(H_h\setminus
                    (\widehat D_q\cup\widehat D_v)\Bigr),\\
 Z_q&=D_q\cup\widehat D_v\cup
                    \Bigl((J\setminus H_h)\setminus\widehat P_q\Bigr).
\end{align*}
We claim that $W\cap A_q$ is the strict working-time prefix of $q$ and
$W\cap Z_q$ its strict working-time suffix.

For the prefix, every earlier high job lies in $E$ and hence in $P_q$;
every earlier low job belongs to $H_h$ and avoids both weak successor
cones. Conversely, a selected job of $W\cap H_h=S_h\subseteq X$ cannot
be at or after $v$, because the center's separator property would put
it in $\widehat D_v$. It therefore lies in $E$, where separation at $q$
excludes its being at or after $q$. The term $P_q$ contains only jobs
before $q$, by feasibility.

For the suffix, $D_q$ and $\widehat D_v$ contain only jobs strictly after
$q$, using $q<v$. A selected high job cannot be in $q$, and if it were
before $q$ the separator on $E$ would put it in $P_q$. Conversely, a
high job after $q$ is selected by the last term. A low job after $q$
but before $v$ belongs to $D_q$ by the local separator, and a low job
at or after $v$ belongs to $\widehat D_v$ by the center separator.
These arguments prove both identities. Notice that $L_h$ restricts
only the low predicate $H_h$, not the term $P_q$: an earlier high job
need not have rank below $h$.

The formula for $A_q$ uses the five occurrences
$P_q,D_a,R_h,\widehat D_q,\widehat D_v$; that for $Z_q$ uses
$D_q,\widehat D_v,D_a,R_h,\widehat P_q$.
If $q=v$, use the center directly: the terminal cutoff is $X$, so
$B_v(H_0)$ and $T_v(H_0)$ give the two sides with three leaves each.
In particular, the displayed formula for $Z_q$ for $q<v$ is not used
at $q=v$, where its weak-center term would include the marked slot.
Choose the prefix formula when this working frame equals the input
frame, and the suffix formula otherwise.

All these are global Boolean formulas whose restrictions to $W$ are
the required temporal prefixes; the formulas need not be global ideals.
Rewriting their cones and rank conditions in the original atomic
language preserves leaf counts. Combining the two types of marks gives
\[
 K_{\mathrm{cut}}\le\max\{m+3,5\}=m+3\qquad(m\ge3).
\]
This completes all five assertions.

\fi
\end{proof}

\paragraph{Completion of the structural theorem.}
We have now obtained two \emph{different} types of globally
valid descriptions: an exact formula for each node set and a
possibly different formula for each marked prefix cut.  The
latter need not describe a node and need not be a global ideal.

\begin{proof}[Proof of Theorem~\ref{af:structure}]
Proposition~\ref{str:annotated} provides a finite hierarchy of
feasible full-slot intervals.  Every nonempty node has a marked
slot and every nonempty child has at most $|W|-m$ jobs.
At the root the set $W=J$ is a one-leaf atom.  Every other
node is a genuine child of its parent, so
Lemma~\ref{str:localization} represents its job set by an
exact global Boolean formula and Lemma~\ref{str:budget}(4)
bounds its leaf count by $\revchange{4m+12}{4m+5}$.
Consequently every node set lies in
$\mathcal F_{\mathrm{node}}=\mathcal F_{\revchange{4m+12}{4m+5}}$, the family
constructed by Algorithm~1.

For a marked full slot $z$ of any node $W$, use
Lemma~\ref{str:budget}(5) in the original input frame.
There is a global formula $C_z$ with at most $\revchange{m+6}{m+3}$ leaves
such that
\[
 W\cap C_z
  =\{x\in W:\tau_{+}(x)<\operatorname{pos}_{+}(z)\}.
\]
Thus $C_z\in\mathcal F_{\mathrm{cut}}=\mathcal F_{\revchange{m+6}{m+3}}$.
Here $\tau_+$ is the time map of the node's chosen complete
witness after it has been oriented in the original input frame.
Different hierarchy nodes may use different complete witnesses;
no single time map is being imposed across all nodes.
If a child's witness was constructed in reversed time,
reverse its full-slot order when interpreting it as the
input-frame witness.  The induced precedence order is then
respected by the definition of simultaneous frame reversal.
The marked slots still define full temporal gaps, although
their order is reversed when necessary.

Definition~\ref{af:def-hierarchy} requires exactly the two
memberships just established: each node set belongs to
$\mathcal F_{\mathrm{node}}$, and each local marked prefix
is the intersection of that node with a global cut in
$\mathcal F_{\mathrm{cut}}$.  It does \emph{not} require
$W\cap C_z$ itself to belong to either family.  Therefore
the annotated hierarchy becomes a hierarchy described by
$(\mathcal F_{\mathrm{node}},\mathcal F_{\mathrm{cut}})$,
which proves Theorem~\ref{af:structure}.

The bounds depend only on $m$, not on the depth of the
hierarchy.  The scheduling algorithm enumerates the two
Boolean-description families; it neither constructs the
normalized witness schedules nor checks the semantic
invariants used in this existence argument.
\end{proof}

\section{Summary and open problem}
\label{rt:improving-runtime}

We have given an explicit deterministic exact algorithm for unit-job
precedence scheduling on every fixed number $m\ge3$ of identical
machines. The structural hierarchy has at most $m-1$ surviving birth
groups and at most $m$ active entries at a node. Exact global
localization and the separate treatment of local flat carries give
the node family $\mathcal F_{\revchange{4m+12}{4m+5}}$ and the prefix-cut family
$\mathcal F_{\revchange{m+6}{m+3}}$. Exact binary tries and an acyclic state-path
evaluation make these structural descriptions algorithmic.

The algorithm decides a deadline, constructs a feasible witness
when one exists, and finds an optimum schedule by deadline search.
Its optimization bound is
\[
 O(L)+m^{O(m^2)}n^{6m^2+13m+1}\log n
\]
on the stated word RAM, and
\[
 O(L)+m^{O(m^2)}n^{6m^2+13m+1}\log^2 n
\]
on a deterministic Turing machine with a fixed number of tapes.
The direct multitape implementation batches exact set queries and
reachability updates using sequential scans, sorting, and merging.
It therefore retains the RAM polynomial degree while accounting
explicitly for sequential memory access.
For $m=3,4,5$, the polynomial degrees in both models are
$94,149,216$, respectively, apart from the displayed logarithmic
factors. A single deadline test saves one logarithmic search factor
in either model. The analysis includes family generation, all
unsuccessful tests, exact lookup, and schedule reconstruction.

\paragraph{Further improvement.}
An extended manuscript, in preparation for posting on
arXiv~\cite{jansen-extended}, uses additional structural, counting, and other arguments. These yield
an improved RAM and multitape bound
\[
 \mathrev{O(L)+2^{O_\gamma(m^2)} n^{3m^2-3m+6+ \frac{m^2}{4(m+1)} + \gamma m}}
\]
For $m=3,4$, and $5$ machines, the respective polynomial degrees (with fixing $\gamma = 0.01$) in $n$ are $24.5925$, $42.84$  and
$67.092$. The improved algorithms and proofs will be presented in an extended version.


\paragraph{Open problem.}
Can the quadratic dependence of the exponent on the number of machines
be reduced to linear dependence? More precisely, does there exist a
computable function $g$ and an absolute constant $c$ such that exact
optimization and schedule construction are possible in
\[
                    O(L)+g(m)\mathrev{n^{cm}}?
\]
\hl{This formulation is equivalent to the one with base }$\mathrev{n+2}$\hl{: the factor }$\mathrev{2^{cm}}$\hl{ can be absorbed into the arbitrary computable function }$\mathrev{g(m)}$\hl{.}
The present description families use $O(m)$ boundary atoms, each
specified by an $m$-job set. Reducing this information, or sharing its
choices across descriptions, is a possible route, but the current
proofs do not give an $O(m)$ exponent.

The charging argument in Lemma~\ref{str:groups} bounds how many
birth groups survive, but does not express the historical boundary
sets defining their entries in terms of a common smaller collection
of choices. These historical sets enter the exact global qualifier
union $Q'$ used in Lemma~\ref{str:localization}. The shorter flat carry
formulas agree with the corresponding qualifier sets only after
restriction to their specified domains, so they cannot simply
replace $Q'$ in the global localization identities. Consequently,
the current enumeration in Lemma~\ref{af:family-count} still contains
the factor $(N+2)^{mK}$, with $K=O(m)$. One possible route is to
construct a smaller sufficient family that shares boundary choices
while preserving the global identities required by the proof.

\paragraph{Parameterized and conditional running-time lower bounds.} Bodlaender and Fellows \cite{bodlaender-fellows1995} proved that precedence-constrained unit-job scheduling is $\mathrm{W[2]}$-hard when parameterized by the number $m$ of machines. A stronger quantitative lower bound follows from the work of Chen, Huang, Kanj, and Xia \cite{chen-huang-kanj-xia2006}, who establish hardness under linear FPT-reductions and apply their framework to this scheduling problem. Their published classification excludes a uniform exact algorithm with running time $f(m)L^{o(m)}$, for any computable function $f$, unless $\mathrm{FPT}=\mathrm{W[1]}$. Here $L$ denotes the input length. This rules out a sublinear dependence on $m$ in the exponent in the uniform parameterized setting, but does not determine the optimal polynomial degree for any particular fixed number of machines.

\paragraph{Preliminary stronger lower bounds.} In separate ongoing work, we investigate quantitative lower bounds for fixed $m$. A new two-epoch reduction from balanced Orthogonal Vectors yields the following conditional claim: for every fixed $m\ge4$ and $\varepsilon>0$, the balanced $(2m-5)$-Orthogonal Vectors hypothesis excludes an exact algorithm with running time $O(L+n^{2m-5-\varepsilon})$. In particular, this gives the proposed thresholds $n^{3-o(1)}$ for four machines and $n^{5-o(1)}$ for five machines. The construction produces precedence graphs of maximum undirected degree three, treewidth at most two, and pathwidth at most four. Related reductions based on a sharp Dominating Set hypothesis lead to the same exponent, while weaker satisfiability hypotheses give conditional restrictions on uniform algorithms with parameter-dependent prefactors. These results are preliminary; their complete verification and proofs are deferred to a separate manuscript.

Determining the optimal exponent remains open. In particular, the known upper bound is quadratic in $m$, whereas the parameterized lower bound excludes sublinear exponents for uniform algorithms under $\mathrm{FPT}\ne\mathrm{W[1]}$. The preliminary Orthogonal Vectors reduction suggests a more precise linear threshold of $2m-5$ for each fixed $m\ge4$, subject to its stated fine-grained hypothesis.


\paragraph{Recent developments in fine-grained complexity.}
The choice of complexity assumptions has become
particularly delicate. Recent preprints by OpenAI
and Anthropic claim deterministic subexponential-time
algorithms for 3-SAT, challenging ETH and consequently
SETH. Other new algorithms challenge the standard
fine-grained hypotheses for 3SUM, odd-arity $k$-SUM,
$k$-Clique, and Hyperclique; see
\cite{alman-williams2026,chan-itakura-ma-yu2026}
and the recent work of Vassilevska Williams and
Zamir, and Fischer \cite{vassilevska-williams-zamir2026,fischer2026}. These developments are recent,
and the most far-reaching claims have not yet
undergone comprehensive independent verification.
Weaker assumptions, such as the Satisfiability
Hypothesis (SH), equivalently Weak ETH, have been
proposed as alternatives \cite{brunner-et-al2026}. 
The status of other
quantitative hypotheses, including suitable
balanced Orthogonal Vectors and even-arity
$k$-SUM hypotheses, must be assessed separately.
Our conditional scheduling lower bounds therefore
state their assumptions explicitly rather than
relying on ETH or SETH.



\Needspace{16\baselineskip}
\begin{samepage}
\section*{Declaration of Use of Generative AI}
\label{ai:methodology}

The development of this work involved extensive and iterative discussions
with OpenAI's ChatGPT. The author formulated the research questions,
contributed mathematical and algorithmic ideas, identified relevant
techniques and connections to previous research, and guided the overall
direction of the investigation. ChatGPT assisted in 
organizing the presentation, and preparing 
portions of the manuscript
in LaTeX. 
The author takes full
responsibility for the scientific content, mathematical correctness,
originality, attribution and conclusions of the manuscript.
\end{samepage}

\end{document}